\documentclass[a4paper,11pt]{article}
\usepackage{jcappub}

\usepackage{hyperref}
\usepackage{verbatim}
\usepackage{tabularx}
\usepackage{orcidlink}

\usepackage{amsmath}
\usepackage{appendix}

\newcommand{\vect}[1]{\boldsymbol{#1}}

\newcommand{\ahod}[0]{\textsc{AbacusHOD} }
\newcommand{\absum}[0]{\textsc{AbacusSummit} }
\newcommand{\compaso}[0]{\textsc{CompaSO} }

\title{\boldmath 
The DESI One-Percent survey: exploring the Halo Occupation Distribution of Emission Line Galaxies with \textsc{AbacusSummit} simulations}

\author[1,\orcidlink{0000-0003-4349-6424}]{Antoine~Rocher}
\author[1,\orcidlink{0009-0000-6063-6121}]{Vanina~Ruhlmann-Kleider}
\author[1]{Etienne~Burtin}
\author[2,\orcidlink{0000-0002-5992-7586}]{Sihan~Yuan}
\author[1]{Arnaud~de~Mattia}
\author[3,4,5]{Ashley~J.~Ross}
\author[6]{Jessica~Aguilar}
\author[7,\orcidlink{0000-0001-6098-7247}]{Steven~Ahlen}
\author[8,\orcidlink{0000-0002-3757-6359}]{Shadab~Alam}
\author[9,\orcidlink{0000-0001-9712-0006}]{Davide~Bianchi}
\author[10]{David~Brooks}
\author[11,\orcidlink{0000-0002-5954-7903}]{Shaun~Cole}
\author[12]{Kyle~Dawson}
\author[13,\orcidlink{0000-0002-1769-1640}]{Axel~de~la~Macorra}
\author[10]{Peter~Doel}
\author[14]{Daniel~J.~Eisenstein}
\author[5,\orcidlink{0000-0003-2371-3356}]{Kevin~Fanning}
\author[15,16,\orcidlink{0000-0002-2890-3725}]{Jaime~E.~Forero-Romero}
\author[17,\orcidlink{0000-0002-9853-5673}]{Lehman~H.~Garrison}
\author[6,\orcidlink{0000-0003-3142-233X}]{Satya~Gontcho~A~Gontcho}
\author[18,19,\orcidlink{0000-0001-9938-2755}]{Violeta~Gonzalez-Perez}
\author[6]{Julien~Guy}
\author[6,20,\orcidlink{0000-0002-2312-3121}]{Boryana~Hadzhiyska}
\author[21,\orcidlink{0000-0003-1197-0902}]{ChangHoon~Hahn}
\author[3,22,5]{Klaus~Honscheid}
\author[6,\orcidlink{0000-0003-3510-7134}]{Theodore~Kisner}
\author[6,\orcidlink{0000-0003-1838-8528}]{Martin~Landriau}
\author[23,\orcidlink{0000-0003-2999-4873}]{James~Lasker}
\author[6,\orcidlink{0000-0003-1887-1018}]{Michael~E.~Levi}
\author[24,\orcidlink{0000-0003-4962-8934}]{Marc~Manera}
\author[25,\orcidlink{0000-0002-1125-7384}]{Aaron~Meisner}
\author[26,24]{Ramon~Miquel}
\author[27,\orcidlink{0000-0002-2733-4559}]{John~Moustakas}
\author[28]{Eva-Maria~Mueller}
\author[29,\orcidlink{0000-0001-8684-2222}]{Jeffrey~A.~Newman}
\author[30,\orcidlink{0000-0001-6590-8122}]{Jundan~Nie}
\author[31,32,33,\orcidlink{0000-0002-0644-5727}]{Will~J.~Percival}
\author[6,34,20]{Claire~Poppett}
\author[35]{Fei~Qin}
\author[36]{Graziano~Rossi}
\author[37,38,39,\orcidlink{0000-0002-1609-5687}]{Lado~Samushia}
\author[40,\orcidlink{0000-0002-9646-8198}]{Eusebio~Sanchez}
\author[6]{David~Schlegel}
\author[41,42]{Michael~Schubnell}
\author[43,\orcidlink{0000-0002-6588-3508}]{Hee-Jong~Seo}
\author[42,\orcidlink{0000-0003-1704-0781}]{Gregory~Tarl\'{e}}
\author[13,\orcidlink{0000-0003-3841-1836}]{Mariana~Vargas-Maga\~na}
\author[25]{Benjamin~A.~Weaver}
\author[44]{Jiaxi~Yu}
\author[38,\orcidlink{0000-0001-6847-5254}]{Hanyu~Zhang}
\author[12,\orcidlink{0000-0003-1887-6732}]{Zheng~Zheng}
\author[30,\orcidlink{0000-0002-4135-0977}]{Zhimin~Zhou}
\author[30,\orcidlink{0000-0002-6684-3997}]{Hu~Zou}

\affiliation[1]{IRFU, CEA, Universit\'{e} Paris-Saclay, F-91191 Gif-sur-Yvette, France}
\affiliation[2]{SLAC National Accelerator Laboratory, Menlo Park, CA 94305, USA}
\affiliation[3]{Center for Cosmology and AstroParticle Physics, The Ohio State University, 191 West Woodruff Avenue, Columbus, OH 43210, USA}
\affiliation[4]{Department of Astronomy, The Ohio State University, 4055 McPherson Laboratory, 140 W 18th Avenue, Columbus, OH 43210, USA}
\affiliation[5]{The Ohio State University, Columbus, 43210 OH, USA}
\affiliation[6]{Lawrence Berkeley National Laboratory, 1 Cyclotron Road, Berkeley, CA 94720, USA}
\affiliation[7]{Physics Dept., Boston University, 590 Commonwealth Avenue, Boston, MA 02215, USA}
\affiliation[8]{Tata Institute of Fundamental Research, Homi Bhabha Road, Mumbai 400005, India}
\affiliation[9]{Dipartimento di Fisica ``Aldo Pontremoli'', Universit\`a degli Studi di Milano, Via Celoria 16, I-20133 Milano, Italy}
\affiliation[10]{Department of Physics \& Astronomy, University College London, Gower Street, London, WC1E 6BT, UK}
\affiliation[11]{Institute for Computational Cosmology, Department of Physics, Durham University, South Road, Durham DH1 3LE, UK}
\affiliation[12]{Department of Physics and Astronomy, The University of Utah, 115 South 1400 East, Salt Lake City, UT 84112, USA}
\affiliation[13]{Instituto de F\'{\i}sica, Universidad Nacional Aut\'{o}noma de M\'{e}xico,  Cd. de M\'{e}xico  C.P. 04510,  M\'{e}xico}
\affiliation[14]{Center for Astrophysics $|$ Harvard \& Smithsonian, 60 Garden Street, Cambridge, MA 02138, USA}
\affiliation[15]{Departamento de F\'isica, Universidad de los Andes, Cra. 1 No. 18A-10, Edificio Ip, CP 111711, Bogot\'a, Colombia}
\affiliation[16]{Observatorio Astron\'omico, Universidad de los Andes, Cra. 1 No. 18A-10, Edificio H, CP 111711 Bogot\'a, Colombia}
\affiliation[17]{Center for Computational Astrophysics, Flatiron Institute, 162 5\textsuperscript{th} Avenue, New York, NY 10010, USA}
\affiliation[18]{Centro de Investigaci\'{o}n Avanzada en F\'{\i}sica Fundamental (CIAFF), Facultad de Ciencias, Universidad Aut\'{o}noma de Madrid, ES-28049 Madrid, Spain}
\affiliation[19]{Instituto de F\'{\i}sica Te\'{o}rica (IFT) UAM/CSIC, Universidad Aut\'{o}noma de Madrid, Cantoblanco, E-28049, Madrid, Spain}
\affiliation[20]{University of California, Berkeley, 110 Sproul Hall \#5800 Berkeley, CA 94720, USA}
\affiliation[21]{Department of Astrophysical Sciences, Princeton University, Princeton NJ 08544, USA}
\affiliation[22]{Department of Physics, The Ohio State University, 191 West Woodruff Avenue, Columbus, OH 43210, USA}
\affiliation[23]{Department of Physics, Southern Methodist University, 3215 Daniel Avenue, Dallas, TX 75275, USA}
\affiliation[24]{Institut de F\'{i}sica d’Altes Energies (IFAE), The Barcelona Institute of Science and Technology, Campus UAB, 08193 Bellaterra Barcelona, Spain}
\affiliation[25]{NSF's NOIRLab, 950 N. Cherry Ave., Tucson, AZ 85719, USA}
\affiliation[26]{Instituci\'{o} Catalana de Recerca i Estudis Avan\c{c}ats, Passeig de Llu\'{\i}s Companys, 23, 08010 Barcelona, Spain}
\affiliation[27]{Department of Physics and Astronomy, Siena College, 515 Loudon Road, Loudonville, NY 12211, USA}
\affiliation[28]{Department of Physics and Astronomy, University of Sussex, Brighton BN1 9QH, U.K}
\affiliation[29]{Department of Physics \& Astronomy and Pittsburgh Particle Physics, Astrophysics, and Cosmology Center (PITT PACC), University of Pittsburgh, 3941 O'Hara Street, Pittsburgh, PA 15260, USA}
\affiliation[30]{National Astronomical Observatories, Chinese Academy of Sciences, A20 Datun Rd., Chaoyang District, Beijing, 100012, P.R. China}
\affiliation[31]{Department of Physics and Astronomy, University of Waterloo, 200 University Ave W, Waterloo, ON N2L 3G1, Canada}
\affiliation[32]{Perimeter Institute for Theoretical Physics, 31 Caroline St. North, Waterloo, ON N2L 2Y5, Canada}
\affiliation[33]{Waterloo Centre for Astrophysics, University of Waterloo, 200 University Ave W, Waterloo, ON N2L 3G1, Canada}
\affiliation[34]{Space Sciences Laboratory, University of California, Berkeley, 7 Gauss Way, Berkeley, CA  94720, USA}
\affiliation[35]{Korea Astronomy and Space Science Institute, 776, Daedeokdae-ro, Yuseong-gu, Daejeon 34055, Republic of Korea}
\affiliation[36]{Department of Physics and Astronomy, Sejong University, Seoul, 143-747, Korea}
\affiliation[37]{Abastumani Astrophysical Observatory, Tbilisi, GE-0179, Georgia}
\affiliation[38]{Department of Physics, Kansas State University, 116 Cardwell Hall, Manhattan, KS 66506, USA}
\affiliation[39]{Faculty of Natural Sciences and Medicine, Ilia State University, 0194 Tbilisi, Georgia}
\affiliation[40]{CIEMAT, Avenida Complutense 40, E-28040 Madrid, Spain}
\affiliation[41]{Department of Physics, University of Michigan, Ann Arbor, MI 48109, USA}
\affiliation[42]{University of Michigan, Ann Arbor, MI 48109, USA}
\affiliation[43]{Department of Physics \& Astronomy, Ohio University, Athens, OH 45701, USA}
\affiliation[44]{Ecole Polytechnique F\'{e}d\'{e}rale de Lausanne, CH-1015 Lausanne, Switzerland}

\emailAdd{antoine.rocher@cea.fr}

\abstract{
The One-Percent survey of the Dark Energy Spectroscopic Instrument collected $\sim 270k$ emission line galaxies (ELGs) at $0.8<z<1.6$. The high completeness 
of the sample allowed the clustering to be measured down to scales never probed before, 
0.04 Mpc/$h$ in $r_p$ for the projected 2-point correlation function (2PCF) 
and 0.17 Mpc$/h$ in galaxy pair separation $s$ for the 
2PCF monopole and quadrupole. The most striking feature of the measurements is a strong signal 
at the smallest scales, below 0.2 Mpc$/h$ in $r_p$ and 1 Mpc$/h$ in $s$.
We analyse these data  
in the halo occupation distribution framework. We consider different distributions for central galaxies, a standard power law for satellites with no condition on the presence of a central galaxy and explore several extensions of these models. For all considered models, the mean halo mass of the sample is found to be $\log_{10} \left\langle M_h \right\rangle \sim 11.9$.
We obtain a satellite mean occupation function which agrees with physically motivated ELG models only if we introduce central-satellite conformity, meaning that the satellite occupation is conditioned by the presence of central galaxies of the same type. 
To achieve in addition a good modelling of the clustering between 0.1 and 1 Mpc$/h$ in $r_p$, 
we allow for 
ELG positioning outside of the halo virial radius 
and find 
$0.5\%$ of ELGs residing
in the outskirts of 
halos. 
Furthermore, 
the satellite velocity dispersion inside halos is found to be 
$\sim 30\%$ larger than 
that of the halo dark matter particles.
 These are the main findings of our work. 
 We investigate assembly bias as a function of halo concentration, local density or  local density anisotropies and observe no significant 
 change in our results. We split the data sample in two redshift bins 
 and report no significant evolution 
 with redshift. Lastly, 
 changing the cosmology in the modelling impacts only slightly 
 our results.
}

\begin{document}
\maketitle
\flushbottom
\clearpage

\section{Introduction}
The large structure of the Universe is a key cosmological probe.
Large scale clustering measurements allow cosmological parameters to be constrained through 
measurements of the baryon acoustic oscillation (BAO) scale and more recently from full shape analyses of 2-point statistics which translate into constraints on both BAO scales and the linear growth rate of structure $f\sigma_8$ through redshift space distortions. At small scales, clustering measurements are invaluable to study the galaxy-halo connection, which is the aim of this paper, and to provide precise measurements of  $f\sigma_8$. They also allow realistic mock catalogues to be produced that 
are used to prepare large scale analyses and to assess their systematic errors related to the complexity of galaxy formation and evolution. As was shown in~\cite{Alam21} for a stage III spectroscopic survey, systematic uncertainties related to the galaxy-halo connection were found at that time to be negligible with respect to other sources of systematic errors. As those are expected to be reduced in DESI, galaxy-halo connection studies are becoming increasingly important to derive robust systematic error budgets for cosmological analyses.

The galaxy-halo connection 
can be studied in various approaches as described in~\cite{Wechsler2018}. 
In our paper, we use the halo occupation distribution (HOD) framework, which provides a simple and empirical method to populate dark matter halos from N-body simulations with galaxies. We rely on the \absum simulation~\citep{AbacusSummit} and apply the HOD method 
to the small scale clustering measurements performed on the ELG sample 
collected by 
DESI 
during the so-called One-Percent Survey, the last two months of survey validation. DESI is a robotic, fibre-fed, highly multiplexed spectroscopic instrument that operates on the Mayall 4-meter telescope at Kitt Peak National Observatory~\citep{Levi2013, DESIOverview2022}. DESI can obtain simultaneous spectra of almost 5000 objects over a $\sim 3$ degree field~\citep{2016arXiv161100037D,Silber2023,DESIcorrector2023} and is currently conducting a five-year survey of about a third of the sky, 
to obtain approximately 40 million galaxies and quasars~\citep{2016arXiv161100036D}. ELGs are the main target class of DESI which should collect 17 million ELG redshifts after five years of observations. ELGs are targeted to measure the growth rate of structure $f\sigma_8$ and the BAO scale at intermediate redshifts $0.6<z<1.6$ with high statistical precision.
In only two months of observation DESI observed $\sim 270k$ ELGs in this range, 
which is the largest ELG spectroscopic sample to date. 
This sample is part of the Early Data Release (EDR) of the DESI collaboration~\citep{EDR}.

The ELG galaxy-halo connection has been previously studied using different approaches \citep[e.g][]{GonzalezPerez18, Avila20, Gao2022, Okumura2021, Lin2023}. From these studies, ELGs are expected to reside in dark matter (DM) halos of mass $\sim 10^{12} M_{\odot}$, and the occupation of DM halos decreases when the halo mass increases. In the literature, a sizeable fraction of ELGs are considered to be satellites. Depending on the galaxy-halo connection modelling, the satellite fraction varies from $\sim 10\%$ to $\sim 30\%$. The purpose of this paper is to study the HOD of the DESI ELG sample from the One-Percent survey. Based on previous work \citep{GonzalezPerez18, Avila20, Alam21}, we study 4 different distributions for central galaxy occupation and allow for different modelling of galaxy satellite velocities. The impact of secondary parameters, such as assembly bias \citep{Gao_White07}, based on the halo concentration, local halo density and density anisotropies is also investigated.  We also test for departures from a pure NFW profile for satellite positioning. Finally, we study variation of the HOD parameters considering 3 different cosmologies. We use a HOD fitting pipeline based on Gaussian processes~\citep{GPpipeline} to derive the best-fitting parameters to DESI ELG data and the corresponding posterior contours. 

This paper is part of a series of papers dedicated to the analysis of the galaxy--halo connection from the DESI One-Percent Survey data. This paper addresses a study of the novel ELG sample in the HOD framework. A companion HOD publication based on \absum simulations deals with the Luminous Red Galaxy (LRG) and quasar (QSO) samples~\citep{abacusLRGQSO}. 
Other methods were also investigated.
\cite{novelAM} studies LRGs and ELGs using a stellar-mass-split abundance matching  applied on \textsc{CosmicGrowth} \citep{2019SCPMA..6219511J}. In parallel, there is a series of Subhalo-abundance matching (SHAM) analyses using different simulations, the \textsc{UNIT} \citep{UNIT} simulation in~\cite{inclusiveSHAM} and the 
\textsc{Uchuu} \citep{2021Ishiyama} one in 
\cite{Uchuu4tracers}. 
These papers present a significant variety of methodologies and mock products appropriate for a large scope of applications. 

The outline of the paper is as follows. Section~\ref{sec:data} presents the data sample studied in this work, introduces the clustering statistics and discusses the clustering measurements used in this work. Sections~\ref{sec:models} and~\ref{sec:simulation} introduce the standard HOD models suitable for ELG clustering and the simulations used to create mock catalogues according to HOD models. Section~\ref{sec:methods} summarises the 
fitting pipeline used in the paper. Section~\ref{sec:standard}
and~\ref{sec:extensions} report the results obtained with the standard HOD models and extensions of these. Section~\ref{sec:zevolution} tests for redshift evolution of the results and Section~\ref{sec:cosmology} explores their dependence with the underlying cosmology. We compare our findings with those of companion papers on the same data sample in Section~\ref{sec:companion} and conclude in Section~\ref{sec:conclusion}.
 
\section{Data}
\label{sec:data}

The ELG data sample studied in this paper was collected during the One-Percent survey of DESI that was conducted at the end of the Survey Validation (SV) campaign in April and May of 2021~\citep{SV} before the start of the main survey operations. 
Before SV, DESI had proven its ability to simultaneously measure spectra at 5000 specific sky locations, with fibres placed accurately using robotic positioners populating the DESI focal plane \citep{Silber2023}. During SV, the DESI data and operation teams proved their ability to optimise operations~\citep{DESIoperations2023} and to efficiently process the spectra through the DESI spectroscopic pipeline \citep{Guy2023}. 
To obtain a high completeness, the footprint of the One-Percent survey was defined as a set of 20 non-overlapping 
regions of the sky, called rosettes in the following, which were observed at least 11 times each. Starting from an initial target list~\citep{Myers2023}, the DESI fibre assignment algorithm~\citep{DESIOverview2022} places each fibre onto a reachable target within a 6~mm patrol radius around the nominal fibre position, so that only a subset of the targets can be observed in every visit. This leads to incompleteness, which decreases rapidly with the number of visits.

The One-Percent survey covered 140 deg$^2$ 
with final target selection algorithms and depths similar to those of the main survey.
The ELG target selection~\citep{ELGpaper} focuses on the redshift range $0.6<z<1.6$ and is designed to select galaxies with strong spectral emission lines. The 
[\small{O} {\rm II}] doublet emission line allows precise redshifts to be measured by DESI.
Higher priority in the spectroscopic measurements is given to objects expected in the interval $1.1<z<1.6$ 
where ELGs are the main tracer of DESI.
Between 0.2 and 1.5 degrees from the 
centre of each rosette, spectra were successfully obtained for 94.5\% of ELG targets, while targets outside these regions were observed with fewer visits and thus lower completeness in fibre assignment. This sample is very appropriate to study 
ELGs inside halos as it provides precise measurements of the galaxy clustering down to very small scales.

\begin{figure}[htbp]
\centering
\includegraphics[width=0.7\textwidth]{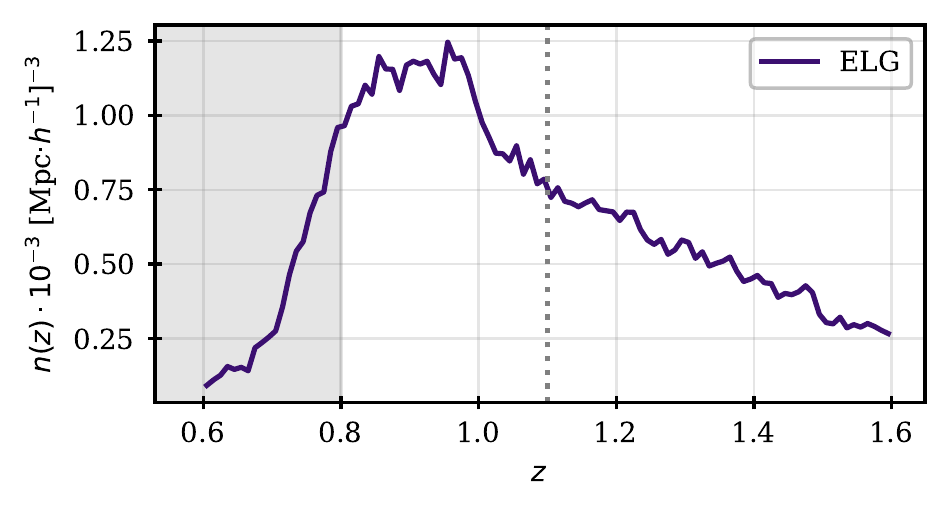}
\caption{\label{fig:density} Number density 
of the DESI One-Percent survey ELG data sample, as a function of redshift (corrected for completeness). The shaded region is not used in this work. The dotted line indicates the mean completeness-weighted redshift of the sample, $\bar z=1.13$. 
 }
 \end{figure}

In the following, we make use of the ELG sample collected during the One-Percent survey and spectroscopically confirmed in the redshift range from 0.8 to 1.6, over which the radial density distribution varies slowly, as shown in Figure~\ref{fig:density}. The region $z<0.8$ is not considered in our final sample as it exhibits dependence of the redshift density with respect to imaging depth.
This sample contains 244k spectra and has a mean density of $7\times10^{-4} (h/{\rm Mpc})^3$. 
Section 4 of \cite{EDR} describes the construction of all EDR large scale structure catalogues, including the random catalogues. We highlight a few details on the random construction here. Random catalogues are first produced by the DESI targeting team \citep{Myers2023} at a fixed density. These randoms are input to the DESI fibre assignment software, which processes them through each observed tile matching the state used during observations. Only the randoms that were identified as observable by this process are kept. Additional vetoes are applied for bright stars and other foregrounds (see \cite{EDR} for the precise details). Therefore, the density and radial distribution of each rosette are the same (within only Poisson fluctuations). Finally, redshifts are assigned randomly in the random catalogue using the redshifts from the galaxy sample to keep the same $n(z)$ distribution.

\subsection{Clustering statistics}
\label{sec:statistics}
The clustering of the selected sample was studied in configuration space with 2-point statistics defined as follows.
We first define the galaxy two-point correlation function in two dimensions, $\xi(r_p,\pi)$, where $\pi$ and $r_p$ are the galaxy pair separation components along and perpendicular to the line-of-sight, respectively. We then introduce the projected correlation function, $w_p(r_p)$, obtained by integrating $\xi(r_p,\pi)$ over the line-of-sight, as well as the monopole 
and quadrupole 
of the two point correlation function $\xi(s,\mu)$, where $s$ is the galaxy pair separation and $\mu$ the cosine of the angle between the line-of-sight and galaxy separation vector:  
\begin{equation}
\begin{split}
w_p(r_p) &=  \int_{\pi_{min}}^{\pi_{max}} \xi(r_p,\pi) d\pi \\
\xi_l(s) &= \frac{2l+1}{2}\int_{-1}^1 \xi(s,\mu) {\it P}_l(\mu) d\mu
\end{split}
\label{stat}
\end{equation}
where $l=[0,2]$ and ${\it P}_l(\mu)$ is the Legendre polynomial of order $l$. 

We rely on \textsc{pycorr}\footnote{\url{https://github.com/cosmodesi/pycorr}}, the DESI implementation of the \textsc{Corrfunc} package~\citep{Corrfunc}, to compute $\xi(r_p,\pi)$ and $\xi(s,\mu)$. For mocks, which are obtained from cubic boxes, they are computed with the natural estimator which compares galaxy pair counts to the expected pair count for a uniform distribution in the box volume. For data, the Landy-Szalay estimator is used~\citep{LS}.  For mocks, the $z$ axis is chosen as line-of-sight for the application of redshift space distortions.

For $\xi(r_p,\pi)$, we use 17 logarithmic bins in $r_p$ between 0.04 and 32 Mpc$/h$ and 80 linear bins in $\pi$
between -40 and 40 Mpc$/h$. The same binning and range are used for $w_p(r_p)$ so that $\pi_{max} = - \pi_{min} = 40\,{\rm Mpc}/h$ in~Eq.\eqref{stat}.
For the multipoles, we use 27 logarithmic bins in $s$ between 0.17 and 32 Mpc$/h$ and 200 linear bins in $\mu$ between -1 and 1. Finally, in the galaxy pair count computation, whether in data or simulation, the fiducial cosmology used to convert galaxy redshift into distances is the Planck 2018 baseline $\Lambda_{\mathrm{CDM}}$ best-fit result~\citep{Planck2018} with $h=0.6736, A_s=2.0830 \times 10^{-9}, n_s=0.9649, \omega_{cdm}=0.12, \omega_{b}=0.02237$ 
and $\sigma_8=0.8079$.

\subsection{Clustering measurements}
\label{sec:clustering}

\begin{figure}[htbp]
\centering
\includegraphics[width=0.8\textwidth] {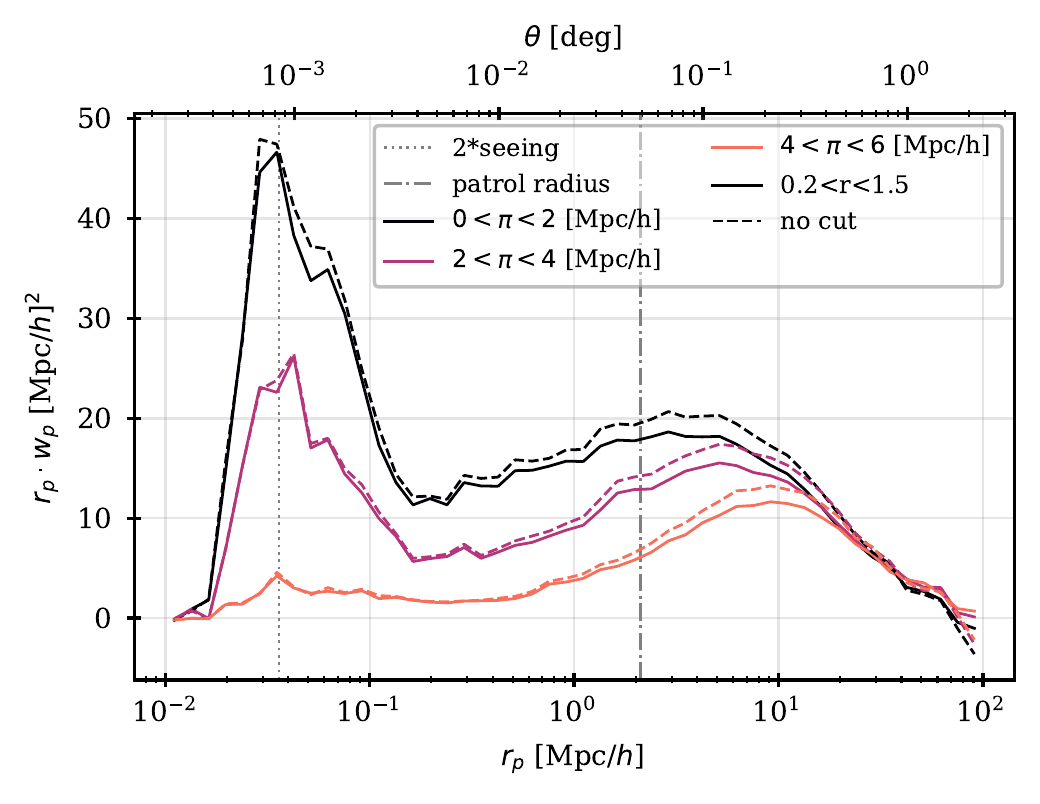} 
\caption{\label{fig:rp_pi} DESI clustering measurements for the One-Percent survey ELG data sample restricted to the redshift range $0.8<z<1.6$. The 2D correlation function in successive bins of $2 {\rm Mpc}/h$ in the galaxy-pair separation along the line-of-sight is shown as a function of the separation perpendicular to the line-of-sight, $r_p$. No correction weight has been applied. Measurements using the whole survey footprint (solid lines) are compared with measurements excluding the inner and outer regions of the rosettes where the survey was less incomplete (dashed lines). Also indicated are the  separation corresponding to the fibre patrol radius (dot-dashed grey line) and the limit corresponding to twice the mean survey seeing (dotted grey line). Below this limit, target blending cannot be resolved, leading to a loss of power. This plot demonstrates that the strong increase in power at small scales (below $0.2{\rm Mpc}/h$) is not due to the (slight) incompleteness of the One-Percent survey.
}
\end{figure} 
The clustering of the One-Percent survey ELG sample is first illustrated in Figure~\ref{fig:rp_pi} which shows the 2D correlation function 
in successive bins in 
$\pi$, as a function of 
$r_p$. 
This figure highlights several key points about the ELG clustering measurement from the One-Percent survey. 
A strong signal at small scales 
is visible at separations larger than $r_p=0.03 {\rm Mpc}/h$, the threshold below which target blending makes clustering measurements unreliable. The strong up-turn in the small-scale clustering appears for transverse separations below $r_p\sim0.2 {\rm Mpc}/h$ and is mostly due to separations along the line-of-sight  below $\pi=3 {\rm Mpc}/h$. In this region, the incompleteness of the survey may bias the clustering measurements due to fibre collisions if the number of visits is limited. 
To illustrate this, measurements corresponding to the complete survey footprint (solid lines) are compared with those excluding regions of lower completeness, outside the interval between 0.2 and 1.5 degrees from the field centre of each rosette (dashed lines). The strong up-turn in the clustering signal appears also in the latter measurements, showing that incompleteness due to fibre collisions is not responsible for the strong ELG clustering at small scales that we observe.

\begin{figure}[htbp]
\includegraphics[width=1.\textwidth]{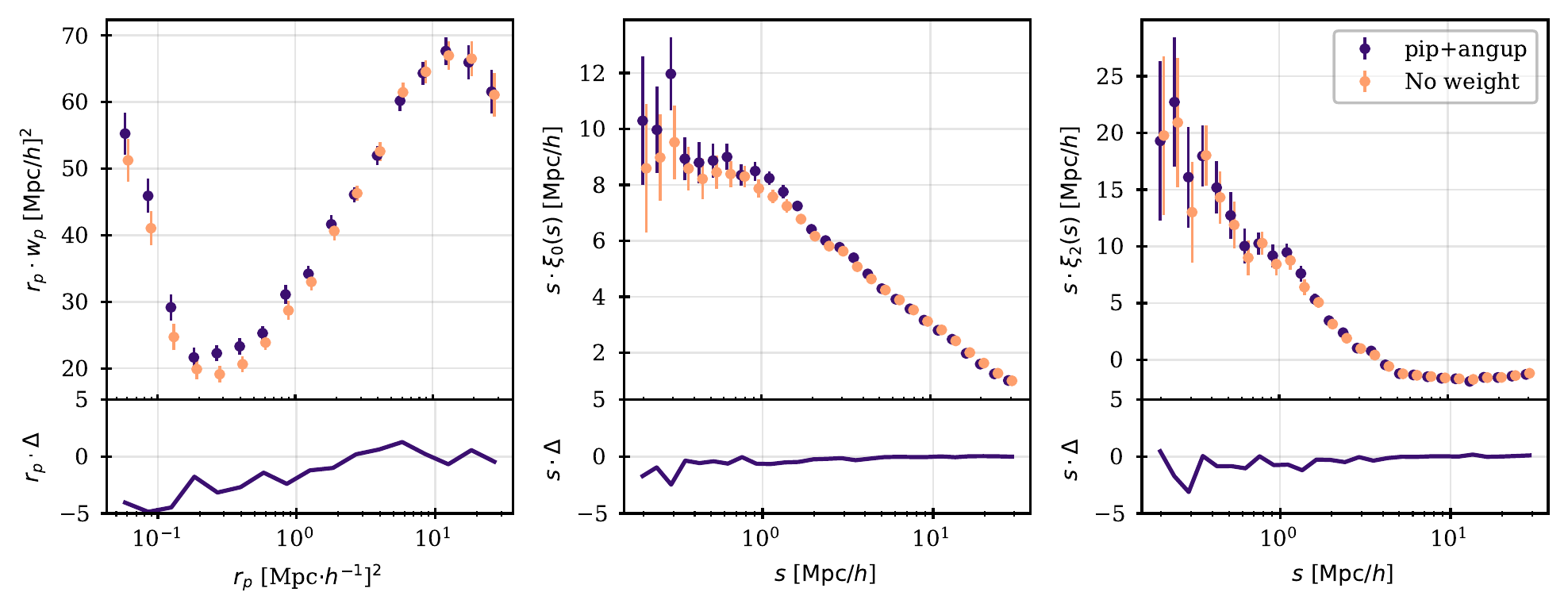}
\caption{\label{fig:weights}  
{\it Top:} DESI clustering measurements for the One-Percent survey ELG data sample restricted to the redshift range $0.8<z<1.6$ and to regions of high completeness. Data are shown without (\textit{orange}) and with (\textit{purple}) tiling incompleteness weights. Errors are jackknife statistical uncertainties. {\it Bottom:} difference between clustering measurements without and with fibre assignment weights applied.
}
\end{figure} 

Incompleteness can however bias the clustering measurements, especially at small scales. To limit that effect, we 
restrict the ELG sample to those targets observed in regions of high completeness, that is between 0.2 and 1.5 degrees from the field centre of each rosette. This reduces the sample size by $12\%$ leaving 215k galaxies. Residual density inhomogeneities in that sample due to residual fibre assignment inefficiencies are corrected with a weighting procedure. This is illustrated in Figure~\ref{fig:weights} which shows how the clustering evolves when fibre assignment corrections are applied.
These corrections are twofold. Incompleteness weights for individual galaxies and for galaxy pairs are computed as inverse probabilities of being targeted in a set of multiple realisations of the actual fibre assignment algorithm, as described in~\cite{Bianchi2017}. These weights are completed by angular up-weighting to treat the case of galaxy pairs with zero selection probability in the previous computation, as described in~\cite{Percival2017}. 
\cite{Mohammad20} showed that this weighting scheme provides an unbiased clustering down to $\sim0.1$Mpc$/h$.
As anticipated from the removal of the regions of lower completeness in the rosettes, the fibre assignment weights have a small impact on the measured clustering, visible essentially at scales lower than $\sim 1{\rm Mpc}/h$, as a result of fibre collisions. 

The clustering measurements can be biased by other systematic effects, such as density inhomogeneities due to imaging conditions or redshift failure rate variations with spectroscopic observing conditions. We checked that the correcting weights associated to these effects have a negligible impact on the small-scale clustering measurements. Besides the completeness weights, we also apply FKP weights~\citep{Feldman1994} that minimise variance in the clustering measurements. We also checked with simulations that the small footprint of the One-Percent Survey has a negligible integral constraint effect~\citep{deMattia2019} on these measurements in the range of separations used in this study. 


Several observational studies of 
ELG clustering at redshifts $z \sim 1$ have already been published, using data from various surveys, such as COSMOS~\citep{Tinker2013}, VIPERS~\citep[e.g][]{Favole2016,Gao2022}, eBOSS~\citep{Guo2019,Avila20,Alam21,Lin2023} or the HSC SSP survey~\citep{Okumura2021}. But the clustering measurements provided by the DESI One-Percent survey are the first redshift space measurements that go down to transverse separation scales as low as $0.03{\rm Mpc}/h$, offering a direct and robust measurement of the one-halo term contribution to the clustering.

In the following, we first use the clustering measurements in the redshift range $0.8<z<1.6$ to test different prescriptions for the ELG HOD modelling and then fit the most promising one to the measurements, splitting the sample in two redshift intervals in Section~\ref{sec:zevolution} to test for a possible HOD parameter evolution with redshift.

\section{Standard ELG HOD models }
\label{sec:models}
The standard HOD formalism describes the relation between galaxies and their dark matter halos as the probability that a halo with mass $M$ hosts $N$ such galaxies. Central and satellite galaxies are considered separately, with $\langle N_{cent}(M)\rangle$ and $\langle N_{sat}(M)\rangle$ their respective mean numbers per halo of a given halo mass.

\subsection{Models for central galaxies}
Based on previous studies of ELG clustering~\citep{Avila20,Alam21}, we retain four possible HOD prescriptions for central galaxies, one with a Gaussian shape and three different functions producing an asymmetric shape:
\begin{itemize}
\item a Gaussian HOD model (GHOD):
\begin{equation}
\label{GHOD}
\left\langle N_{cent}(M)\right\rangle = \frac{A_c}{\sqrt{2\pi}\sigma_{\textsc{m}}} \cdot e^{-\frac{(\log_{10}M-\log_{10}M_{c})^2}{2\sigma_{\textsc{m}}^2}}
\equiv \left\langle N^{GHOD}_{cent}(M)\right\rangle 
\end{equation}
\item a LogNormal HOD model (LNHOD): defining $x=\log_{10}M - (\log_{10}M_c-1)$, the prescription for central galaxies is:
\begin{equation}
\label{LNHOD}
\left\langle N_{cent}(M)\right\rangle = 
\frac{A_c}{\sqrt{2\pi}\sigma_{\textsc{m}}\cdot x} \cdot e^{-\frac{(\ln x)^2}{2\sigma_{\textsc{m}}^2}} \; \;\text{for $x>0$, and 0 otherwise} 
\end{equation}
\item  a Star Forming HOD model (SFHOD):
\begin{equation}
\left\langle N_{cent }(M)\right\rangle  = 
\begin{cases}
\left\langle N^{GHOD}_{cent}(M) \right\rangle 
&  M \leq M_{c} \\
\frac{A_c}{\sqrt{2 \pi} \sigma_{\textsc{m}}} 
\cdot\left(\frac{M}{M_c}\right)^\gamma &  M > M_{c}\end{cases}
\end{equation}
\item a modified High Mass Quenched model (mHMQ):
\begin{equation}
\label{HMQ}
\left\langle N_{cent}(M) \right\rangle =
\left\langle N^{GHOD}_{cent}(M)\right\rangle \cdot
\left[1+\operatorname{erf}\left(\frac{\gamma (\log_{10}M-\log_{10}M_{c})}{\sqrt{2}\sigma_{\textsc{m}}}\right)\right] 
\end{equation}
Note that this model is derived from the High Mass Quenched model of~\cite{Alam21} setting the quenching factor to infinity to only retain the asymmetric shape of the central distribution.

\end{itemize}

In the above formulas, $A_c$ sets the size of the central galaxy sample, $M_c$ is the characteristic mass 
for a halo to host a central galaxy, $\sigma_{\textsc{m}}$ is the width of the distribution and $\gamma$, if present, controls its asymmetry. 
A Bernoulli distribution with mean equal to $\left\langle N_{\mathrm{cent}}(M) \right\rangle$ is used to generate either 0 or 1 central galaxy per halo.

\subsection{Baseline model for satellite galaxies}
For satellite galaxies, we adopt the following HOD~\citep{Avila20,Alam21}:
\begin{equation}
\label{eq:power_law}
\left\langle N_{sat}(M)\right\rangle = A_s\bigg(\frac{M-M_0}{M_1}\bigg)^{\alpha}
\end{equation}
where $A_s$ sets the size of the satellite galaxy sample, $M_{0}$ is the cut-off halo mass from which satellites can be present and $\alpha$ controls the increase in satellite richness with increasing halo mass. $M_1$ is introduced for normalisation purpose and corresponds to the halo mass at which 1 satellite is expected if $A_s=1$ and $M_0$ is negligible w.r.t. $M_1$. 
Note that the above form (without the normalisation factor $A_s$) was first introduced in~\cite{Kravtsov2004} based on N-body simulations and in~\cite{Zheng2005} based on semi-analytical models and hydrodynamical simulations of galaxy formation, for it was found to provide a very good description of the occupation distribution of satellites predicted in this framework. The normalisation factor $A_s$ was introduced in later works as a way to model the incompleteness of the satellite sample. In this paper, we use both $A_c$ and $A_s$ to impose a density constraint to our HOD models, as explained in the next section. 

Throughout the paper, unless stated otherwise, the actual number of satellite galaxies as a function of halo mass is drawn from a Poisson distribution with mean equal to $\left\langle N_{sat}(M)\right\rangle$.
By default, several satellites can thus be present in the same halo, and satellites can be present even if there is no central galaxy in the halo.  We note that in such a case, classifying them as satellites may appear inappropriate, but is no more than a convenience to refer to the parametrisation used.
Beyond the above functional forms for the mean numbers of central and satellite galaxies as a function of halo mass, a prescription must be chosen to define how satellite positions and velocities are distributed. This is described in Section~\ref{sec:methods}. 

From the above equations, derived parameters can be calculated analytically, such as the expected total number density of the galaxy sample:
\begin{equation}
\bar{n}_{\mathrm{gal}} =\int \frac{\mathrm{d} n(M)}{\mathrm{d} M}\left[\left\langle N_{cent}(M)\right\rangle+\left\langle N_{sat}(M)\right\rangle\right] \mathrm{d} M
\label{ngal}
\end{equation}
the fraction of satellites:
\begin{equation}
f_{sat} = \frac{1}{\bar{n}_{\mathrm{gal}}} \int \frac{\mathrm{d} n(M)}{\mathrm{d} M}\left\langle N_{sat}(M)\right\rangle \mathrm{d} M
\label{fracsat}
\end{equation}
or the average halo mass of the sample:
\begin{equation}
\left\langle M_h \right\rangle = \frac{1}{\bar{n}_{\mathrm{gal}}} \int \frac{\mathrm{d} n(M)}{\mathrm{d} M}\left[\left\langle N_{cent}(M)\right\rangle+\left\langle N_{sat}(M)\right\rangle\right] M \mathrm{d} M
\label{meanmass}
\end{equation}
where $\frac{\mathrm{d} n(M)}{\mathrm{d} M}$ is the halo mass function, taken from the N-body simulation.
We also define an effective $M'_1$ mass parameter that is equivalent to the $M_1$ mass scale in the original parametrisation for satellite occupation without the $A_s$ parameter:
\begin{equation}
M'_1 \equiv \frac{M_1}{A_s^{1/\alpha}}
\end{equation}
$M'_1$ is the halo mass scale to have one satellite on average if $M_0$ is negligible w.r.t. $M'_1$.

\subsection{HOD free parameters and density constraint}
The HOD parameters are 
$A_c, M_c, \sigma_\mathrm{M}$ (and possibly $\gamma$)  for central galaxies and $A_s, M_0, \alpha, M_1$ for satellite galaxies. $M_1$ being degenerate with $A_s$ and
$\alpha$ cannot be constrained in the fits. Unless otherwise stated, it is fixed to a value of $10^{13} M_{\odot}/h$ in the fits described in this paper. 
The normalisation parameters $A_c$ and $A_s$ are used to impose a density constraint in the fitting procedure to match the density in DESI data, as explained below. All other parameters are left free to vary. 

The galaxy sample number density in Eq.~\eqref{ngal} is governed by both $A_c$ and $A_s$ and the fraction of satellites in Eq.~\eqref{fracsat} is controlled by their ratio. All other conditions being equal, the same clustering is obtained  whatever $A_c$ and $A_s$ values, provided their ratio is fixed. The density constraint is introduced in the following way. At each point in the HOD parameter space, we set $A_c$ to an initial value, while $A_s$ is sampled from a flat prior range. We compute the total number density in Eq.~\eqref{ngal} for these initial values of $A_c$ and $A_s$ and rescale them by the same factor (to preserve the clustering) in order to normalise the galaxy density to $10^{-3}(h/\rm{Mpc)}^{3}$, close to that of the DESI ELG sample. 
In our tables, we report $A_c$ initial values, best-fit values of $A_s$ which are unrescaled and we provide the corresponding rescaling factor used to set the density of the mocks to that of data. 
This factor is applied for the derived parameters and for mock creation.

\section{Simulation}
\label{sec:simulation}
To create mock catalogues from simulations according to the above HOD models, we rely on the \absum suite of high-accuracy cosmological N-body %
simulations~\citep{AbacusSummit} 
designed for the clustering analyses of DESI. We use the cleaned halo catalogues obtained with the \compaso 
algorithm~\citep{CompaSO} 
applied to these simulations.
The suite is defined primarily in the base Planck 2018 $\Lambda_{\rm CDM}$ best-fit 
cosmology~\citep{Planck2018} 
but contains also several variants, and proposes different resolutions and cubic box sizes. 

\begin{table}[thbp]
\centering
\resizebox{\columnwidth}{!}{
\begin{tabular}{|l|c|c|c|c|}\hline 
  usage  &  cosmology & box size &  resolution & realisations \\ \hline 
  baseline modelling & Planck 2018 $\Lambda_{\rm CDM}$ & 1.185 ${\rm Gpc}/h$ & 4096$^3$ & 1 \\  
  correlation matrix & Planck 2018 $\Lambda_{\rm CDM}$ & 0.5 ${\rm Gpc}/h$ & 1728$^3$ & 1800 \\
  cosmic variance & Planck 2018 $\Lambda_{\rm CDM}$ & 2 ${\rm Gpc}/h$ & 6912$^3$ & 25 \\
  high $N_{\mathrm{eff}}$ & $N_{\mathrm{eff}}=3.7$ & 1.185 ${\rm Gpc}/h$ & 4096$^3$  & 1 \\
  high $N_{\mathrm{eff}}$ cosmic variance & $N_{\mathrm{eff}}=3.7$ & 2 ${\rm Gpc}/h$ & 6912$^3$  & 6 \\
  low $\sigma_8$ & Planck 2018 with $\sigma_8=0.75$ & 1.185 ${\rm Gpc}/h$ & 4096$^3$  & 1  \\  
  low $\sigma_8$ cosmic variance & Planck 2018 with $\sigma_8=0.75$ & 2 ${\rm Gpc}/h$ & 6912$^3$  & 6  \\
  \hline 
\end{tabular}  
}
\caption{\label{tab:simulation} Cosmology, box size and mass resolution of the \absum simulations used in this work. The mass resolution is given as the number of particles in the box. The first column indicates the use of each set of simulations: baseline HOD modelling, correlation matrix for data, cosmic variance for the model covariance matrix. The last four sets are used to explore different cosmologies but with identical simulation initial conditions as in the baseline modelling.
}
\end{table} 

Table~\ref{tab:simulation} presents the subset of simulations used in this work. They all have the same resolution, that is
6912$^3$ particles in a box of 2 ${\rm Gpc}/h$ length, which corresponds to a particle mass of about 2$\times 10^9 M_{\odot}/h$. This ensures that halos are well resolved down to $10^{11}M_{\odot}/h$  giving $\sim$50 particles/halo~\citep{AbacusSummit}. Besides, the halos 
corresponding to best fitting results obtained in this work have a mass larger than $3\times10^{11}M_{\odot}/h$ which corresponds to 150 particles/halo. 
Note that throughout the paper, we define the halo mass as the number of particles in the halo multiplied by the particle mass.

\section{Fitting Methodology}
\label{sec:methods}
The HOD fitting pipeline used in this work is described in~\cite{GPpipeline}. It proceeds in two steps, HOD mock generation and HOD parameter fitting, based on Gaussian processes. 
The 
main features of the pipeline are summarised hereafter and the construction of the covariance matrix used in 
the fits is described afterwards.

\subsection{Pipeline based on Gaussian processes}
\label{sec:GPpipeline}
The fitting pipeline uses Gaussian processes (GP) to obtain a surrogate model of the likelihood surface describing the comparison of clustering measurements between data and HOD mocks. At each point of the HOD parameter space of a given model, DM halos from the 
baseline \absum 1.18~Gpc/$h$ cubic box (see Table~\ref{tab:simulation}) are populated with galaxies according to the HOD parameters, in order to generate mock catalogues. As explained in~\cite{GPpipeline}, mocks are created with a fixed galaxy density of $10^{-3}(h/$Mpc$)^{3}$, close to that of the actual ELG sample. In this process, the HOD prescriptions for the mean numbers of central and satellite galaxies described in Section~\ref{sec:models} are complemented by the following assumptions. 
Central galaxies are positioned at the centre of their halos. Satellite positions 
obey a Navarro-Frenk-White profile~\citep{NFW} 
using $r_{25}$, the radius of a sphere that contains 25$\%$ of the halo particles, as a proxy for $r_s$, the scale radius of the profile. Since the mass enclosed in a sphere of radius $r$ is divergent for the NFW profile, we further apply a cut-off at the halo virial radius $r=r_\mathrm{vir}$, taking $r_{98}$, the radius of a sphere containing 98$\%$ of the halo particles, as a proxy for $r_\mathrm{vir}$.  The above proxies were chosen because we observed that they provide a predicted clustering which is very close to that obtained with satellite positioning using DM particles, as illustrated in Appendix~\ref{app:proxies}. 
Satellite velocities are normally distributed around their mean halo velocity, with a dispersion equal to that of the halo dark matter particle velocities, rescaled by an extra free parameter denoted $f_{\sigma_v}$, following~\cite{Alam21}.

At each point of the parameter space,
we generate 20 mock catalogues and compare their clustering to that of data to produce one $\chi^2$ value per mock. 
The 20 $\chi^2$ values are then averaged and both the mean $\chi^2$ value and the standard deviation of the mean are fed into the GP.
The covariance matrix entering the $\chi^2$ definition contains a data component 
and a model component 
that accounts for the stochastic noise of the mock creation and the cosmic variance to be expected for 1.18~Gpc/$h$ cubic boxes. These two covariance matrices are discussed further in Section~\ref{sec:covmat}.  
In the $\chi^2$ computation, each of the covariance matrix components is corrected for the Hartlap effect~\citep{Hartlap07},
which predicts a biased estimate of the inverse covariance matrix entering the $\chi^2$ computation if there is a large difference between the number of measurements and the number of mocks used to estimate the covariance matrix.

Initial training of the GP is obtained from the $\chi^2$ values and errors computed on a given set of points. Based on the conclusions of~\cite{GPpipeline}, the training sample is obtained from Hammersley sampling of the HOD parameter space using flat priors. 
After initial training, the GP model of the likelihood surface is further improved by an iterative procedure adding one point to the training sample at each iteration. The added point is randomly chosen in Monte Carlo Markov chains (MCMC) sampling the GP prediction. This allows us to obtain both an accurate minimisation of the $\chi^2$ and reliable error contours in the HOD parameter space. We refer the reader to~\cite{GPpipeline} for the results of repeatability and accuracy tests of the method performed on simulations.

In the following, we use an initial training sample of 800 points from Hammersley sampling, followed by 800 iterations 
and check the fit convergence during the iterative step by means of the Kullback-Leibler (KL) divergence \citep{KL} between the MCMC chains. As shown in~\cite{GPpipeline}, instabilities in the GP likelihood surface estimate can be generated in the course of the iterative procedure, due to learning phases triggered in small regions of the parameter space by the addition of the extra point. These instabilities do not strongly impact the iterative evolution of the marginalised parameter values but affect the KL divergence. To check the fit convergence, we thus compute the KL divergence from cleaned MCMC chains, where points with  uncertainties above 10 in the $\chi^2$ value predicted by the GP have been removed. We consider a fit to be converged when the KL divergence is below 0.1 in a set of 20 consecutive iterations and we define the final iteration as the last iteration of the last such set of iterations. When no such set of consecutive iterations is found, the final iteration is the last iteration with a KL divergence below 0.1. 
The fit results are defined by the marginalised HOD parameter values
at that final iteration, with statistical uncertainties given by the  [$0.16-0.84$] quantiles of the parameter posteriors at that same iteration.

As 2-point statistics, the GP pipeline uses the projected correlation function, $w_p(r_p)$, as well as the monopole $\xi_0(s)$ and quadrupole $\xi_2(s)$ 
of the two point correlation function, as introduced in Section~\ref{sec:statistics}.

\subsection{Covariance matrix for data and model}
\label{sec:covmat}

\begin{figure}[tbp]
\includegraphics[width=1.\textwidth]{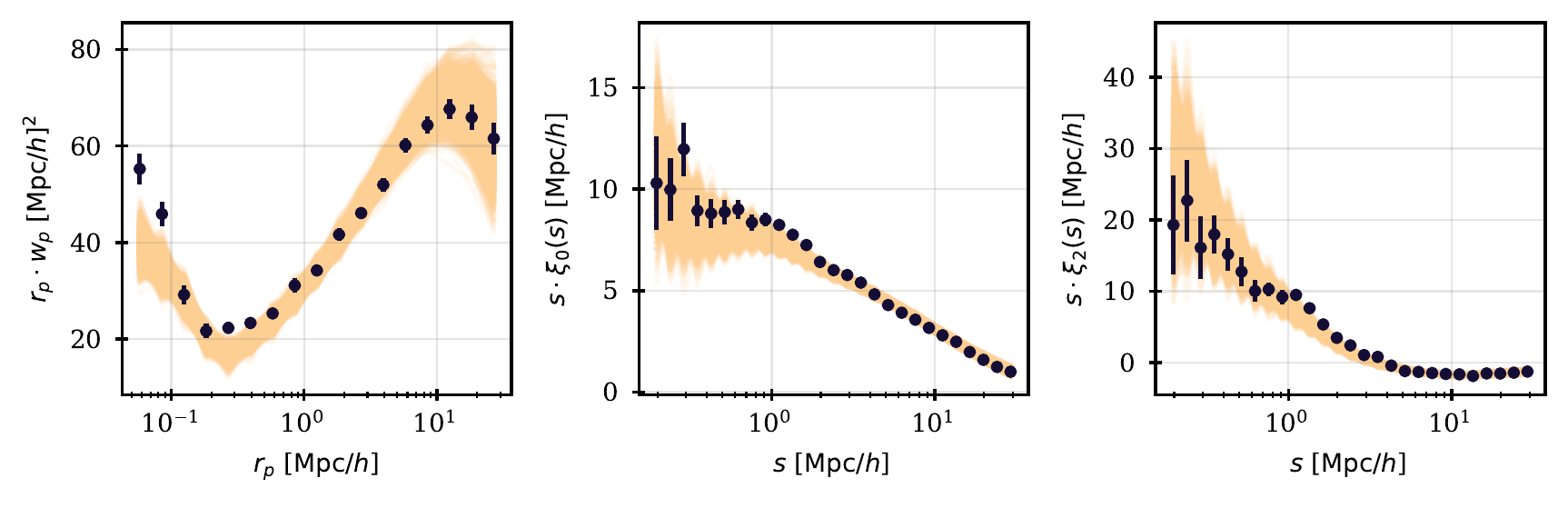}
\caption{\label{fig:HODforCovMat} DESI ELG clustering measurements from the One-Percent survey data sample. From left to right, we show the projected correlation function, the monopole and quadrupole of the correlation function. Data (dots with error bars) are compared to expectations (solid lines) from 1800 realisations of a HOD model obtained from a preliminary fit to these data using a pure Jackknife covariance matrix. Uncertainties are Jackknife errors. 
}
\end{figure}

\begin{figure}[htbp]
\centering
\includegraphics[width=0.8\textwidth]{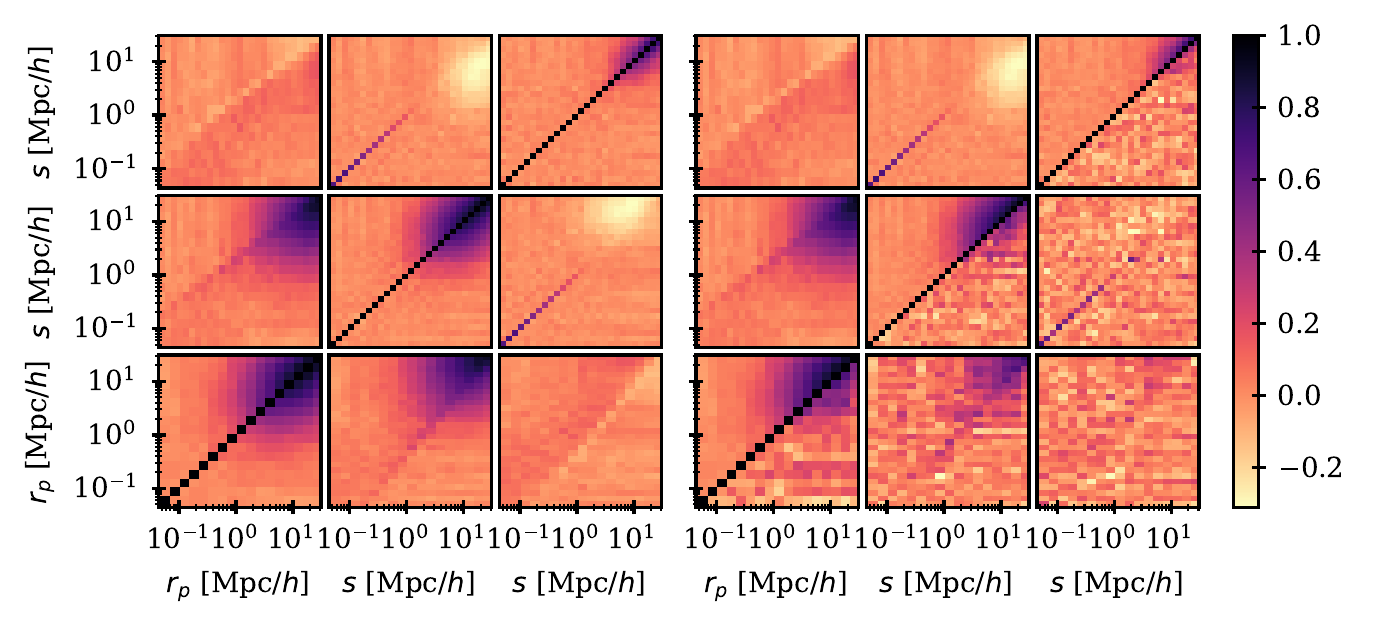}
\caption{\label{fig:MockBasedCovMat} {\it Left:} correlation matrix derived from 1800 mocks built from the HOD model in Figure~\ref{fig:HODforCovMat}. {\it Right:} correlations from the mock-based matrix (above the diagonal) compared with those from the pure Jackknife covariance matrix (below the diagonal).
}
\end{figure}

A data covariance matrix appropriate for the ELG clustering measurements used by the GP pipeline was derived applying the delete-one Jackknife method to the One-Percent survey footprint divided into 128 independent regions, the maximum number of large enough regions given the small extent of the footprint. 
The jackknife regions were defined using a K-means sampler that cuts the footprint into regions of similar size in RA/DEC, as implemented in the DESI package \texttt{pycorr}.
To recover an unbiased estimate of the covariance matrix, correction terms were applied as described in~\cite{Mohammad22}. 
As the off-diagonal terms of that matrix were affected by noise, a smooth correlation matrix was derived from simulation to replace that from data. For that purpose, we resorted to the 1800 small boxes from the \absum simulations in Table~\ref{tab:simulation} that allow cosmic variance to be included with good statistical precision. 

The result of a preliminary HOD fit to the data using the Jackknife covariance matrix was first used to populate the small box halos, with a density identical to that in the data sample.
Figure~\ref{fig:HODforCovMat} compares the clustering from the data with the calculated clustering for the mocks used to determine the correlation matrix. The binning of the three statistics is that defined in Section~\ref{sec:statistics}.
The off-diagonal terms of the mock-based correlation matrix are much smoother than the ones calculated from the data as shown in Figure~\ref{fig:MockBasedCovMat}.
In the following, we define the data covariance matrix of the HOD fits,  $\vect{C}_{data}$ from the mock-based correlation matrix, using the Jackknife diagonal errors to appropriately normalise variances and covariances.

The GP pipeline also considers a covariance matrix for the model, $\vect{C}_{model}$, as described in~\cite{GPpipeline}. To build it, correlations are assumed to have small variations over the HOD parameter space and we first compute a fixed correlation matrix from 1000 realisations of the HOD model under test, at a given reference point in the parameter space. When scanning the HOD parameter space, the model covariance matrix at each point is then obtained by normalising all terms of the previous correlation matrix by the quadratic sum of two sets of diagonal errors. The first set contains the variances of the clustering measurements over the 20 realisations drawn to compute the $\chi^2$ at the current point, and thus accounts for stochastic noise. The second set contains the variances of the clustering measurements obtained from 48 realisations of the HOD model at the reference point, each drawn from a different sub-cube of 1 Gpc$/h$ length cut out of 25 realisations of the same simulation box (see Table~\ref{tab:simulation}, third line) and corrected for volume effects. This set accounts for cosmic variance. 

In the GP pipeline, the $\chi^2$ computed at each point of the HOD parameter space is thus defined as follows:
\begin{equation}
\chi^2 = (\vect{\xi}_{data}-\vect{\xi}_{model})^\top\big[\vect{C}_{data}/(1-D_{data})+\vect{C}_{model}/(1-D_{model})\big]^{-1}(\vect{\xi}_{data}-\vect{\xi}_{model})
\end{equation}
where $\vect{\xi}$ is a vector of clustering measurements, 
$\vect{C}$ the corresponding covariance matrix and $D$ the 
Hartlap correction factor~\citep{Hartlap07} based on the number of mocks used to derive the corresponding correlation matrix. This is averaged over 20 HOD realisations.

\section{Standard HOD results}
\label{sec:standard}
\begin{figure}[tbp]
\includegraphics[width=1.\textwidth]
{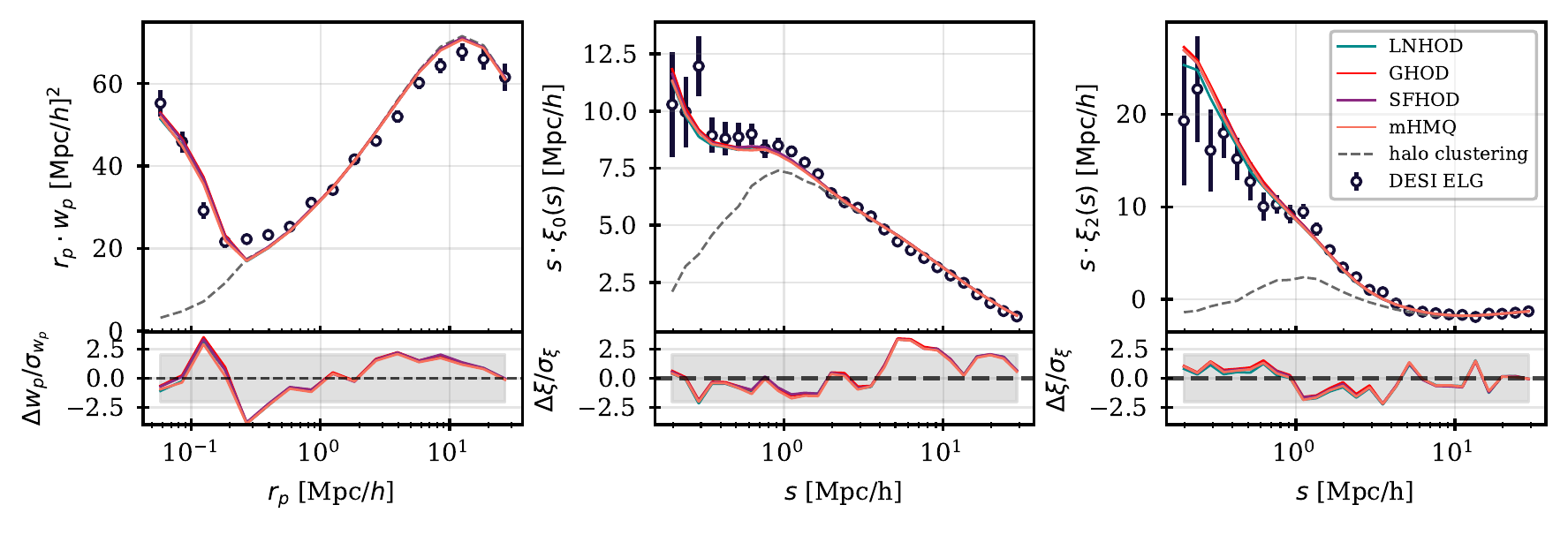}
\caption{\label{fig:standard_HOD} {\it Top:} DESI ELG clustering measurements from the One-Percent survey data sample, compared to best fitting standard HOD models obtained with the GP pipeline. 
The models in solid line correspond to different prescriptions for the central galaxies, keeping the standard power-law prescription for satellites. The model in dashed line is the pure halo clustering, showing that pairs of galaxies from the one-halo term have a strong impact on the clustering at the lowest scales.
Errors are Jackknife uncertainties only. {\it Bottom:} Fit residuals normalised by the diagonal errors of the full covariance matrix, that comprise Jackknife uncertainties for the data as well as stochastic noise and cosmic variance for the model, but no Hartlap factor corrections.
}
\end{figure}

Best fitting clustering from the GP pipeline are presented in Figure~\ref{fig:standard_HOD} and the corresponding best-fit values of the HOD parameters are summarised in Table~\ref{tab:standard_HOD}. We test the four prescriptions for central galaxies of Section~\ref{sec:models}, keeping the standard prescription for satellites (see Eq.~\eqref{eq:power_law} and Section~\ref{sec:methods}). 
Except for $M_1$ which is kept fixed in the fits, we used flat priors for all other HOD parameters. The exact prior ranges depend on the HOD models tested but we took care to choose them wide enough to get reasonably enclosed contours. Examples of prior ranges are shown in Appendix~\ref{app:contours}.

\begin{table}[htbp]
\centering
\begin{tabular}{|l|c|c|c|c|}\hline
  parameter \& $\chi^2$ &  LNHOD & GHOD & SFHOD & mHMQ \\ \hline
$A_c$ (resc.)  & $1$ (0.08) & $1$ (0.08)& $1$ (0.08)& $1$ (0.08) \\
$\log_{10}M_0$ & $11.78^{+0.04}_{-0.05}$ & $11.72^{+0.03}_{-0.04}$  & $11.73^{+0.03}_{-0.03}$ & $11.70^{+0.03}_{-0.03}$ \\
$A_s$ & $0.09^{+0.01}_{-0.01}$  & $0.08^{+0.04}_{-0.01}$ & $0.09^{+0.04}_{-0.02}$ &  $0.10^{+0.04}_{-0.03}$ \\ 
$\log_{10}M_{c}$ & $11.87^{+0.01}_{-0.01}$ & $11.89^{+0.02}_{-0.02}$ & $11.87^{+0.03}_{-0.03}$ & $11.72^{+0.06}_{-0.04}$\\  
$\alpha$ & $-0.28^{+0.03}_{-0.03}$ & $-0.31^{+0.08}_{-0.05}$ & $-0.28^{+0.06}_{-0.04}$ & $-0.26^{+0.08}_{-0.08}$ \\ 
$f_{\sigma_v}$ & $1.29^{+0.07}_{-0.06}$ & $1.23^{+0.06}_{-0.06}$ & $1.27^{+0.07}_{-0.07}$ & $1.27^{+0.07}_{-0.06}$ \\ 
$\sigma_M$ & $0.08^{+0.02}_{-0.01}$ & $0.11^{+0.02}_{-0.02}$ & $0.07^{+0.04}_{-0.02}$ & $0.22^{+0.08}_{-0.11}$\\
$\gamma$ & - & - & $-4.42^{+0.99}_{-0.76}$ & $7.06^{+1.33}_{-1.97}$ \\
\hline
$\log_{10} M'_{1}$ & 5.37 & 6.12 & 5.57 & 4.77  \\ 

$f_{sat}$ & $0.10^{+0.02}_{-0.02}$ & $0.12^{+0.03}_{-0.02}$ & $0.11^{+0.02}_{-0.02}$ & $0.12^{+0.02}_{-0.02}$ \\
$f_{1h}$ & $0.041^{+0.007}_{-0.005}$ & $0.040^{+0.005}_{-0.006}$ & $0.039^{+0.005}_{-0.006}$ & $0.039^{+0.009}_{-0.008}$ \\
$\log_{10} \left\langle M_h \right\rangle $ &  $11.87^{+0.01}_{-0.01}$ & $11.87^{+0.01}_{-0.01}$ & $11.88^{+0.01}_{-0.01}$ & $11.87^{+0.02}_{-0.01}$ \\ \hline
$\chi^2$ (ndf) &  $156.0 \pm 1.0$ (65)& $157.6 \pm 1.3$ (65) & $155.5 \pm 1.2$ (64) & $158.2 \pm 1.0$ (64) \\  \hline 
\end{tabular} 
\caption{\label{tab:standard_HOD} Results of standard HOD fits to the DESI ELG clustering measurements from the One-Percent survey. The first line provides the initial fixed value of $A_c$ and the rescaling factor applied to impose the density constraint in the fits. The following 
six or seven parameters are the free HOD parameters, 
the next four are derived parameters.
$\log_{10}M_1'$ is given for best-fit values of $\alpha$ and $A_s$ (the latter after rescaling). 
$f_{sat}$ is the fraction of galaxies which are satellites and $f_{1h}$ is the 
fraction of galaxies which are not alone in their halos. All masses are in units of $(M_{\odot}/h)$. 
}
\end{table} 

The four best fitting models provide similar expectations for the ELG clustering, which agree reasonably well with data. Features difficult to model correctly are the slope of the projected correlation function between $0.2$ and $10$ Mpc/$h$ and the bump at $s \sim 1-2$ Mpc/$h$ in the monopole and quadrupole. This partially explains the high $\chi^2$ values which average at $\sim$157 for 65 degrees of freedom, depending on the model. Since all models behave similarly, it implies that there are ingredients missing in the standard HODs for ELGs. This will be studied in the following sections.

Also shown in Figure~\ref{fig:standard_HOD} is the expected clustering computed from halos only, regardless of the galaxies they contain (dashed line). This highlights the fact that pairs of galaxies inside the same halo contribute, as expected, only at low scales in the three statistics.
This contribution constitutes the so-called one-halo term of the galaxy-halo connection and is essential to reproduce the strong clustering measured at small scales in our data, notably the strong up-turn of the projected correlation function at $r_p<0.3$ Mpc$/h$. Note however that between 0.3 and  $\sim$1 $\mathrm{Mpc}/h$ in $r_p$, the measured clustering is above the predicted halo clustering, meaning that the one-halo contribution arising from the NFW profile is not sufficient to describe the data in this region.

\begin{figure}[tbp]
\centering
\begin{tabular}{cc}
\includegraphics[width=0.47\textwidth]{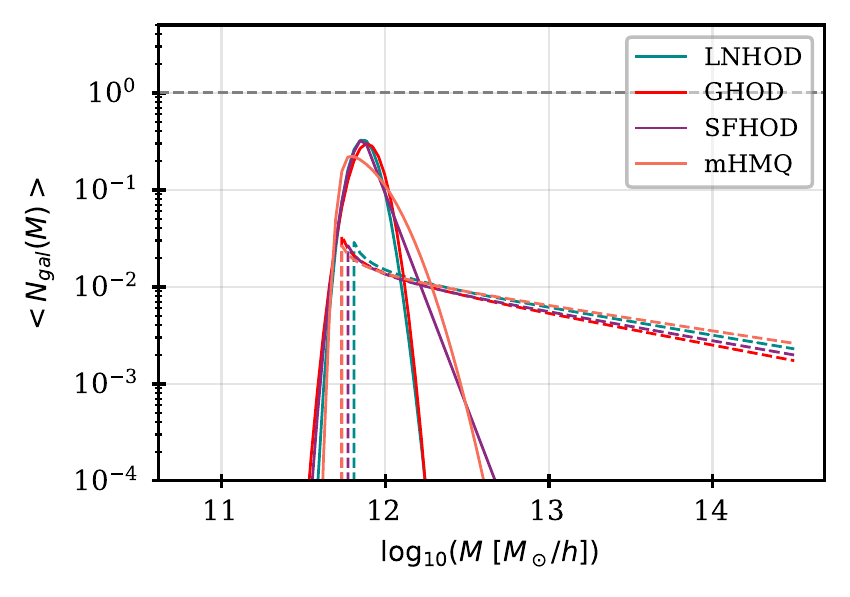} &
\includegraphics[width=0.5\textwidth]{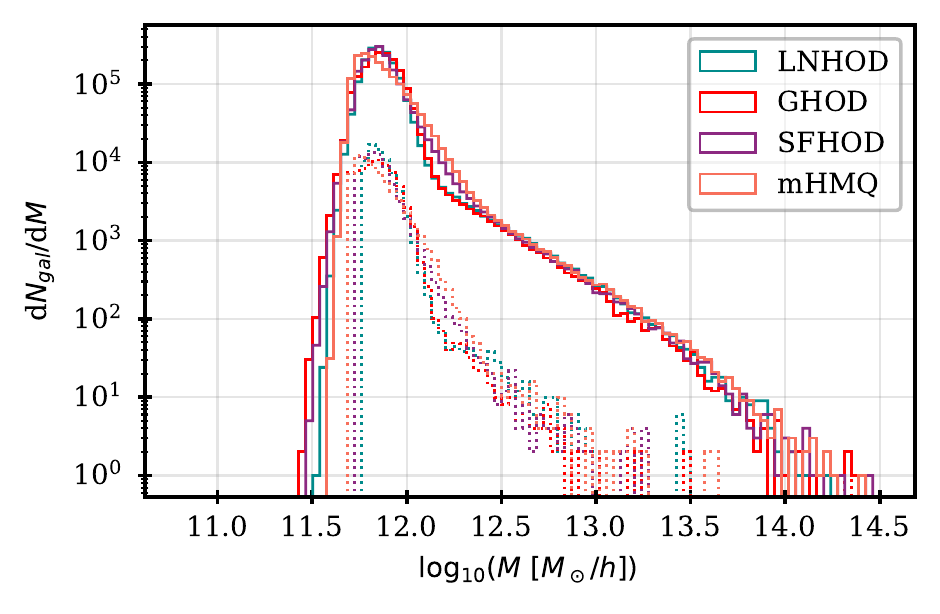}
\end{tabular}
\caption{\label{fig:standard_HOD2} {\it Left:} Best fitting HOD models to the DESI One-Percent ELG sample obtained with standard prescriptions for central (solid lines) and satellite (dashed lines) galaxies. We recall that satellites can populate halos even if no central galaxy is present. Four models for central galaxies were used and give similar results. Most noticeably, the satellite average number decreases with increasing halo mass.
{\it Right:} 
Number of galaxies per halo mass bin for halos populated according to the four HOD models on the left. The simulation box volume is $1.66\,($Gpc$/h)^3$. The full distributions are in solid lines. The dashed lines show the contribution of halos hosting more than one galaxy, that is the one-halo component of the full distributions. 
The four prescriptions for central galaxies lead to similar results, both for the full distribution or for its one-halo component.}
\end{figure} 

These results are further illustrated in Table~\ref{tab:standard_HOD}, which provides the best-fit values of the model parameters, and in Figure~\ref{fig:standard_HOD2}, which shows 
the four best fitting HOD models and the 
distributions of the number of galaxies per halo mass bin for halos populated according to these HOD models.
The four models exhibit similar features. The HOD for centrals peak at a mass slightly below $10^{12}M_{\odot}/h$ and span a short interval of halo masses as shown by the low values of $\sigma_M$. The minimal mass to populate halos with satellites is slightly below $M_{c}$ and the satellite HOD has a negative power-law index. Both features reflect the need to have close pairs of galaxies in low mass halos in order to reproduce the ELG clustering at small scales, and translate into a one-halo component of the 
distribution of the number of galaxies per populated halo mass bin 
that peaks at low halo mass, as shown in the right-hand plot in Figure~\ref{fig:standard_HOD2}. The mean halo mass of the galaxy sample and the satellite fraction are analytically calculated using Equations~\eqref{fracsat} and~\eqref{meanmass} from the [16-84] quantiles of the best-fit Markov chains. The fraction of galaxies which are not alone in their halos, $f_{1h}$, is found to be about $4\%$ in the four models tested. This fraction is computed numerically from 50 mocks generated from random HOD parameters drawn from the $1\sigma$ errors of the best fitting HOD models. Note that the value of $f_{1h}$ depends on the number density of the mocks, which is constrained to be that of our data sample $\sim 10^{-3}(h/$Mpc$)^{3}$.


Previous small-scale clustering studies of 
ELG samples at redshifts $\sim1$ were performed in different frameworks, either HOD ~\citep[e.g][]{Tinker2013,Avila20,Okumura2021}, Abundance Matching~\citep[e.g][]{Favole2016,Gao2022,Lin2023} or conditional stellar mass function method~\citep{Guo2019}. 
They find consistent results about the mean mass of halos hosting such galaxies, 
$\log_{10} \left\langle M_{h} \right\rangle \sim12$
They reported satellite fractions ranging from 13 to 22$\%$ for standard HOD prescriptions but extended ones can increase significantly these numbers~\citep{Avila20} showing that the satellite fraction does not provide a robust way to make precise comparisons between different analyses. For these two parameters, our findings are similar, namely 
$\log_{10} \left\langle M_{h} \right\rangle \sim 11.9$ and $f_{sat} \sim 12\%$.

All previous HOD studies also reported a satellite HOD that increases at high halo mass~\citep{Avila20,Okumura2021,Gao2022,Lin2023}, 
or possibly becomes uniform~\citep{Guo2019}, while we find a significant decrease (see Figure~\ref{fig:standard_HOD2}). This decrease is also responsible for the meaningless values of the effective $\log_{10}M'_1$ parameter reported in Table~\ref{tab:standard_HOD}, as the mass scale for having one satellite on average cannot be found at high halo mass.
As pairs of galaxies from the one-halo term dominates the clustering at small-scales (see Figure~\ref{fig:standard_HOD}), we attribute this decrease to the strong signal observed by DESI in a range of scales which were not previously probed and that we can model only with pairs of galaxies preferentially in low mass halos. 

However, physically motivated models of 
ELGs, either based on semi-analytical modelling~\citep[e.g][]{GonzalezPerez18,Contreras2019,Favole2020,Gonzalez-Perez2020} or hydrodynamical simulations~\citep[e.g][]{Hadzhiyska2021} do predict 
an increasing satellite HOD at high halo mass for 
ELGs at redshifts $\sim 1$.
We thus interpret our negative index result as a sign of an inadequate HOD model to describe DESI ELGs. In the next section we modify the model to include central-satellite conformity, 
that is the fact that satellite occupation may be conditioned by the presence of central galaxies of the same type, an hypothesis corroborated by hydrodynamical simulations~\citep{Hadzhiyska2022b}. 
Note that indications of conformity between central and satellite galaxies related to their types have already been reported in the literature~\citep{Weinmann2006}.

\begin{figure}[tbp]
\includegraphics[width=1\textwidth]{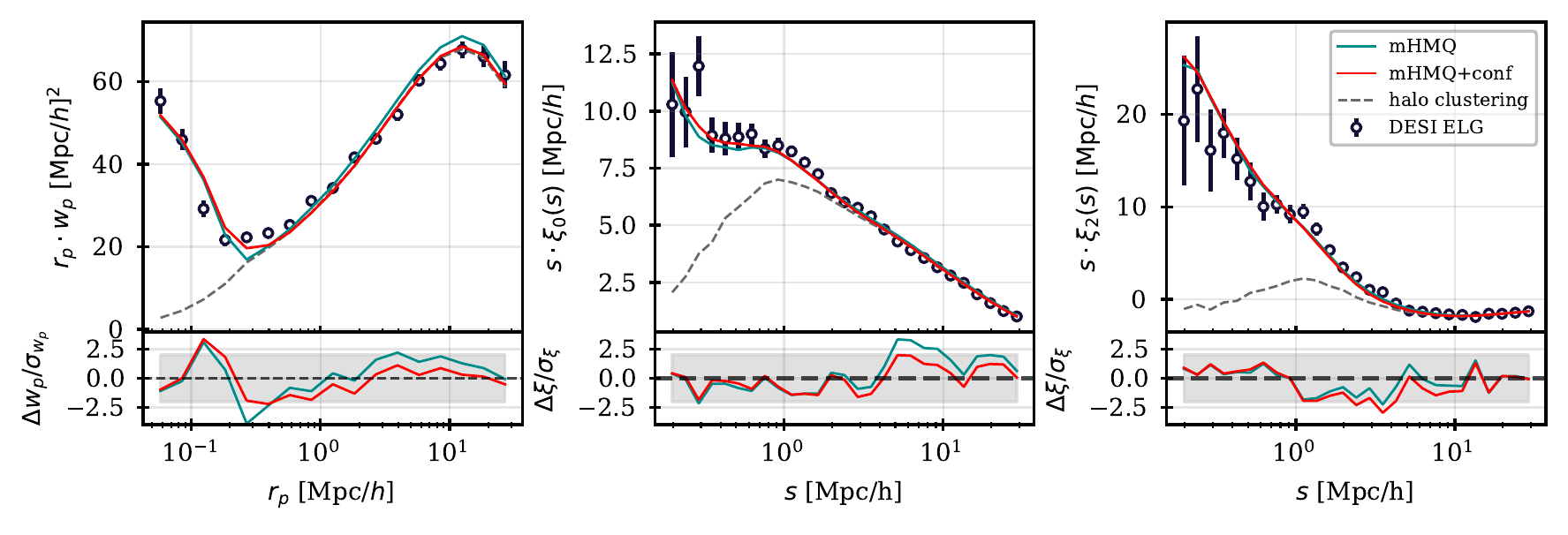} \\

\caption{\label{fig:ELG_conformity} 
{\it Top:} DESI ELG clustering measurements from the One-Percent survey data sample, compared to best fitting mHMQ models obtained with the GP pipeline, without (green line) and with (red line) strict conformity bias. The dashed line is the pure halo clustering. The agreement between data and expectations is slightly improved by requiring strict conformity, that is by conditioning satellite occupation to the presence of a central galaxy. Errors are Jackknife uncertainties only. 
{\it Bottom:} Fit residuals normalised by the diagonal errors of the full covariance matrix, that comprise Jackknife
uncertainties for the data as well as stochastic noise and cosmic variance for the model, but no Hartlap factor corrections.
}
\end{figure}

\begin{figure}[htbp]
\centering
\begin{tabular}{cc}
\includegraphics[width=0.47\textwidth]{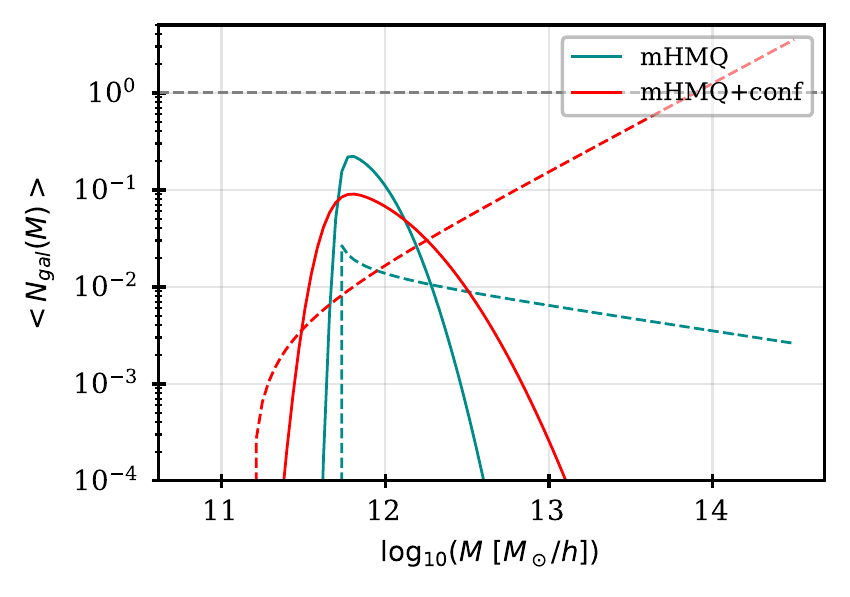} &
\includegraphics[width=0.5\textwidth]{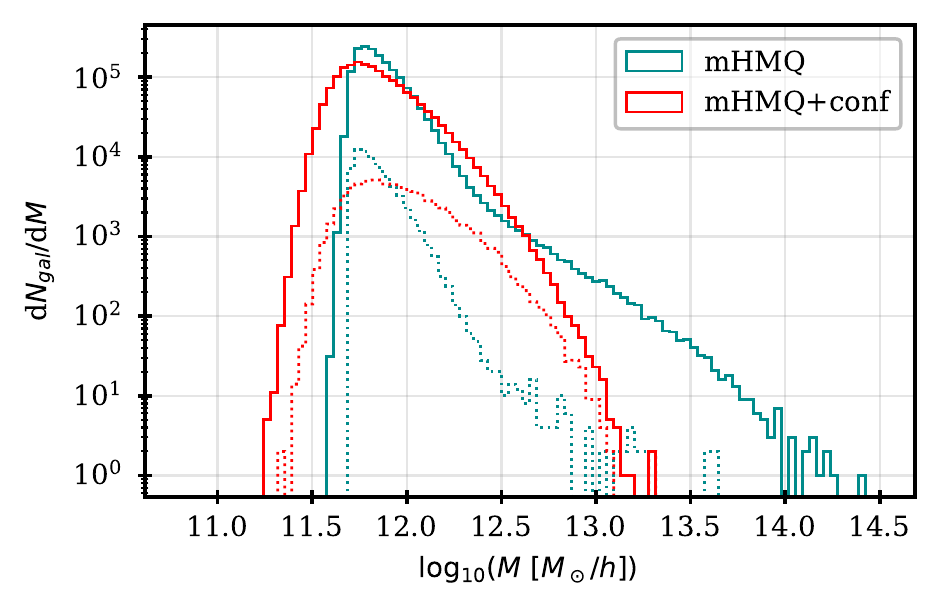}
\end{tabular}
\caption{\label{fig:LNHOD_conformity} {\it Left:} Best fitting HOD models to the DESI One-Percent ELG sample obtained without (green line) and with (red line) strict conformity bias between central (solid lines) and satellite (dashed lines) galaxies. The mHMQ prescription for centrals is used.
{\it Right:} 
Number of galaxies per halo mass bin for halos populated according to the two HOD models on the left. The simulation box volume is $1.66\,($Gpc$/h)^3$. The full distributions are in solid lines. The dashed lines show the contribution of halos hosting more than one galaxy, that is the one-halo component of the full distributions. Requiring strict conformity drastically changes the HOD models and 
the distributions of the number of galaxies per populated halo mass bin: satellites are forced to populate only halos with central galaxies and thus are spread over a wider range of halo masses.
}
\end{figure}

\section{Results in extended HOD models}
\label{sec:extensions}
In this section, we modify the standard prescription for satellite occupation. We first test conformity bias, as suggested by the results previously described and by studies from hydrodynamical simulations. We then test other possible changes in an attempt to better fit the clustering measurements in the problematic regions spotted in the above section.
Throughout this section, 
the baseline model for central galaxies is the mHMQ prescription. We also describe a cross-check of our results using the \ahod pipeline as an alternative fitting tool.

\subsection{Strict conformity bias}
\label{sec:conformity}

Best fitting clustering with strict central-satellite conformity from the GP pipeline are presented in Figure~\ref{fig:ELG_conformity} and compared to previous results without conformity. In this extended model, satellites can populate a halo only if a central galaxy is already present. Best-fit values of the model parameters are reported in~Table~\ref{tab:LNHOD_conformity}.
Strict conformity only slightly improves the agreement with data and the best-fit $\chi^2$ value. On the other hand, the shape of the HOD model and 
that of the distribution of galaxies per populated halo mass bin are significantly modified, as shown in Figure~\ref{fig:LNHOD_conformity}. With strict conformity, pairs of satellites in halos with no central galaxy are forbidden. To obtain the strong one-halo term needed to reproduce the small scale clustering, when conformity is required, pairs of galaxies are distributed over a wider range of halo masses at both low and high halo mass, as can be seen in the right-hand panel of Figure~\ref{fig:LNHOD_conformity} (see distributions in dashed lines). This translates into a satellite HOD that increases linearly ($\alpha=0.91^{+0.14}_{-0.11}$) with halo mass, as expected in physically motivated models. Note also that the mass scale for having one satellite on average is now obtained at large halo mass, as can be seen directly on the left panel in Figure~\ref{fig:LNHOD_conformity} and from the value of the effective $\log_{10}M'_1$ parameter in Table~\ref{tab:standard_HOD}.
We recall that even though the HOD of satellites increases with halo mass, strict conformity can only populate halos that already have a central. Hence, the satellite 
distribution per populated halo mass bin cannot exceed that of the centrals (see Figure \ref{fig:LNHOD_conformity}).

As a consequence, with strict conformity, the HOD parameters are all changed, except for the velocity dispersion parameter, $f_{\sigma_v}$ which we discuss further in section~\ref{sec:velocity}. The fraction of satellites $f_{sat}$ is five times smaller than without conformity, as a result of trading satellites alone in their halos for central-satellite pairs.
On the other hand the one-halo term fraction $f_{1h}$ and the mean halo mass remain very close to their values without conformity. This shows that these are model-independent characteristics that can be constrained by clustering measurements. As such they provide suitable quantities to compare results from analyses done in different frameworks. Note that, by definition, in the case of strict conformity, $f_{1h}=2f_{sat}$ for halos hosting one central and one satellite, which is typical of our ELG sample, cases with more than 1 satellite being rare.

\begin{table}[htbp]
\centering
\begin{tabular}{|l|c|c|}\hline
parameter \& $\chi^2$ &  mHMQ & mHMQ+conformity \\ \hline
$A_c$ (resc.)  & $1$ (0.08) & $0.1$ (0.63) \\
$\log_{10}M_0$ & $11.70^{+0.03}_{-0.03}$ &  $11.19^{+0.12}_{-0.10}$\\
$A_s$ & $0.10^{+0.04}_{-0.03}$  & $0.31^{+0.15}_{-0.08}$ \\ 
$\log_{10}M_{c}$ & $11.72^{+0.06}_{-0.04}$ &  $11.64^{+0.04}_{-0.04}$\\  
$\alpha$ & $-0.26^{+0.08}_{-0.08}$ &  $0.91^{+0.14}_{-0.11}$ \\ 
$f_{\sigma_v}$ & $1.27^{+0.07}_{-0.06}$ &  $1.34^{+0.08}_{-0.08}$ \\ 
$\sigma_M$ & $0.22^{+0.04}_{-0.02}$ &  $0.39^{+0.08}_{-0.10}$ \\ 
$\gamma$ & $7.06^{+1.33}_{-1.97}$ &  $4.50^{+1.49}_{-1.29}$ \\ 
\hline
$\log_{10} M'_{1}$ & 4.77 & 13.78  \\ 
$f_{sat}$ & $0.12^{+0.02}_{-0.02}$ & $0.024^{+0.030}_{-0.017}$ \\
$f_{1h}$ & $0.039^{+0.009}_{-0.008}$ & $0.048^{+0.010}_{-0.012}$ \\
$\log_{10} \left\langle  M_h \right\rangle $ &  $11.87^{+0.02}_{-0.01}$ &  $11.86^{+0.02}_{-0.02}$\\ \hline
$\chi^2$ (ndf=64) &  $156.0 \pm 1.0$ & $152.5 \pm 1.1$ \\  \hline 
\end{tabular} 
\caption {\label{tab:LNHOD_conformity} Results of mHMQ fits without and with strict conformity bias between central and satellite galaxies. The first line provides the initial fixed value of $A_c$ and the rescaling factor applied to impose the density constraint in the fits. The following
seven parameters are the free HOD parameters, 
the next four are derived parameters.
$\log_{10}M_1'$ is given for best-fit values of $\alpha$ and $A_s$ (the latter after rescaling). 
$f_{sat}$ is the fraction of galaxies which are satellite galaxies. $f_{1h}$ is the fraction of galaxies which are not alone in their halos. All masses are in units of $(M_{\odot}/h)$. 
}
\end{table} 

\begin{figure}[htbp]
\centering

\begin{tabular}{cc}
\includegraphics[width=0.47\textwidth]{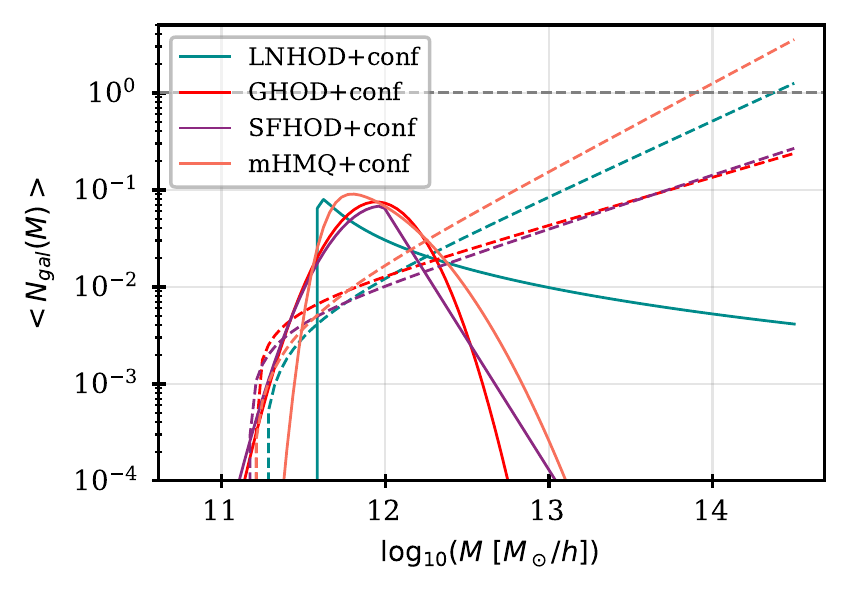} &
\includegraphics[width=0.5\textwidth]{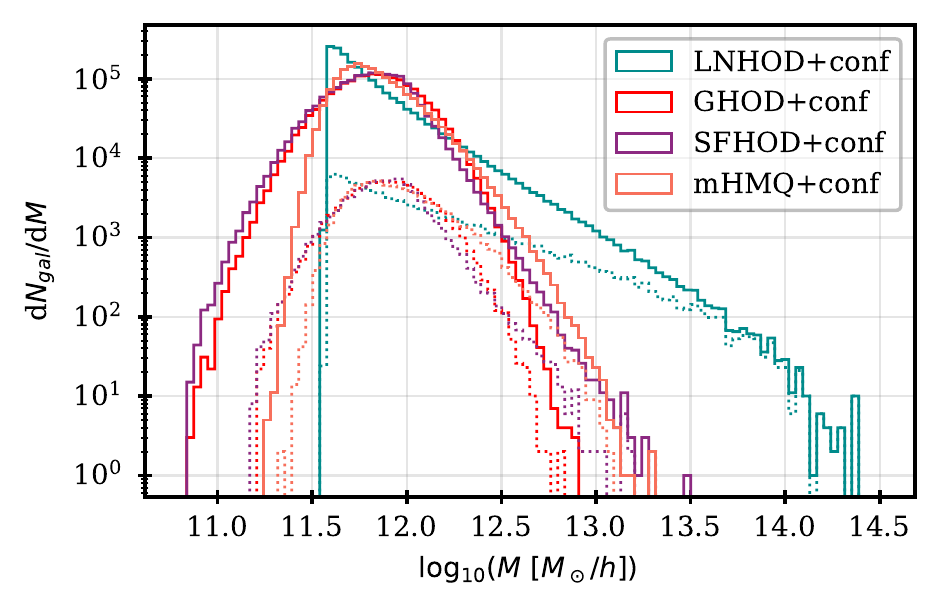}
\end{tabular}

\caption{\label{fig:HOD_conformity}  {\it Left:} Best fitting HOD models to the DESI One-Percent ELG sample obtained with strict conformity bias between central and satellite galaxies. Four prescriptions for central galaxies are used. They reproduce clustering data equally well but give different HOD shapes.
{\it Right:} 
Number of galaxies per halo mass bin for halos populated according to the four HOD models on the left. The simulation box volume is $1.66\,($Gpc$/h)^3$. The full distributions are in solid lines. The dashed lines show the one-halo component of the full distributions. The four models show an increase of the satellite HOD with increasing halo mass. The LNHOD model converges towards a triangular shaped HOD for centrals showing a sharp cut-off in mass for halos to host centrals, $\log_{10}M_h > \log_{10}M_c-1$ which originates from the large best-fit value of $\sigma_M$.
} 
\end{figure}

Finally, Figure~\ref{fig:HOD_conformity} presents the best fitting HOD models and 
distributions of the number of galaxies per populated halo mass bin
for the four prescriptions we can use for central galaxies.
The four models show an increase of the satellite HOD with increasing halo mass. The LNHOD model converges toward a triangular shaped HOD for centrals showing a sharp cut-off in mass for halos to host centrals, $\log_{10}M_h > \log_{10}M_c-1$ which originates from the large best-fit value of $\sigma_M$. This deviates substantially from physically inspired ELG models although the shape of the best fitting clustering statistics is almost indiscernible from the three other HOD models. This is reflected in the best-fit $\chi^2$ values that are similar in the four models, $\sim152$ for mHMQ and LNHOD, $\sim156$ for GHOD and $\sim161$ for SFHOD. 
With strict conformity, the minimal halo mass to host satellites, $\log_{10} M_0$ is around $\sim 11.2$ for the four models, compared to $11.7$ without conformity. This decrease is to be expected since with conformity the value of $M_0$ is driven by the minimum mass of halos hosting a central galaxy and reflects the need for having galaxy pairs in low-mass halos. We emphasise that all conformity models strongly favour putting satellites as soon as the halo is populated by central galaxies. The values of the characteristic halo mass for centrals, $\log_{10}M_{c}$, which were similar in the four models without conformity, are around 11.8 except for the LNHOD model which gives 12.6 due to the skewness of the HOD shape.

\subsection{Velocity bias}
\label{sec:velocity}
In the GP pipeline, satellite velocities are normally distributed around their halo velocity, computed as the mean halo dark matter particle velocities. The satellite velocity dispersion is that of the particle velocities rescaled by the $f_{\sigma_{v}}$ parameter which is left free to vary in the fits, namely:
\begin{equation}
\label{eq:dispersion}
\vec{v}_\mathrm{sat} \sim \mathcal{N}(\vec{v}_{h},f_{\sigma_{v}} \cdot \sigma_{v_h})
\end{equation}
This parameter represents a simple way to make ELG satellites hotter or cooler than dark matter particles, an hypothesis which was tested in studies of the eBOSS ELG sample~\citep{Avila20,Alam21}. 
The GP pipeline results previously presented show that, without or with conformity bias, and whatever the HOD prescription for central galaxies, the best-fit value for $f_{\sigma_{v}}$ is significantly higher than 1, which is in line with what was reported in~\cite{Avila20}. The best-fit values range from 1.2 to 1.5 depending on the model, with an error around $\pm0.1$. 

We check the impact of satellite velocities on this result using two other prescriptions. Instead of drawing satellite velocities according to~\eqref{eq:dispersion}, we set them to the halo velocity and add a circular velocity drawn from a NFW profile as defined in~\cite{NFW}:
\begin{equation}
\label{eq:vNFW}
\vec{v}_\mathrm{sat}(r) = \vec{v}_{h} + \sqrt{\frac{GM_h}{r_{vir}}} \sqrt{\frac{g(c_h \cdot r)}{r \cdot g(c_h)}}\vec{u}_{circ}
\;\;\;\;\;   \text{with} \;\;\; g(x) = \ln{(1+x)} - \frac{x}{(1+x)}
\end{equation}
$\vec{u}_{circ}$ is a unitary vector perpendicular to the vector joining the halo centre to the satellite position and whose orientation in this plane is randomly chosen. In the above equation,
$r$ is the satellite radial position, $r_{vir}$ is the virial halo radius and $c_h$ its concentration. 
As mentioned in Section~\ref{sec:GPpipeline}, we take $r_{98}$ as a proxy for 
$r_{vir}$ and $r_{25}$ as a proxy for $r_s$, so that 
$c_h \equiv r_{vir}/r_s =r_{98}/r_{25}$.
As a second choice, we first draw a satellite velocity $\vec{u}_{sat}$ according to~\eqref{eq:dispersion} and add to it a common infall velocity $\vec{v}_\mathrm{infall}$ defined along the line between the satellite position to the halo centre:
\begin{equation}
\label{eq:infall}
\vec{u}_\mathrm{sat} \sim \mathcal{N}(\vec{v}_{h},\sigma_{v_h}) \;\;\;\; \text{then} \;\;\;\;
\vec{v}_\mathrm{sat} = \vec{u}_\mathrm{sat} + \vec{v}_\mathrm{infall} \;\;\;
\text{with} \;\;\; 
\vec{v}_\mathrm{infall} = 
v_\mathrm{infall} \cdot \frac{\vec{r}_h-\vec{r}_\mathrm{sat}}{\vert \vec{r}_h-\vec{r}_\mathrm{sat} \vert}
\end{equation}
This model is a good approximation of the 
prediction presented 
in~\cite{AnguloOrsi2018} based on semi-analytical models of star-forming galaxies. The latter predict that among star-forming galaxies those which were accreted the latest could have a net infall velocity towards the halo centres.

\begin{figure}[tbp]
\includegraphics[width=1\textwidth]{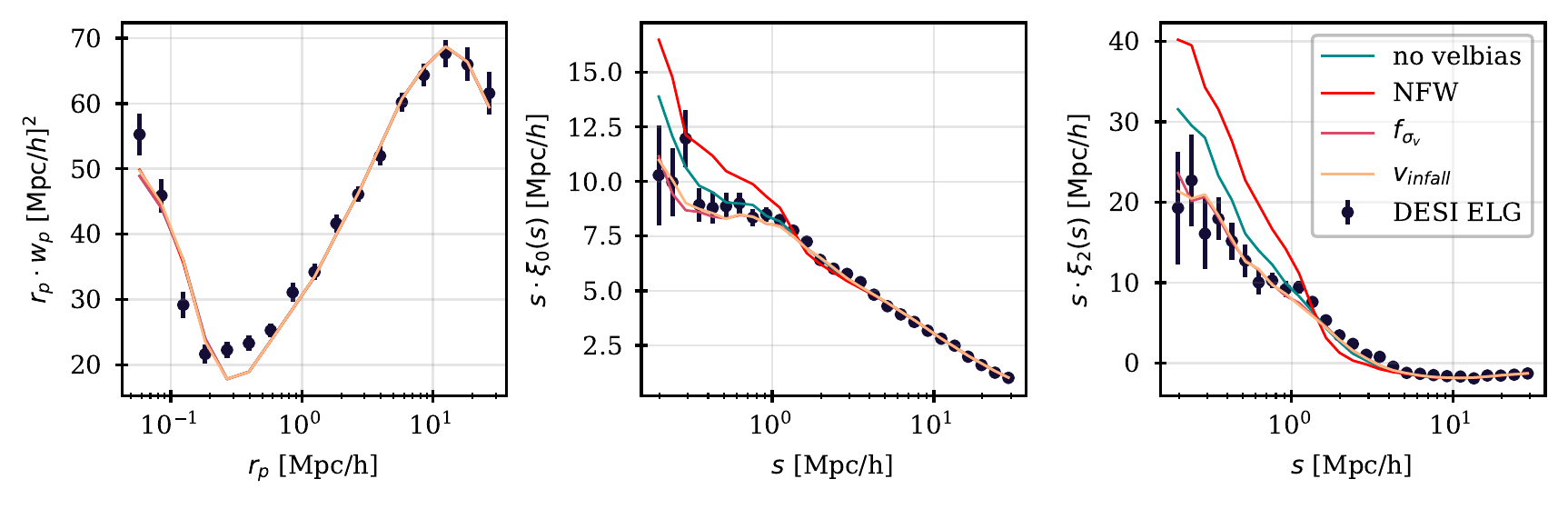}
\caption{\label{fig:velocity_bias} 
DESI ELG clustering measurements from the One-Percent survey data sample compared to HOD models differing only by their satellite velocity prescriptions. We show the best fitting mHMQ model with conformity bias found in Section~\ref{sec:conformity} (purple) which corresponds to rescaling the dispersion of dark matter particles by a factor $f_{\sigma_v}$=1.34 to describe the satellite velocities. Other models correspond to the following changes: $f_{\sigma_v}$=1 (green), drawing satellite velocities from a NFW profile (red) and assuming a common infall velocity (yellow) of $v_{infall}=170$~km$/s$. Errors are Jackknife uncertainties. The four models give exactly the same projected clustering but produce differences in the 2-point correlation function multipoles at small-scales. We note that the same clustering can be obtained by rescaling the particle velocity dispersion or assuming a common infall velocity.
}
\end{figure}

In order to illustrate the impact of satellite velocities on the clustering statistics, Figure~\ref{fig:velocity_bias} compares the DESI data clustering to the best fitting mHMQ model with strict conformity bias found in the previous section (purple curve)  and to predictions from that model where we modify the satellite velocity prescription, keeping the other HOD parameters fixed, without refitting the data.
The satellite velocity is modified according to equations~\eqref{eq:vNFW} and ~\eqref{eq:infall}, using a value of 170 km$/s$ for $v_{infall}$ in the latter case. Also shown is the 
predicted clustering with $f_{\sigma_{v}}$ set to 1 to remove any velocity bias (green curve). The four models predict the same projected clustering, as expected since velocities have no effects on this statistic. Taking a NFW profile for velocities (red curve) does not provide a good modelling of the 2-point correlation functions multipoles. Not rescaling the velocity dispersion (green curve) provides a good modelling of the monopole only, while up-scaling the dispersion (purple curve) allow both multipoles to be correctly modelled. Last, there is practically no difference in the predicted clustering between an up-scaling of the particle velocity dispersion with a factor of 1.34 and a net infall velocity of 170km$/s$ added to velocities normally distributed around the halo velocity with a dispersion equal to that of the particle velocities (orange curve). These two models, although different, have quite similar impact on the clustering and cannot be disentangled with the statistics we are using. Note that random errors in the ELG redshift determination~\citep{inclusiveSHAM} are equivalent to a 60 km$/s$ velocity dispersion along the line of sight and only accounts for 0.03 on the observed shift in $f_{\sigma_v}$ w.r.t. 1. We conclude that a velocity dispersion larger than that of DM particles is needed to reproduce the clustering of the DESI One-Percent survey ELG sample. Interestingly, a velocity bias was also reported by SDSS for main galaxies at low redshifts ($z<0.2$) and LRGs at intermediate ($z\sim0.5)$ redshifts~\citep{Guo2015a, Guo2015b} but the effect goes in the opposite direction, with satellites moving more slowly than particles by a factor that depends on the galaxy luminosity, the bias being stronger for more luminous galaxies. 


In the following, we continue with the baseline prescription for satellite velocities, expressed as a rescaling of dark matter particle velocities by a factor $f_{\sigma_{v}}$.

\subsection{Comparison to \ahod pipeline}
At this point, we cross-check our results with the \ahod pipeline~\citep{Yuan22}, which is particle-based and highly efficient. Designed specifically for multi-tracer analyses and HOD-cosmology combined analyses, it takes advantage of the large volume and precision of the \absum simulations by optimising computational efficiency. We refer the reader to~\cite{Yuan22} for a detailed description of this pipeline. 

The baseline HOD prescription for ELG central galaxies in \ahod is the HMQ model of~\cite{Alam21}. In the present work, we restrict to the simpler mHMQ model of section~\ref{sec:models} (with $A_c$ renamed to $p_{max}$). For the satellite galaxies, we adopt the baseline power law model of Eq.~\eqref{eq:power_law} except we reparametrise $M_0 = \kappa M_c$. Central galaxies are assigned the position and velocity vector of the centre of mass of the largest sub-halo while satellite galaxies are assigned to DM particles of the halo with equal probabilities. Each halo can only host at most one central galaxy and each particle can also host at most one satellite.

The \ahod implementation of ELG central-satellite conformity introduces one extension parameter to the standard satellite HOD to modulate the strength of the conformity effect. Specifically, we modulate the $M_1$ parameter, which controls the overall amplitude of satellite occupation, by whether the halo hosts a central ELG or not:
\begin{equation}
    \left\langle N_{sat}(M)\right\rangle = 
    \begin{cases}
    \left(\frac{M-\kappa M_\mathrm{c}}{M_\mathrm{1, EE}}\right)^\alpha & \textrm{if ELG central}\\
    \left(\frac{M-\kappa M_\mathrm{c}}{M_\mathrm{1}}\right)^\alpha & \textrm{if not}.\\
    \end{cases}
    \label{equ:elgconfhod}
\end{equation}
where $M_{1,EE}$ is the new parameter that modulates the ELG-ELG conformity strength. If there is no conformity, then $M_{1,EE} = M_1$, and if there is maximal conformity, i.e. ELG satellites only occupy halos with ELG centrals, then $M_{1,EE} \ll M_1$. In principle, we also implement another conformity term between ELG satellites and LRG centrals. However, it is beyond the scope of this analysis, and we refer the reader to~\cite{abacusLRGQSO} for a conformity analysis of the ELGxLRG cross-correlation functions. 

Velocity bias prescriptions are different between the two pipelines.
For the GP pipeline, bias on velocities are changed only for satellites, through the scaling parameter $f_{\sigma_v}$, as described in Eq.~\eqref{eq:dispersion}. The \ahod pipeline allows both for central  
and satellite velocity biases, through parameters $\alpha_c$ and $\alpha_s$, respectively. Those impact velocities as $v_\mathrm{cent} = v_h + \alpha_c\delta_v(\sigma_{vh})$ for centrals, where $\delta_v(\sigma_{vh})$ is the Gaussian scatter of the velocity dispersion of the halo, and $v_\mathrm{sat} = v_\mathrm{particles} + \alpha_s(v_\mathrm{particles} - v_h)$, as described in equations 8 \& 9 in \cite{Yuan22}.

\begin{figure}
\centering
\includegraphics[width=\textwidth]{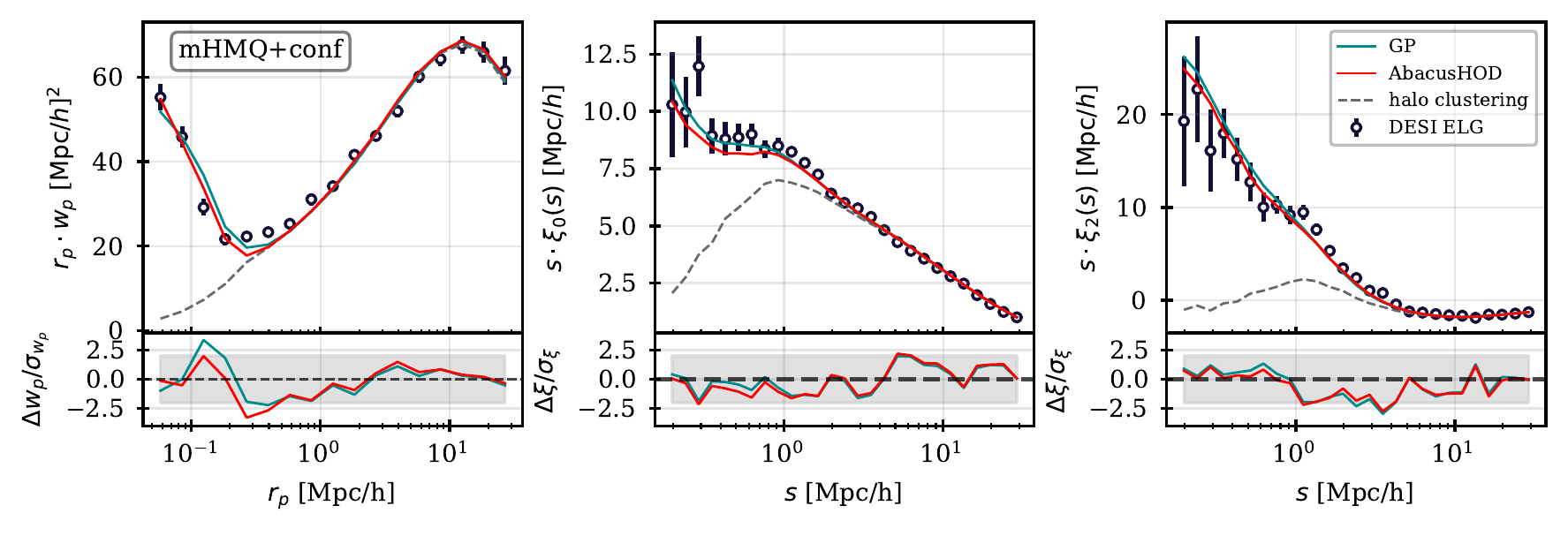}
\caption{{\it Top}: DESI ELG clustering measurements from the One-Percent survey data sample compared to best fitting mHMQ models with parametrised central-satellite conformity in the \ahod pipeline ({\it red}) and with strict conformity bias in the GP pipeline ({\it green}). Bottom: Fit residuals normalised by the diagonal errors of the full covariance matrix, that comprise Jackknife uncertainties for the data as well as stochastic noise and cosmic variance for the model, but no Hartlap factor corrections.}    
\label{fig:comp_ahod_GP}
\end{figure}

Although the baseline statistics of the \ahod pipeline is the galaxy two-point correlation function in two dimensions, for this cross-check it is run using the same 2-point statistics (see Section~\ref{sec:statistics}) and the same data covariance matrix as the GP pipeline matrix (see Section~\ref{sec:covmat}). The model covariance matrix of the GP pipeline is ignored in the fits but is used to compute best-fit $\chi^2$ values provided below.
We compare the mHMQ best fitting results from the two pipelines 
in Figure~\ref{fig:comp_ahod_GP} for the predicted clustering and in Table~\ref{tab:comp_ahod_GP} for the HOD and derived parameters. Best-fit parameters from the \ahod pipeline are derived using global optimisation chains using Gaussian priors so no error bars are provided. The two pipelines produce quite similar best fitting clustering predictions and goodness of fit results, despite the completely different nature of the pipelines and their different prescriptions for some parameters of the mHMQ model. 

\begin{figure}
\centering
\includegraphics{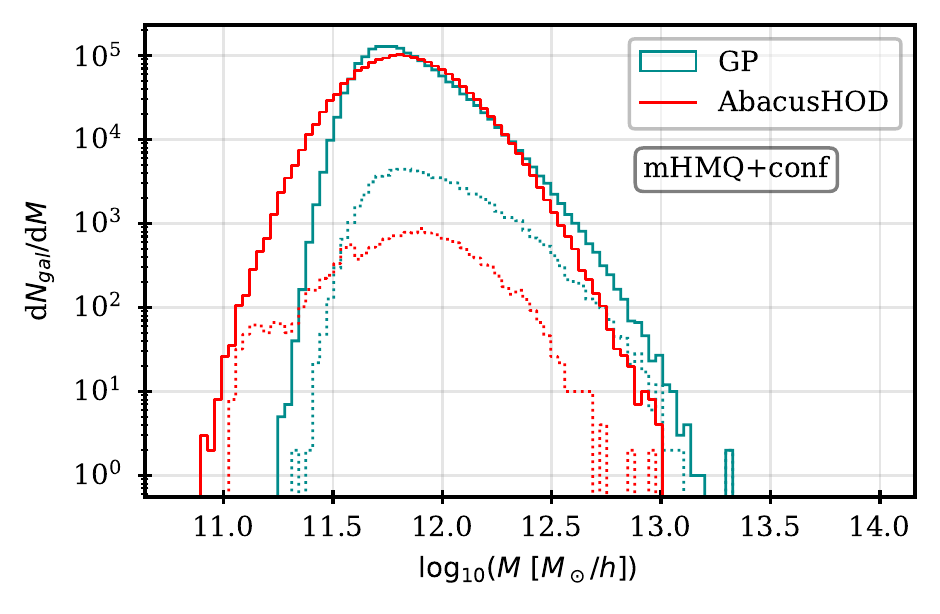}
\caption{
Number of galaxies per halo mass bin for halos populated according to the best fitting mHMQ models to the DESI One-Percent ELG sample, from \ahod with parametrised conformity bias (red) and from the GP pipeline with strict conformity bias (green). The simulation box volume is $1.66\,($Gpc$/h)^3$. The full distributions are in solid lines and the dashed lines show the one-halo component of the full distributions.}
\label{fig:comp_ahod_GP2}
\end{figure}

Most parameters treated in the same way in both pipelines have similar best fitting values, except for the $\gamma$ parameter that controls the asymmetry of the central HOD. This difference is reflected in the shape of the 
distribution of the number of galaxies per populated halo mass bin,
whose asymmetry is more pronounced for the GP pipeline result, as can be seen in Figure~\ref{fig:comp_ahod_GP2}. On the other hand, the $\gamma$ parameter is hardly constrained in the fits (see error bars in Table~\ref{tab:comp_ahod_GP} and $\gamma$ posteriors in Appendix~\ref{app:contours}), which means that our clustering statistics are not very sensitive to the asymmetric character of the HOD distributions, so that 
distributions of the number of galaxies per populated halo mass bin
as different as those in Figure~\ref{fig:comp_ahod_GP2} can produce very similar clustering signals (see Figure~\ref{fig:comp_ahod_GP}). 

Although the velocity bias prescriptions 
are different, both pipelines end up with the same conclusion, namely that the satellite velocity dispersion is higher than that of halo particles. As for central velocities, the \ahod pipeline result shows that allowing for a velocity dispersion of centrals is not really mandatory. 
As for central-satellite conformity, the \ahod parametrised bias indicates clearly a preference for conformity  since $M_{1,EE}$ is lower than $M_1$ by more than 5 units, making the strict conformity of the GP pipeline implementation a good approximation. Remarkably, both pipelines agree well on the derived parameters, the satellite fraction, one-halo term fraction and the mean halo mass value of the sample.
Finally, we note that the $\chi^2$ of the \ahod result is slightly better than that of the GP pipeline 
but does not significantly improve the goodness of fit. 

This means that the reason for the poor goodness of fit of our results so far is not to be found in the fitting methodology but rather in the HOD model itself. In the following, we test other extensions of the model in the GP pipeline to check whether an improvement can be found.

\begin{table}
\centering
\begin{tabular}{|l|c|c|}\hline
parameter \& $\chi^2$ &  \ahod pipeline & GP pipeline \\ \hline
$p_{max} = A_c$ (resc.) & 0.08 & 0.10 (0.63) \\
$\log_{10} M_0$ & 11.03 ($\kappa =0.19$)  &  $11.19^{+0.12}_{-0.10}$\\
$A_s$ & 1 (fixed)  & $0.31^{+0.15}_{-0.08}$ \\ 
$\log_{10}M_{c}$ & 11.75 &  $11.64^{+0.04}_{-0.04}$\\  
$\alpha$ & $0.72$ &  $0.91^{+0.14}_{-0.11}$ \\ 
$\alpha_c,\alpha_s$ or $f_{\sigma_v}$ & $0.19, 1.49$ & $1.34^{+0.08}_{-0.08}$ \\
$\log_{10} M_1$ & 19.83 & 13 (fixed) \\
$\sigma_M$ & $0.31$ &  $0.39^{+0.08}_{-0.10}$ \\ 
$\gamma$ & $1.39$ &  $4.50^{+1.49}_{-1.29}$ \\
$\log_{10} M_{1,EE}$ & 14.25 & - \\
\hline
$f_{sat}$ & 0.020 & $0.024^{+0.030}_{-0.017}$ \\
$f_{1h}$ & 0.040 & $0.048^{+0.010}_{-0.012}$ \\
$\log_{10} \left\langle M_h \right\rangle $ &  11.89 &  $11.86^{+0.02}_{-0.02}$\\ \hline
$\chi^2$ (ndf) &  143.53 (62) & $152.5 \pm 1.1$ (64) \\  \hline 
\end{tabular} 
\caption {\label{tab:comp_ahod_GP} Results of mHMQ fits with parametrised central-satellite conformity from the \ahod pipeline ({\it left}) and with strict conformity bias from the GP pipeline ({\it right}). The upper ten rows list HOD parameters, the next three give derived parameters. $f_{sat}$ is the fraction of galaxies which are satellite galaxies. $f_{1h}$  is the fraction of galaxies which are not alone in their halos. All masses are in units of $(M_{\odot}/h)$.}
\end{table} 

\subsection{Assembly bias}
HOD modelling is primarily a function of halo mass only but semi-analytical models and hydrodynamical simulations predict dependencies in other properties that are referred to as secondary biases in the literature. In this section, we explore assembly bias which introduces a dependence related to the halo assembly history. 
We test dependencies either in halo concentration, local halo density or local halo density anisotropies, using the parametrisation suggested in~\cite{Hadzhiyska2022c}:

\begin{align}
 \left\langle N'_{cent}(M)\right\rangle &= 
[1 + a_{cent} f_a (1-\left\langle N_{cent}(M)\right\rangle)]
\left\langle N_{cent}(M)\right\rangle \\
\left\langle N'_{sat}(M)\right\rangle &= 
[1 + a_{sat} f_a ] \left\langle N_{sat}(M)\right\rangle
\end{align}
where $\left\langle N_{cent}(M)\right\rangle$ and $\left\langle N_{sat}(M)\right\rangle$ are given in Section~\ref{sec:models}. In the above equations, $f_a$ is introduced to materialise the property of each halo in a normalised way. In a given halo mass bin, halos are first ranked by decreasing values of the halo property and each halo is attributed a different value of $f_a$, assuming that the latter decreases linearly between $0.5$ and $-0.5$ when going from the top ranked halo to the last one.

The halo properties we consider are the halo concentration, $c_h=r_{98}/r_{25}$ and the halo environment that we first characterise by the local halo density. To compute the latter, we project all halos in the simulation box onto a grid of 5~Mpc$/h$ mesh using a count-in-cell resampling algorithm and calculate the density in each grid cell. Each halo is then attributed the local density of the grid cell it belongs to. As a third halo property, we consider local halo density anisotropies deduced from the so-called adaptive halo shear, computed from the smoothed local density field as described in~\cite{Hadzhiyska2022c}, using a smoothing scale of 1.5~Mpc$/h$.

\begin{figure}[tbp]
\includegraphics[width=1.\textwidth]{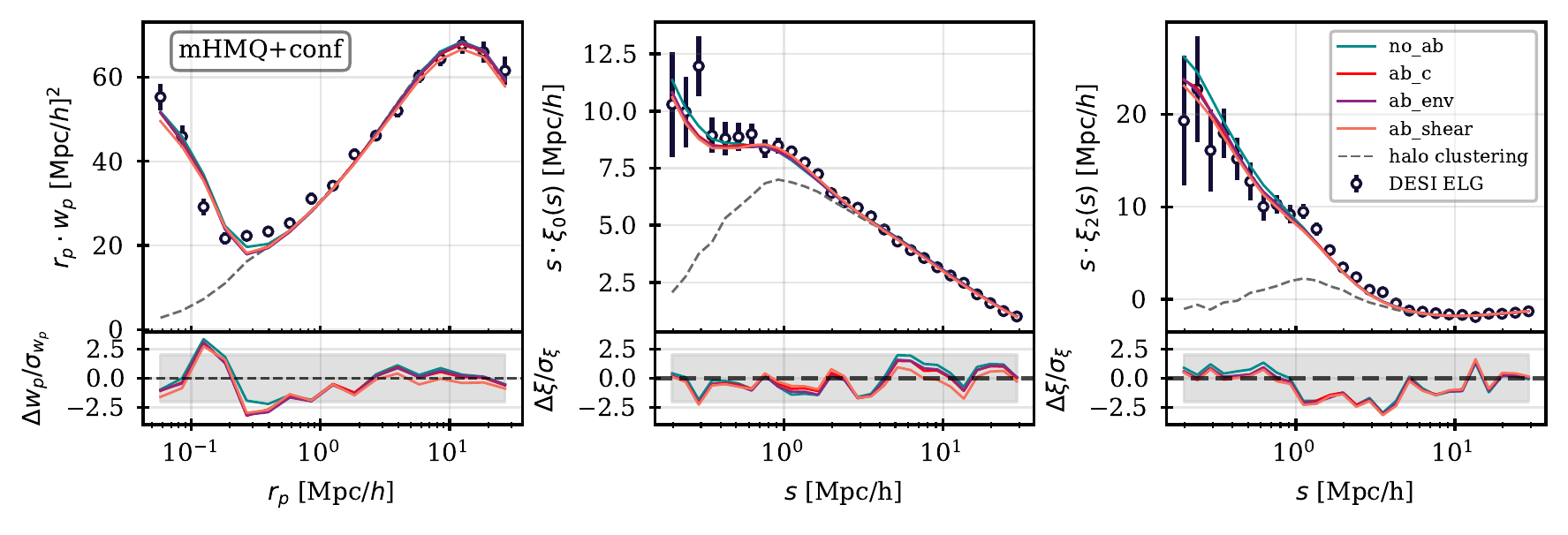}
\caption{\label{fig:assembly_bias} {\it Top:}
DESI ELG clustering measurements from the One-Percent survey data sample compared to different best fitting mHMQ models  with strict conformity bias: baseline result found in Section~\ref{sec:conformity} (green), model with assembly bias for both centrals and satellites as a function of concentration (red), 
local halo density (purple) and local halo density anisotropies (orange). 
Errors are Jacknnife uncertainties only. {\it Bottom:} Fit residuals normalised by the diagonal errors of the full covariance matrix, that comprise Jackknife
uncertainties for the data as well as stochastic noise and cosmic variance for the model, but no Hartlap factor corrections.}
\end{figure}

\begin{table}[!h]
\centering
\resizebox{\columnwidth}{!}{
\begin{tabular}{|l|c|c|c|c|}\hline
parameter \& $\chi^2$ &  no assembly & $c_h$ assembly & $\rho$ assembly & 'shear' assembly \\ \hline
$A_c$ (resc.) & $0.1$ (0.63) & $0.1$ (0.60) & $0.1$ (0.64)& $0.1$ (0.69)  \\
$\log_{10} M_0$ &   $11.19^{+0.12}_{-0.10}$& $11.17^{+0.13}_{-0.10}$ & $11.19^{+0.12}_{-0.11}$ & $11.19^{+0.13}_{-0.11}$\\
$A_s$ & $0.31^{+0.15}_{-0.08}$ & $0.35^{+0.10}_{-0.07}$ & $0.28^{+0.07}_{-0.05}$ & $0.27^{+0.07}_{-0.05}$\\ 
$\log_{10}M_{c}$  &  $11.64^{+0.04}_{-0.04}$& $11.63^{+0.04}_{-0.03}$ & $11.61^{+0.04}_{-0.03}$ & $11.66^{+0.03}_{-0.03}$ \\  
$\alpha$ &  $0.91^{+0.14}_{-0.11}$ & $0.93^{+0.10}_{-0.07}$ & $0.86^{+0.06}_{-0.06}$ & $0.92^{+0.08}_{-0.10}$\\ 
$f_{\sigma_v}$ & $1.34^{+0.08}_{-0.08}$& $1.35^{+0.08}_{-0.09}$ & $1.34^{+0.10}_{-0.09}$ & $1.31^{+0.08}_{-0.08}$ \\ 
$\sigma_M$ &  $0.39^{+0.08}_{-0.10}$&  $0.39^{+0.08}_{-0.09}$ &  $0.44^{+0.13}_{-0.11}$ & $0.41^{+0.10}_{-0.08}$\\ 
$\gamma$  &  $4.50^{+1.49}_{-1.29}$ & $4.54^{+1.20}_{-0.87}$ & $5.76^{+1.13}_{-1.19}$ & $6.05^{+1.04}_{-1.13}$\\ 
$a_{cen}$ & - & $0.75^{+0.12}_{-0.25}$& $-0.02^{+0.22}_{-0.24}$ & $0.10^{+0.05}_{-0.05}$\\ 
$a_{sat}$ & - & $-0.32^{+0.59}_{-0.42}$ & $0.02^{+0.63}_{-0.65}$&  $0.00^{+0.61}_{-0.57}$\\
\hline
$\log_{10} M'_{1}$ & 13.78 & 13.72 & 13.87 & 13.79 \\ 
$f_{sat}$ &  $0.024^{+0.030}_{-0.017}$ & $0.022^{+0.024}_{-0.015}$ & $0.021^{+0.024}_{-0.015}$ & $0.021^{+0.022}_{-0.017}$ \\
$f_{1h}$ &  $0.048^{+0.010}_{-0.012}$& $0.044^{+0.009}_{-0.013}$ &  $0.042^{+0.013}_{-0.008}$ & $0.042^{+0.015}_{-0.01}$ \\
$\log_{10} \left\langle M_h \right\rangle $ &  $11.86^{+0.02}_{-0.02}$ &  $11.84^{+0.02}_{-0.02}$ & $11.83^{+0.02}_{-0.02}$ & $11.82^{+0.02}_{-0.02}$\\ \hline
$\chi^2$ (ndf) & $152.5 \pm 1.1$ (64) & $144.8 \pm 1.0$ (62) & $150.4 \pm 1.4$ (62) & $147.98 \pm 1.14$ (62)\\  \hline 
\end{tabular} 
}
\caption {\label{tab:LNHOD_assembly} Results of mHMQ fits with strict conformity bias between central and satellite galaxies without (left) and with assembly bias as a function of halo concentration ($c_h$), local density ($\rho$) and local density anisotropies ('shear'). The first line provides the initial fixed value of $A_c$ and the rescaling factor applied to impose the density constraint in the fits. The following seven or nine parameters are the free HOD parameters, the next four are derived parameters.
$\log_{10}M_1'$ is given for best-fit values of $\alpha$ and $A_s$ (the latter after rescaling). 
 $f_{sat}$ is the fraction of galaxies which are satellite galaxies. $f_{1h}$ is the fraction of galaxies which are not alone in their halos. 
All masses are in units of $(M_{\odot}/h)$. 
}
\end{table} 

Figure~\ref{fig:assembly_bias} presents the clustering predicted by the best fitting mHMQ models with strict conformity bias obtained without and with the three assembly bias prescriptions (see Table~\ref{tab:LNHOD_assembly} for HOD and derived parameters). 
The mHMQ goodness of fit does not improve significantly when adding assembly bias based on halo concentration, local density or local density anisotropies. HOD parameters and derived parameters are within $1\sigma$ of their values in the model without assembly bias. As a result all models exhibit similar clustering (almost indistinguishable). We note that the model with assembly bias using halo concentration slightly improves the $\chi^2$ value 
with a preference for highly concentrated halos, $a_{cen}$ being close to 1. 
This constitutes a mild preference for assembly bias, but as the effect on clustering statistics is small, this preference cannot be established unambiguously. 
The best fit for assembly bias using halo local density points towards no dependence with the density as $a_{cen}$ is found to be compatible with 0. 
In the case of local density anisotropies, best-fit results indicate a preference for halos with a slightly positive shear, $a_{cen}$ being positive, but this preference is weaker than that for halo concentration.
Lastly, the $a_{sat}$ parameter is consistent with 0 and poorly constrained in the three models as a consequence of the fact that the satellite fraction with strict conformity bias is small ($\sim 2\%$ ). 

\subsection{Satellite positioning with a modified NFW profile}
None of the extensions of the HOD model studied in the previous sections 
succeeds in producing extra pairs of galaxies at scales $r_p=\left[0.1,1\right]$Mpc$/h$ as required by data.

\begin{figure}[thbp]
\includegraphics[width=1\textwidth]{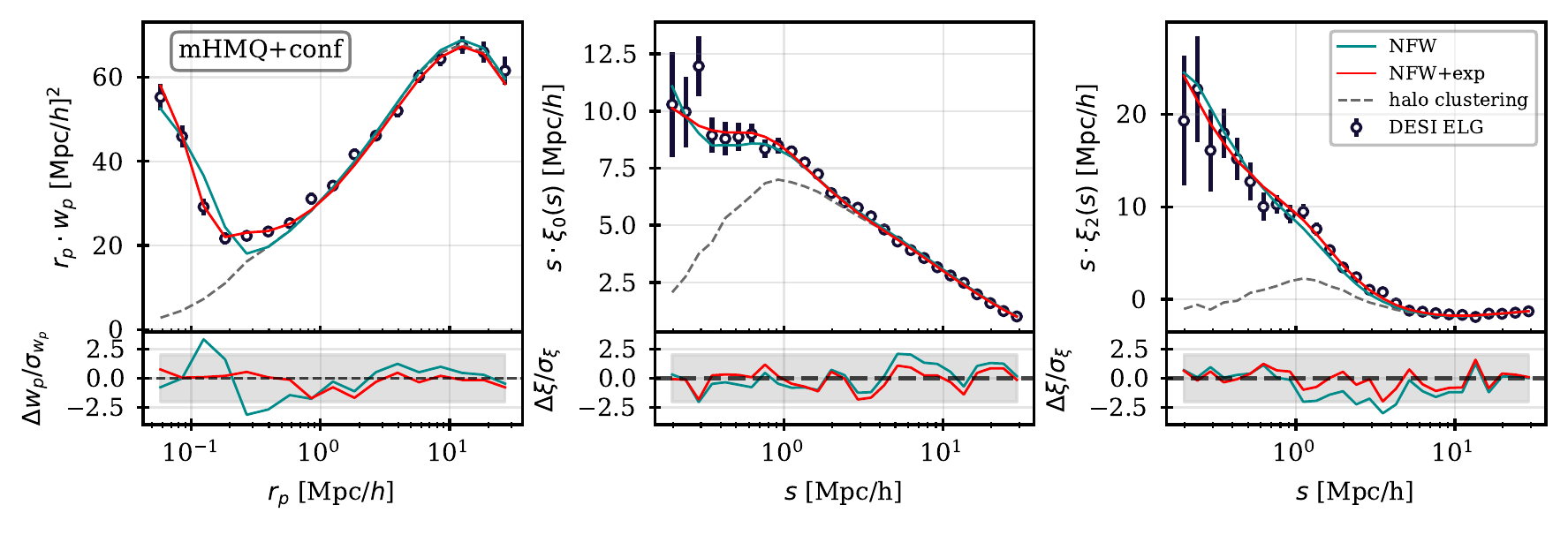}
\caption{\label{fig:modifiedNFW} 
{\it Top:}
DESI ELG clustering measurements from the One-Percent survey data sample compared to best fitting mHMQ models with strict conformity bias: baseline result found in Section~\ref{sec:conformity} (green) and model with satellite positioning according to a modified NFW profile (red). Errors are Jacknnife uncertainties only. {\it Bottom:} Fit residuals normalised by the diagonal errors of the full covariance matrix, that comprise Jackknife
uncertainties for the data as well as stochastic noise and cosmic variance for the model, but no Hartlap factor corrections.}
\end{figure}

Nevertheless, it is possible to overcome this by changing the radial profile of satellites. 
\cite{AnguloOrsi2018} suggest that, whatever the halo mass, ELGs populate preferentially the outskirts of their host halos, galaxies accreted more recently being found further away from the halo centre. This is explained by the fact that satellite galaxies can present high star formation rates only for a short period once the galaxy gas has been depleted by tidal and ram pressure stripping. As a consequence, star-forming satellite ELGs are expected to be preferentially located in the outskirts of their halo where recently accreted subhalos free of the above processes can be found. On the observational side, results showing that the quenched fraction of the specific star formation rate distribution of galaxies is radially dependent within a halo were already reported for SDSS galaxies~\citep{Blanton2007,Wetzel2012}.

\begin{table}[htbp]
\centering
\begin{tabular}{|l|c|c|}\hline
parameter \& $\chi^2$ &  NFW profile & modified profile \\ \hline
$A_c$ (resc.) & $0.1$ (0.63) & $0.1$ (0.51) \\
$\log_{10} M_0$   &  $11.19^{+0.12}_{-0.10}$ & $11.20^{+0.11}_{-0.09}$ \\
$A_s$   & $0.31^{+0.15}_{-0.08}$ & $0.41^{+0.10}_{-0.15}$ \\ 
$\log_{10}M_{c}$ &  $11.64^{+0.04}_{-0.04}$ & $11.64^{+0.04}_{-0.04}$\\  
$\alpha$  &  $0.91^{+0.14}_{-0.11}$ & $0.81^{+0.08}_{-0.14}$\\ 
$f_{\sigma_v}$ &  $1.34^{+0.08}_{-0.08}$ & $1.63^{+0.11}_{-0.10}$\\ 
$\sigma_M$  &  $0.39^{+0.08}_{-0.10}$ &  $0.30^{+0.09}_{-0.07}$\\ 
$\gamma$  &  $4.50^{+1.49}_{-1.29}$ &  $5.47^{+1.37}_{-1.58}$\\ 
$f_{exp}$ & - &$0.58^{+0.06}_{-0.05}$  \\
$\tau$ & - & $6.14^{+1.11}_{-1.20}$\\
$\lambda_{NFW}$ & - & $0.67^{+0.06}_{-0.06}$\\
\hline
$\log_{10} M'_{1}$ & 13.78 & 13.84 \\ 
$f_{sat}$ & $0.024^{+0.030}_{-0.017}$ & $0.034^{+0.010}_{-0.012}$ \\
$f_{1h}$  & $0.048^{+0.010}_{-0.012}$ & $0.069^{+0.020}_{-0.024}$\\
$\log_{10} \left\langle M_h \right\rangle $  &  $11.86^{+0.02}_{-0.02}$ & $11.86^{+0.03}_{-0.03}$\\ \hline
$\chi^2$ (ndf) & $152.5 \pm 1.1$ (64) & $ 87.91\pm1.84 $ (61) \\  \hline 
\end{tabular} 
\caption {\label{tab:modifiedNFW} Results of mHMQ fits with strict conformity bias using a standard NFW profile for satellite positioning (left) and our modified profile (right).  The first line provides the initial fixed value of $A_c$ and the rescaling factor applied to impose the density constraint in the fits. The following seven or ten
parameters are the free HOD parameters, 
the next four are derived parameters.
$\log_{10}M_1'$ is given for best-fit values of $\alpha$ and $A_s$ (the latter after rescaling). 
$f_{sat}$ is the fraction of galaxies which are satellite galaxies. $f_{1h}$ is the fraction of galaxies which are not alone in their halos. All masses are in units of $(M_{\odot}/h)$. 
}
\end{table} 

Inspired by the above publications, we test a modified NFW profile to position ELG satellites. The number of satellites for a given halo is first drawn according to the standard prescription in Eq.~\eqref{eq:power_law}. A fraction of them, $f_{exp}$ have radial positions drawn from an exponential law:
\begin{equation}
\label{eq:exponential}
\frac{dN(r)}{dr} =  e^{-r / (\tau \cdot r_s) }
\end{equation}
where $r$ is the distance between the satellite and the halo centre, and $\tau$ governs the slope of the exponential and acts on the extension of the profile. Radial positions of the remaining satellites obey a NFW profile with the same proxy for $r_{vir}$ as in Section~\ref{sec:GPpipeline} but squeezing the proxy for $r_s$ by a factor $\lambda_{NFW}$, namely $r_s \rightarrow r_s/\lambda_{NFW}$. This is almost equivalent to 
extending the profile cut-off with respect to $r_{vir}$ into $r_{cut off}=\lambda_{NFW} \cdot r_{vir}$ and allows for modifications of the profile extension.
The three parameters $f_{exp}, \tau$ and $\lambda_{NFW}$ are left free to vary in the fits. Note that galaxies positioned beyond the halo virial radius are improperly called satellites but we keep that denomination here to reflect the HOD parametrisation component they come from.

The best fitting mHMQ results with strict conformity and the above prescription are compared with the baseline results using a pure NFW profile in Figure~\ref{fig:modifiedNFW} for the clustering predictions and in Table~\ref{tab:modifiedNFW} for the HOD and derived parameters.
The modified positioning of satellites translates into a significant improvement of the agreement between data and predictions, with a $\chi^2$ value dropping from $\sim$152 to $\sim$88 (p-value of 1.4$\%$). The improvement is most notable in the region of the up-turn of the projected correlation function (see residuals in Figure~\ref{fig:modifiedNFW}) showing that additional pairs of galaxies have been generated at these scales with the extended profile, with no degradation of the agreement elsewhere. 

\begin{figure}[thbp]
\centering
\includegraphics[width=0.6\textwidth]{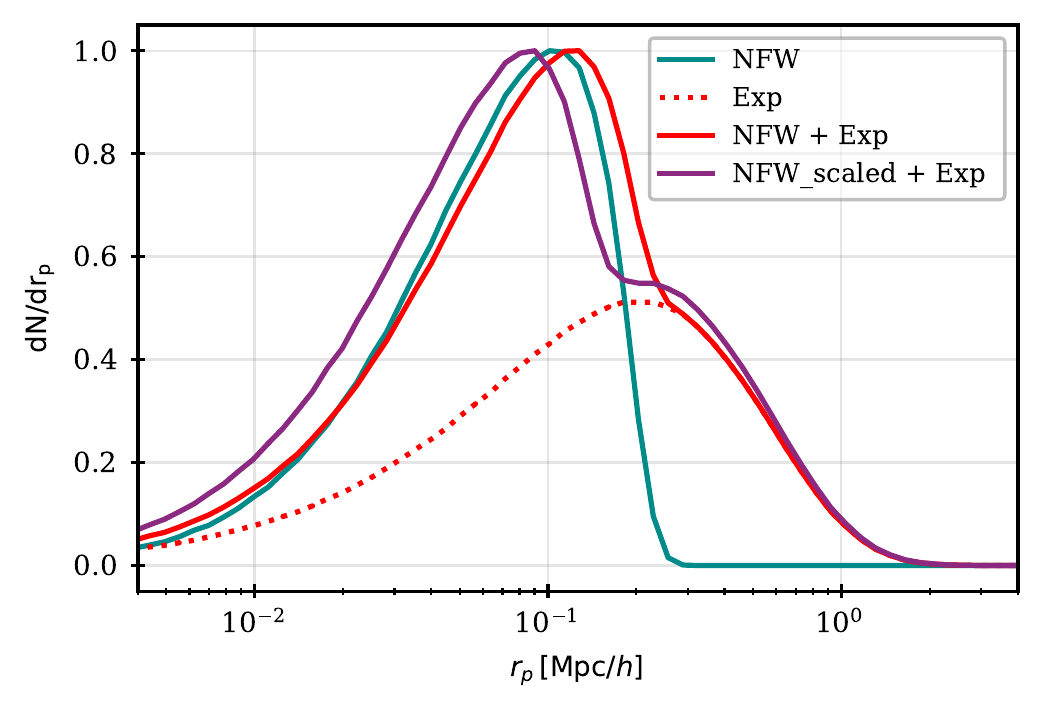}
\caption{\label{fig:profile} 
Normalised satellite density profile for best-fit parameters in the mHMQ model with strict conformity and our modified NFW profile prescription for satellites, 
as a function of the projected galaxy-halo centre distance perpendicular to the line of sight. 
Once this profile is embedded into a HOD model, this distance is also the projected separation of central-satellite pairs.
In this example, we consider a halo of concentration 5 and $r_s=0.06$Mpc$/h$ (corresponding to halo masses around $10^{12}M_{\odot}/h$, close to the mean halo mass value of our sample). Curves (all normalised at a maximal value of 1) are for the NFW profile (blue), the added exponential law (dotted), the combination of the two with no scaling of the NFW cut-off (green) and the complete modified model (red). }
\end{figure}

\begin{figure}[htbp]
\centering
\begin{tabular}{cc}
\includegraphics[width=0.45\textwidth]{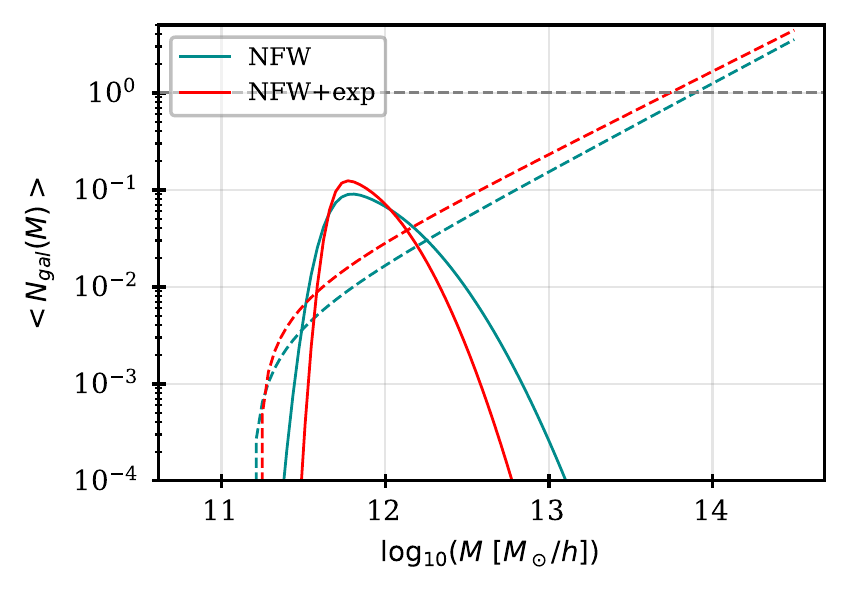}   &
\includegraphics[width=0.48\textwidth]{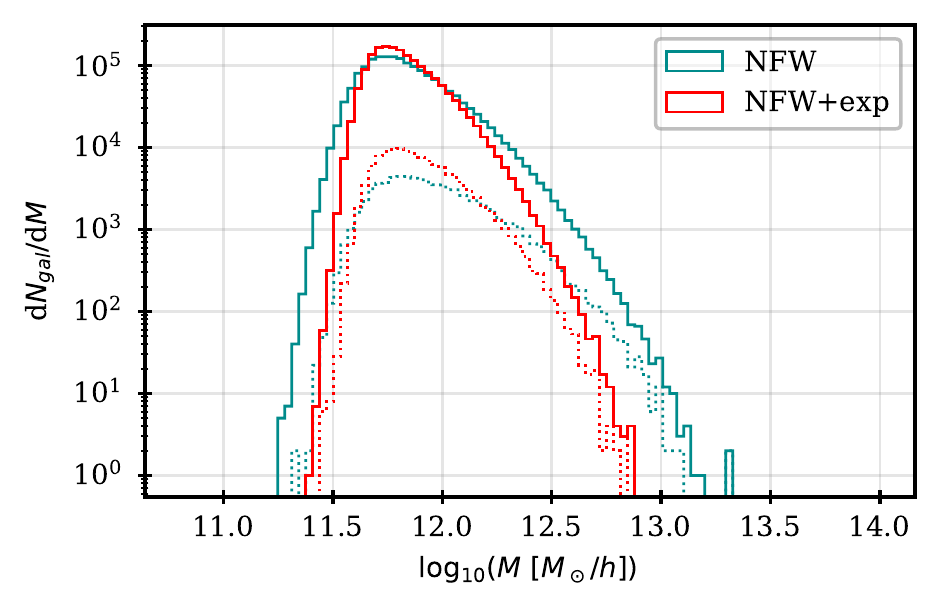}   \\
\end{tabular}
\caption{\label{fig:HMF_modifiedNFW} {\it Left:} Best fitting HOD models to the DESI One-Percent ELG sample with strict conformity bias, obtained with a standard NFW profile for satellites (green) and with our modified NFW profile (red). Solid (resp. dashed) lines represent central (resp. satellite) galaxies. The mHMQ prescription is used for centrals. {\it Right:} 
Number of galaxies per halo mass bin for halos populated according to the mHMQ models on the left. The simulation box volume is $1.66\,($Gpc$/h)^3$. The full distributions are in solid lines. The dashed lines show the one-halo component of the full distributions. The two satellite profiles produce similar 
results, with a larger scatter in populated halo masses for the modified profile.
}
\end{figure} 

An example of satellite density profile corresponding to the best fitting parameters is represented in Figure~\ref{fig:profile} as a function of 
the radial position of the satellites with respect to the halo centre projected perpendicular to the line of sight. The profile clearly shows that the exponential component acts at projected scales between 0.03 and 1 Mpc$/h$, the region of the up-turn in $w_p$. 
Note that the scales covered by our clustering measurements are more sensitive to the region close to the halo virial radius (hence to the cut-off applied to the NFW profile) than to the shape of the profile deep in the halo core.

Table~\ref{tab:modifiedNFW} shows that the HOD parameters as well as the derived parameters are similar between the two models, except for a 20$\%$ increase of the value of $f_{\sigma_v}$, meaning that the extended profile of satellites leads to a higher satellite velocity dispersion. 
This provides a coherent picture as recently-accreted subhalos in the outskirts of halos are expected to have higher velocities than the virial velocity of the halo.
The comparison between the two models is further illustrated in Figure~\ref{fig:HMF_modifiedNFW}, which presents the HOD and 
the distribution of the number of galaxies per populated halo mass bin
of the two models. The only difference is a larger scatter in populated halo masses for 
with the modified NFW profile.

The profile parameters, $f_{exp}, \tau$ and $\lambda_{NFW}$, are all well constrained by data and their best fitting values are in favour of a departure from a standard NFW profile. We find that the exponential profile contains around $60\%$ of the satellites and a fraction of these (approximately 12$\%$ of the total number of satellites, as measured in the mocks at best fitting HOD parameters) are placed beyond our proxy for the halo virial radius (see Figure~\ref{fig:profile}). 
The above modified profile is empirical and can most probably be replaced by a more physics driven modelling. Nevertheless, our main finding is that the ELG clustering measured by the DESI One-Percent survey clearly favours a fraction of ELGs residing in the outskirts of halos, as suggested by~\cite{ Blanton2007, Wetzel2012} and~\cite{AnguloOrsi2018}.

\section{Testing for redshift evolution}
\label{sec:zevolution}
\begin{figure}[tbp]
\centering
\includegraphics[width=\textwidth]{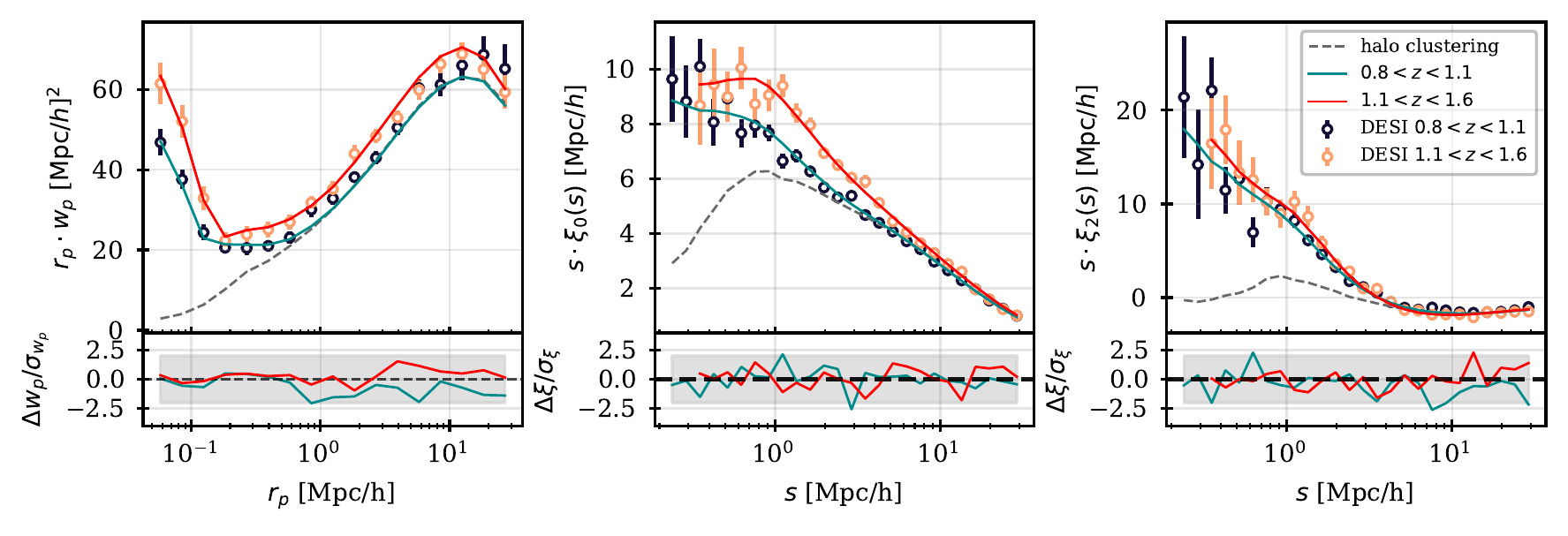}   
\caption{\label{fig:clustering_z} {\it Top:} DESI ELG clustering measurements from the One-Percent survey data sample in two different redshift bins from 0.8 to 1.1 (black) and 1.1 to 1.6 (orange), compared to best fitting mHMQ models with strict conformity bias 
and our modified NFW profile for satellite positioning, for the redshift 0.8 to 1.1 (green) and 1.1 to 1.6 (red). The dashed line is the pure halo clustering for the low redshift bin. Errors are Jackknife uncertainties. {\it Bottom:} Fit residuals normalised by the diagonal errors of the full covariance matrix (calculated for each redhsift bin), that comprise Jackknife uncertainties for the data
as well as stochastic noise and cosmic variance for the model, but no Hartlap factor corrections.}
\end{figure}

The ELG clustering measurements are produced in two separate redshift bins, from 0.8 to 1.1 and 1.1 to 1.6, with completeness-weighted redshifts of 0.95 and 1.32, respectively. The clustering measurements for the two redshift bins including completeness and FKP weights are  shown in Figure~\ref{fig:clustering_z}. 
Measurements in the two redshift bins agree for most separations but exhibit significant differences in the monopole up to 10~Mpc$/h$ and in the projected correlation function around the up-turn scale of 0.3~Mpc$/h$. It is thus interesting to fit the two bins in redshift separately to see how the agreement between HOD modelling and data evolves. 
The HOD model in each bin is calculated from the N-body simulation snapshot closest to 
the mean completeness-weighted redshift of the bin (i.e. snapshots at z=0.95 and z=1.325, respectively).

\begin{table}[htbp]
\centering
\resizebox{\columnwidth}{!}{%
\begin{tabular}{|l|c|c|c|c|}\hline
parameter \& $\chi^2$ &  $0.8<z<1.1 \;\;\; \bar{z}=0.95$ & $1.1<z<1.6 \;\;\; \bar{z}=1.325$ & $0.8<z<1.6 \;\;\; \bar{z}=1.1$ \\ \hline
$A_c$ (resc.) & $0.1$ (0.43) &  $0.1$ (0.51) & $0.1$ (0.51) \\
$\log_{10} M_0$ & $11.10^{+0.05}_{-0.04}$  &  $11.23^{+0.16}_{-0.14}$ & $11.20^{+0.11}_{-0.09}$ \\
$A_s$ & $0.38^{+0.04}_{-0.04}$  & $0.47^{+0.13}_{-0.13}$ & $0.41^{+0.10}_{-0.15}$ \\ 
$\log_{10}M_{c}$ & $11.62^{+0.02}_{-0.04}$ & $11.67^{+0.04}_{-0.04}$ & $11.64^{+0.04}_{-0.04}$\\  
$\alpha$  & $0.74^{+0.07}_{-0.05}$ &$0.85^{+0.08}_{-0.10}$ & $0.81^{+0.08}_{-0.14}$\\ 
$f_{\sigma_v}$ & $1.71^{+0.11}_{-0.14}$ &  $1.71^{+0.20}_{-0.15}$ & $1.63^{+0.11}_{-0.10}$\\ 
$\sigma_M$  & $0.21^{+0.10}_{-0.05}$  & $0.29^{+0.11}_{-0.08}$ & $0.30^{+0.09}_{-0.07}$\\ 
$\gamma$  & $6.49^{+0.69}_{-1.39}$ & $5.10^{+1.51}_{-1.20}$ & $5.47^{+1.37}_{-1.58}$\\ 
$f_{exp}$ &  $0.70^{+0.10}_{-0.09}$ & $0.55^{+0.10}_{-0.09}$  &$0.58^{+0.06}_{-0.05}$ \\
$\tau$ & $5.69^{+1.72}_{-2.00}$ & $7.22^{+1.77}_{-3.14}$ & $6.14^{+1.11}_{-1.20}$\\
$\lambda_{NFW}$ & $0.60^{+0.09}_{-0.09}$ & $0.67^{+0.07}_{-0.07}$ & $0.67^{+0.06}_{-0.06}$\\
\hline
$\log_{10} M'_{1}$ & 14.05  & 13.73 & 13.84 \\ 
$f_{sat}$   &$0.026^{+0.005}_{-0.005}$ & $0.035^{+0.010}_{-0.011}$ & $0.034^{+0.010}_{-0.012}$ \\
$f_{1h}$  &$0.053^{+0.009}_{-0.009}$ & $0.069^{+0.019}_{-0.021}$ & $0.069^{+0.020}_{-0.024}$\\
$\log_{10} \left\langle M_h \right\rangle $ &  $11.78^{+0.03}_{-0.04}$ & $11.86^{+0.05}_{-0.05}$ &$11.86^{+0.03}_{-0.03}$\\ \hline
$\chi^2$ (ndf) & $89.78 \pm 0.66$ (59) & $58.35 \pm 0.41$ (55) & $ 87.91\pm 0.84 $ (61)\\  \hline 
\end{tabular} 
}
\caption {\label{tab:HOD_z} Results of mHMQ fits with strict conformity bias and our modified NFW profile for satellite positioning, presented separately in two redshift bins and compared to the results with the whole redshift bin (right). 
The first line provides the initial fixed value of $A_c$ and the rescaling factor applied to impose the density constraint in the fits. The following ten
parameters are the free HOD parameters, 
the next four are derived parameters.
$\log_{10}M_1'$ is given for best-fit values of $\alpha$ and $A_s$ (the latter after rescaling).
$f_{sat}$ is the fraction of galaxies which are satellite galaxies. $f_{1h}$  is the fraction of galaxies which are not alone in their halos. The number of degrees of freedom is different in the three bins and indicated in brackets in the $\chi^2$ row. All masses are in units of $(M_{\odot}/h)$. 
}
\end{table} 

\begin{figure}[htbp]
\centering
\begin{tabular}{cc}
\includegraphics[width=0.45\textwidth]{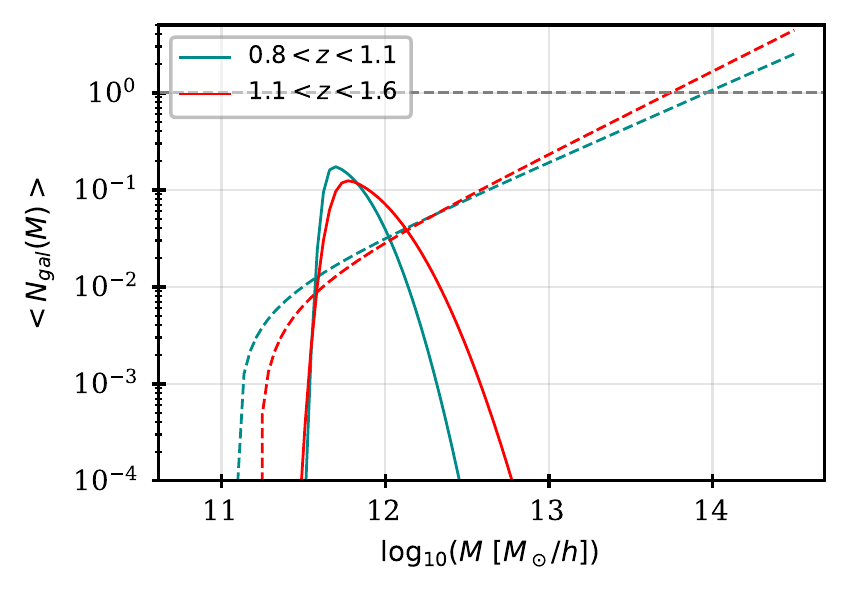}   &
\includegraphics[width=0.48\textwidth]{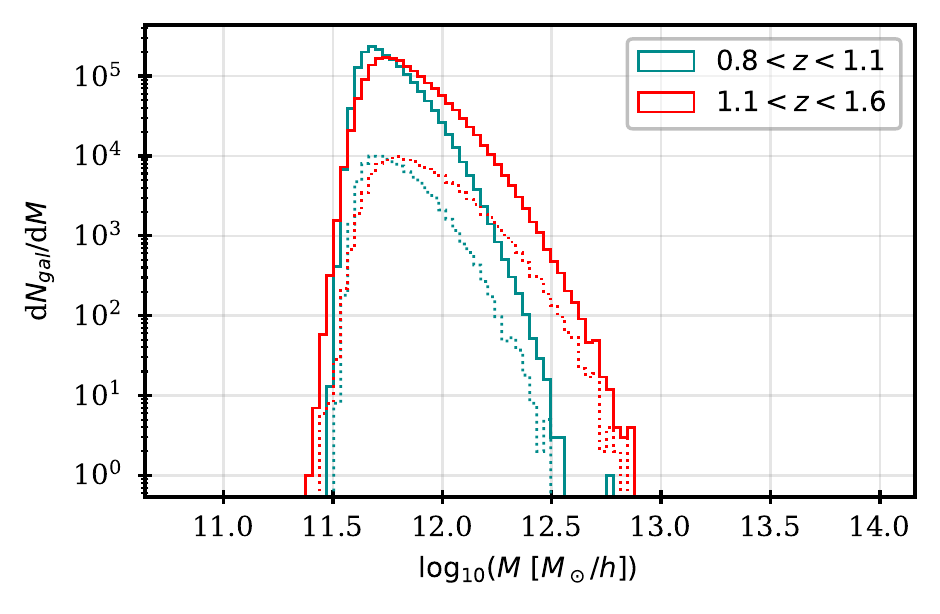}   \\
\end{tabular}
\caption{\label{fig:HMF_z} {\it Left:} Best fitting HOD 
models to the DESI One-Percent ELG sample split in two separate redshift bins, 0.8-1.1 (green) and 1.1 to 1.6 (red), with strict conformity bias and our modified NFW profile for satellite positioning. Solid (resp. dashed) lines represent central (resp. satellite) galaxies. The mHMQ prescription is used for centrals. {\it Right:} 
Number of galaxies per halo mass bin for halos populated according to the mHMQ model on the left. The simulation box volume is $1.66\,($Gpc$/h)^3$. The full distributions are in solid lines. The dashed lines show the one-halo component of the full distributions. The two redshift bins exhibit similar 
distributions, 
the higher redshift bin (1.1 to 1.6) showing a larger scatter towards higher populated halo masses.
}
\end{figure} 

Best fitting results in the two redshift bins from the mHMQ model with strict conformity and our modified NFW profile for satellite positioning are presented in Figure~\ref{fig:clustering_z} and summarised in Table~\ref{tab:HOD_z}. With respect to results obtained in the full redshift bin (see right column in Table~\ref{tab:HOD_z}, p-value of $1.4\%$), the goodness of fit is similar in the low redshift bin (p-value of $0.6\%$) and much better in the high redshift bin (p-value of $35\%$). 
Variations of the HOD parameters and the derived parameters with redshift 
appear to be moderate, parameter values in the two redshift bins being all within $1\sigma$. The same is true for the 
distribution of the number of galaxies per populated halo mass bin
as shown in Figure~\ref{fig:HMF_z}. In the high redshift bin, there is a small increase of the scatter in the latter 
towards higher populated halo masses, which is reflected in the higher value of $\left\langle M_h \right\rangle$, the average halo mass 
of the ELG sample, $11.86^{+0.05}_{-0.05}$ vs 
$11.78^{+0.03}_{-0.04}$ but the difference is at the level of $1\sigma$.
For completeness, we show in Appendix~\ref{app:contours} the contour plots of the mHMQ fits with strict conformity and our modified NFW profile for satellite positioning obtained at final iteration in the two redshift bins. 

To conclude, 
changes of the ELG sample with redshift in terms of the mean halo mass or in the one-halo term fraction are at the level of $1\sigma$ and thus cannot be considered as significant.
In the companion paper~\cite{novelAM}, the ELG sample of the DESI One-Percent survey was split in narrower redshift bins but did not show a significant variation with redshift of the characteristic halo mass hosting ELGs either. 
In a second companion paper~\cite{inclusiveSHAM}, the luminosity of that sample (from  [\small{O} {\rm II}] emission) was also found to evolve very mildly with redshift (see their Figure 9).
We discuss further the results from the companion analyses in Section~\ref{sec:companion}.
Using a sample of 
[\small{O} {\rm II}]  emitters at $z>1$ in the Subaru HSC survey, \cite{Okumura2021} also found a constant mass across redshifts bins, in agreement with our findings.

\section{Testing for cosmology dependence}
\label{sec:cosmology}
In this section, we study how the previous results evolve when changing the reference cosmology both in the simulation box (used for the modelling) and in the fiducial cosmology (used to convert redshifts to distances).
We test one cosmology with a high $N_{\mathrm{eff}}$ value and one with a low $\sigma_8$ value (see Table~\ref{tab:cosmology} for the complete list of cosmological parameter values). In this section, we continue with the mHMQ model with strict conformity bias and the extended NFW profile for satellite positioning but perform fits in the full redshift bin. 

\begin{table}[htbp]
\centering
\begin{tabular}{|c|c|c|c|c|c|c|c|}\hline
Cosmologies & $\Omega_{cdm} h^2$ & $\Omega_b h^2$ & $\sigma_8$ & $n_s$ & $h$ & $w_0$ & $w_a$ \\ \hline
baseline & 0.1200 & 0.02237 & 0.811355 & 0.9649 & 0.6736 & -1 & 0  \\
high $N_{\mathrm{eff}} (c003) $& 0.1291 & 0.02260 & 0.855190 & 0.9876 & 0.7160 &-1 & 0\\
low $\sigma_8$ (c004) & 0.1200 & 0.02237 & 0.753159 & 0.9649 & 0.6736 & -1 & 0  \\ \hline
\end{tabular} 
\caption {\label{tab:cosmology} Parameter values of the three cosmologies used in Section \ref{sec:cosmology}. Indicated are the present-day densities of cold dark matter and baryons, the normalisation today of the linear power spectrum in spheres of radius $8\mathrm{Mpc}/h$, the spectral index of the primordial matter power spectrum, the reduced value of the Hubble constant and the dark energy equation of state parameters.
}
\end{table} 

\begin{figure}[tbp]
\centering
\includegraphics[width=\textwidth]{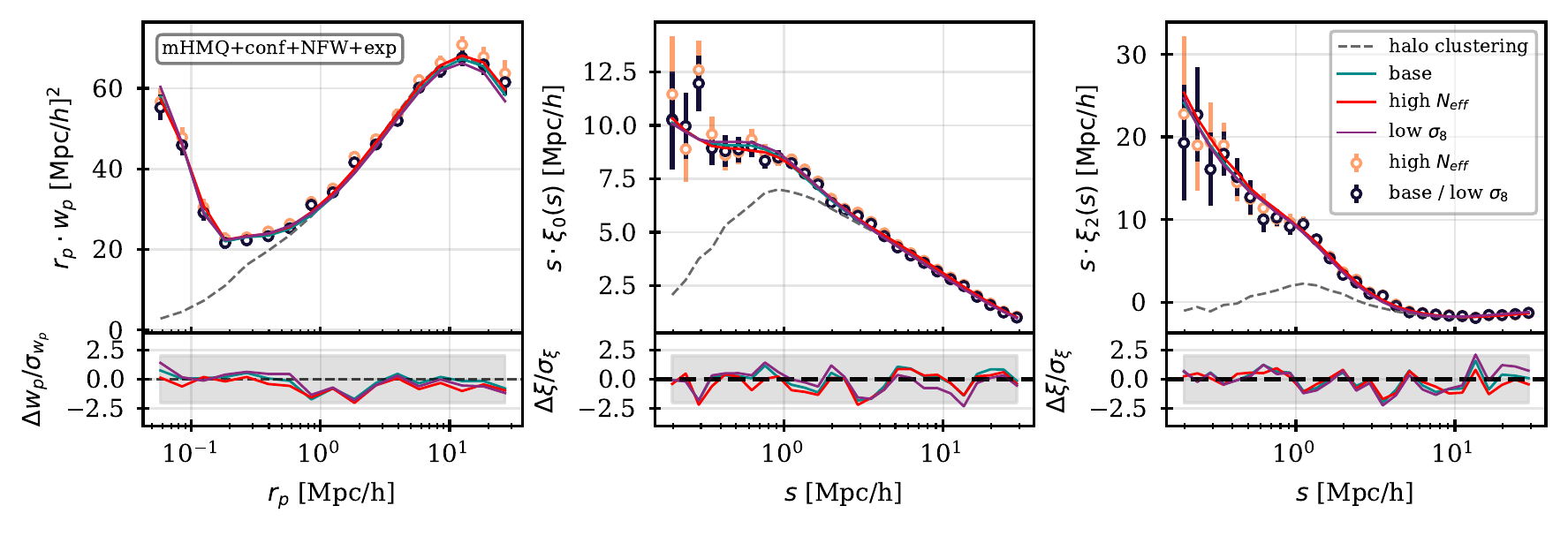}
\caption{\label{fig:clustering_cosmo} {\it Top:} DESI ELG clustering measurements from the One-Percent survey data sample in different cosmologies, high $N_{eff}$ (orange dots) and low $\sigma_8$ (dark blue dots). The distance-redshift relation in the low $\sigma_8$ cosmology is the same as in the baseline cosmology. Data are compared to best fitting HOD models obtained in the baseline (green), low $\sigma_8$ (purple) and high $N_{eff}$ cosmologies. The HOD model is the mHMQ model with strict conformity bias and our modified NFW profile for satellite positioning.  The dashed line is the pure halo clustering. Errors are jackknife uncertainties. {\it Top:} high $N_{\mathrm{eff}}$ cosmology {\it Bottom:} Fit residuals normalised by the diagonal
errors of the full covariance matrix (calculated for each cosmology), that comprise Jackknife uncertainties for the data as well
as stochastic noise and cosmic variance for the model, but no Hartlap factor corrections.}
\end{figure}

\begin{figure}[htbp]
\centering
\begin{tabular}{cc}
\includegraphics[width=0.45\textwidth]{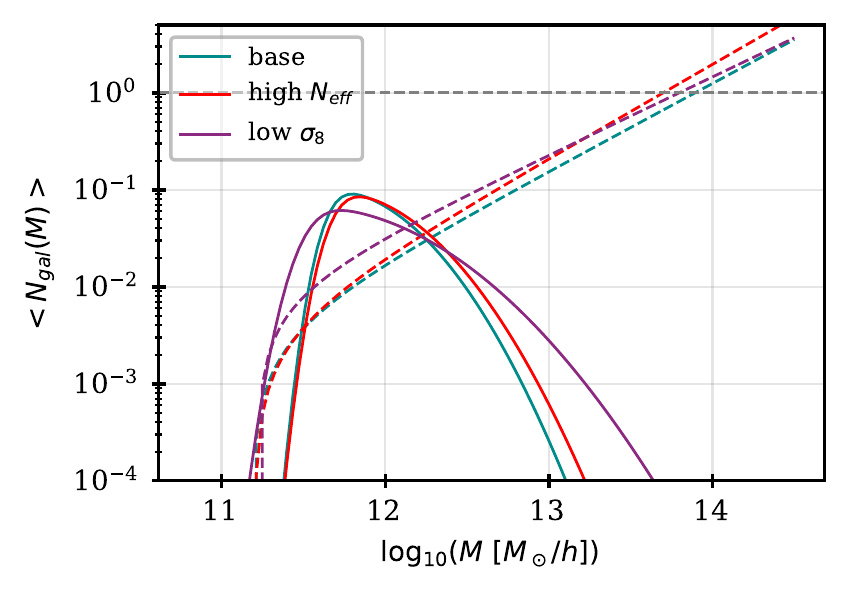}   &
\includegraphics[width=0.48\textwidth]{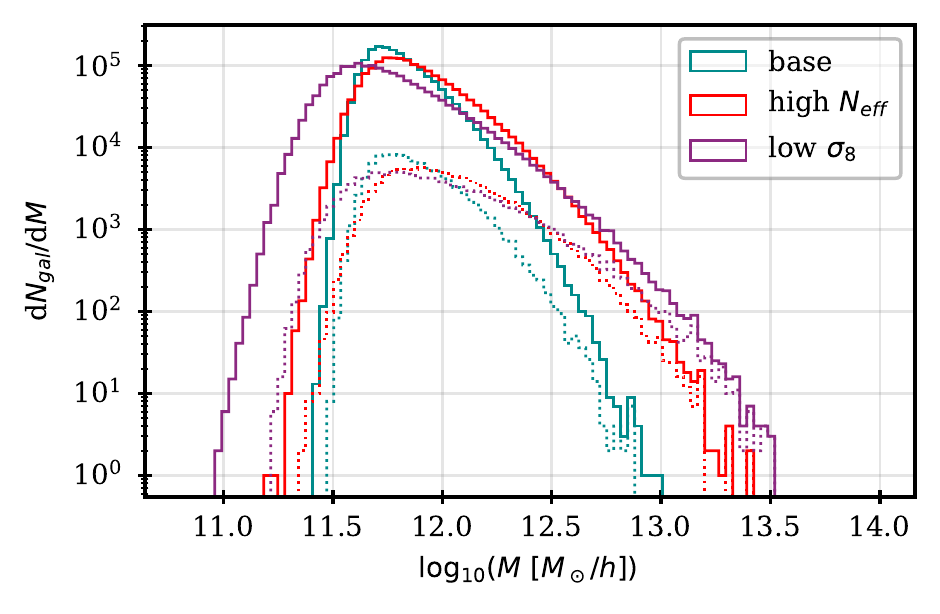}   \\
\end{tabular}
\caption{\label{fig:HMF_cosmo} 
{\it Left:} Best fitting HOD 
models to the DESI One-Percent ELG sample with strict conformity bias and our modified NFW profile for satellite positioning, obtained with different cosmologies: baseline (green), high $N_{eff}$ (red) and low $\sigma_8$ (purple). Solid (resp. dashed) lines represent central (resp. satellite) galaxies. The mHMQ prescription is used for centrals. 
{\it Right:} Number of galaxies per halo mass bin for halos populated according to the best fitting HOD models on the left. The simulation box volume is $1.66\,($Gpc$/h)^3$. The full distributions are in solid lines. The dashed lines show the contribution the one-halo component of the full distributions. }
\end{figure}

Best fitting results are presented in Figure~\ref{fig:clustering_cosmo} and summarised in Table~\ref{tab:LNHOD_cosmo}. Despite the change of cosmology, the data clustering can be modelled with similar goodness of fit as in the baseline cosmology, showing that the tested changes have a negligible impact on clustering. Changing the cosmology does not lead to significant changes for most HOD and derived parameters.
The largest changes are for $\log_{10} M_c$ and $f_{\sigma_v}$, with shifts between 1 and 2$\sigma$. For the derived parameters, both the satellite and one-halo fractions have consistent values. As a consequence of the variation of $\log_{10} M_c$, the mean halo mass, $\log_{10} \left\langle M_h \right\rangle $, varies by at most $2.7\sigma$ (0.08 dex) with the cosmological changes tested. 
Figure~\ref{fig:HMF_cosmo} shows the distribution of the number of galaxies per populated halo mass bin at best fit for the three cosmologies. 
The spread of the distribution is different in the three cases, the largest spread being observed for the low $\sigma_8$ cosmology. The baseline and low $\sigma_8$ cosmologies differ only by their values of the $\sigma_8$ parameter which has a direct impact on structure formation. A higher $\sigma_8$ value is expected to generate fewer small-mass halos and more large-mass halos at the same redshift and thus may explain the reduced spread at smaller halo masses for the baseline cosmology. At large mass, the spread evolves in the opposite direction to that expected from $\sigma_8$ values and may be governed more by the clustering to be modelled. The same argument holds for the $N_{eff}$ cosmology, although in that case, several other parameters have different values than in the baseline cosmology (lower $\Omega_{m}$ and higher $n_s, h$) which have a different effect on structure formation that can partly compensate for the effect of $\sigma_8$.

\begin{table}[htbp]
\centering
\begin{tabular}{|l|c|c|c|}\hline
parameter \& $\chi^2$ &  high $N_{\mathrm{eff}}$ & low $\sigma_8$ &  Planck 2018  \\ \hline
$A_c$ (resc.) & $0.1$ (0.63) &  $0.1$ (0.49) & $0.1$ (0.51) \\
$\log_{10} M_0$ & $11.20^{+0.17}_{-0.13}$  &  $11.22^{+0.12}_{-0.12}$ & $11.20^{+0.11}_{-0.09}$ \\
$A_s$  & $0.42^{+0.14}_{-0.11}$   & $0.46^{+0.11}_{-0.12}$ & $0.41^{+0.10}_{-0.15}$  \\ 
$\log_{10}M_{c}$ & $11.67^{+0.03}_{-0.02}$ & $11.51^{+0.02}_{-0.02}$ & $11.64^{+0.04}_{-0.04}$\\  
$\alpha$  & $0.97^{+0.12}_{-0.14}$ &$0.80^{+0.09}_{-0.11}$ & $0.81^{+0.08}_{-0.14}$\\ 
$f_{\sigma_v}$ & $1.40^{+0.09}_{-0.10}$ &  $1.65^{+0.13}_{-0.17}$ & $1.63^{+0.11}_{-0.10}$\\ 
$\sigma_M$  & $0.42^{+0.10}_{-0.08}$  & $0.59^{+0.09}_{-0.08}$ & $0.30^{+0.09}_{-0.07}$\\ 
$\gamma$  & $4.36^{+0.9}_{-0.88}$ & $5.02^{+1.18}_{-1.38}$ & $5.47^{+1.37}_{-1.58}$\\ 
$f_{exp}$ &  $0.57^{+0.07}_{-0.07}$ & $0.55^{+0.05}_{-0.05}$  &$0.58^{+0.06}_{-0.05}$ \\
$\tau$ & $6.01^{+1.07}_{-1.04}$ & $8.05^{+1.18}_{-1.62}$ & $6.14^{+1.11}_{-1.20}$\\
$\lambda_{NFW}$ & $0.64^{+0.06}_{-0.06}$ & $0.63^{+0.07}_{-0.06}$ & $0.67^{+0.06}_{-0.06}$\\
\hline
 $\log_{10} M'_{1}$ & 13.60  & 13.81 & 13.84 \\ 
$f_{sat}$   &$0.034^{+0.009}_{-0.009}$ & $0.034^{+0.008}_{-0.010}$ & $0.034^{+0.010}_{-0.012}$ \\
$f_{1h}$  &$0.067^{+0.018}_{-0.017}$ & $0.067^{+0.019}_{-0.016}$ & $0.069^{+0.020}_{-0.024}$\\
$\log_{10} \left\langle M_h \right\rangle $ &  $11.94^{+0.03}_{-0.03}$ & $11.84^{+0.02}_{-0.02}$ &$11.86^{+0.03}_{-0.03}$\\ \hline
$\chi^2$ (ndf=61) & $78.23 \pm 0.90$ & $93.80 \pm 0.83$ & $ 87.91\pm 0.84 $ \\  \hline 
\end{tabular} 
\caption {\label{tab:LNHOD_cosmo} Results of mHMQ fits with strict conformity bias between central and satellite galaxies in our baseline cosmology (right), in the high $N_{\mathrm{eff}}$ cosmology (left) and in the low $\sigma_8$ cosmology (middle). 
The first line provides the initial fixed value of $A_c$ and the rescaling factor applied to impose the density constraint in the fits. The following ten
parameters are the free HOD parameters, 
the next four are derived parameters.
$\log_{10}M_1'$ is given for best-fit values of $\alpha$ and $A_s$ (the latter after rescaling).
$f_{sat}$ is the fraction of galaxies which are satellite galaxies. $f_{1h}$ is the fraction of galaxies which are not alone in their halos. All masses are in units of $(M_{\odot}/h)$. 
}
\end{table} 

\section{Comparing to companion papers}
\label{sec:companion} 
Two companion papers studied the clustering of the One-Percent DESI ELG sample in the same redshift range as in the present paper, but with different methodologies, SHAM in~\cite{inclusiveSHAM} and a novel abundance matching method based on the stellar-halo mass relation (SHMR-AM) in~\cite{novelAM}. Despite differences in methodology, N-body simulation, reference cosmology, clustering statistics and separation ranges included in the analysis, their findings on 
the mean halo mass scale of the DESI ELG sample, $11.90\pm0.06$ in the SHAM paper and $\sim 12.07$ in the SHMR-AM one, agree with ours, $11.86^{+0.02}_{-0.01}$. 

The satellite fraction we find without central-satellite conformity - that is allowing for satellite ELG galaxies with no central ELG galaxy in their halo - is $12\%\pm2\%$. This result becomes $3.4\%\pm1.0\%$ with central-satellite conformity. 
Note that both companion SHAM analyses include satellite galaxies (living in subhalos) with no ELG central galaxy in the main halo, which is comparable to no central-satellite conformity. 
The SHMR-AM analysis uses measurements of $w_p$ above 0.1 Mpc$/h$ in $r_p$ and multipole measurements on scales above 0.3 Mpc$/h$ and measures a satellite fraction $\sim 15\%$, which is consistent with our result. The SHAM analysis uses multipole measurements on scales above 5 Mpc$/h$ and thus can only derive a predicted fraction of satellites. Their result is $3.4\%\pm2.0\%$, which does not agree with the above results, most probably as a result of too high a threshold on scales included in their fits. This high threshold also makes it impossible to achieve a good modelling of the $w_p$ up-turn at small-scales (see Figure C4 in~\cite{inclusiveSHAM}). Despite their using small-scale measurements in their fits, the SHMR-AM analysis also struggles to correctly reproduce the $w_p$ clustering at the smallest scales (see Figure 11 in~\cite{novelAM}). Work is underway to include central-satellite conformity in the SHMR-AM analysis, which should improve the results. 

\begin{figure}[thbp]
\centering
\includegraphics[width=0.8\textwidth]{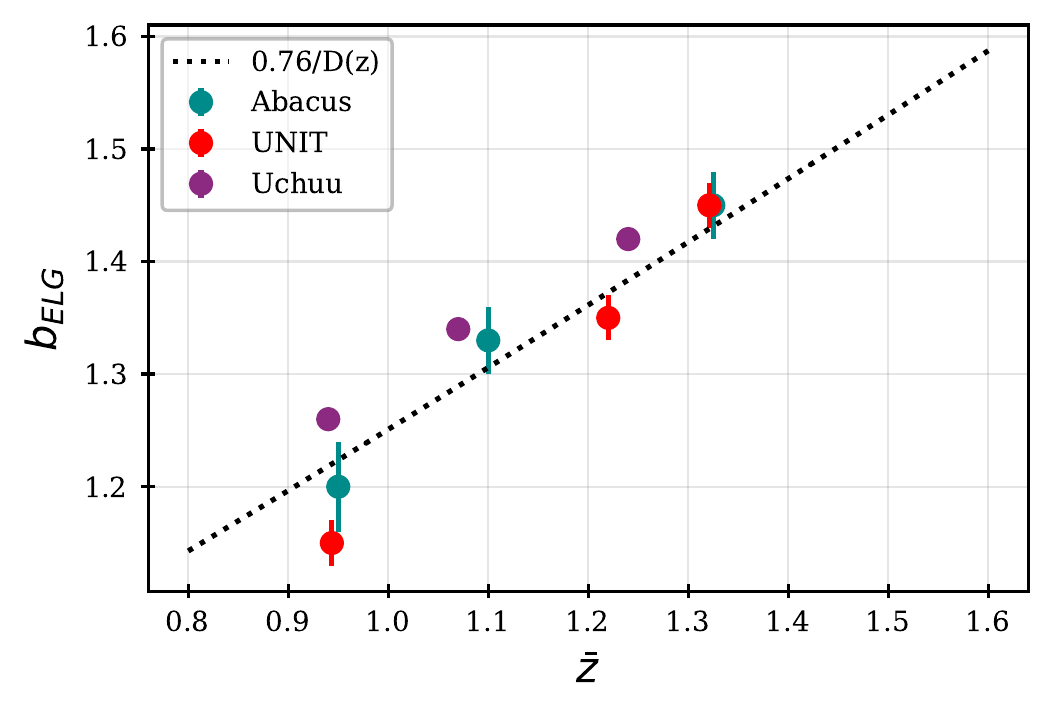}  \\
\caption{\label{fig:bias_z} 
Linear bias factor of the DESI One-Percent survey ELG sample as a function of redshift, as found in this paper (green dots with errors) and in two companion papers  which explored the galaxy-halo connection with a different methodology (red diamonds
from~\cite{inclusiveSHAM} and blue dots from~\cite{Uchuu4tracers}). Errors for the green dots and red diamonds (resp. blue points) include (resp. do not include) statistical errors from the measured clustering. The dashed line is the predicted evolution of the inverse of the linear growth factor $D(z)$  (in the baseline cosmology of our paper) arbitrarily normalised.
 }
\end{figure}

Using our best fitting mHMQ model with strict conformity bias and our modified NFW profile, we also compute the predicted linear bias factor of the DESI One-Percent ELG galaxy sample. To do so, we produce 
100 mocks with HOD parameters randomly selected in the MCMC chains at the fit final iteration
, convert them to real-space and compare the 2PCF from these mocks to the predicted real space 2PCF from linear theory (at the same cosmology), which are related by the squared value of the linear bias factor of the galaxy sample:
\begin{equation}
    \xi^{r}_{mocks}(s) = b^2 \xi^{r}_{linear}(s)
\end{equation}
Using this equation for $s$ between 40 and 80 Mpc/h, we fit the value of $b$ for each mock and average them over all mocks. In order to propagate the uncertainties from the measured clustering and the fitting methodology (which are reflected in the pool of HOD parameter values used to produce the mocks), the dispersion over the mocks is taken as the error on the reported value of b.
Our results are presented in Figure~\ref{fig:bias_z} as a function of the redshift of the simulation snapshot used for the modelling. 
We find the following values: $b_{0.95}=1.20^{+0.04}_{-0.04} $ for the low redshift bin, $b_{1.1}=1.33_{-0.03}^{+0.03}$ for the complete redshift bin and $b_{1.325}=1.45^{+0.03}_{-0.03} $ for the high redshift bin.
We also indicate the evolution with redshift of the inverse of the linear growth factor, with arbitrary normalisation. The bias deduced from our HOD study 
has an evolution consistent at the $1\sigma$ level with that of the growth factor.

Figure~\ref{fig:bias_z} also presents the results derived in two companion papers, both SHAM analyses, the first one already mentioned~\cite{inclusiveSHAM} based on the \textsc{UNIT} simulation, and the second one~\cite{Uchuu4tracers} using the \textsc{UCHUU} simulation and the ELG data sample restricted to the redshift range between 0.8 and 1.34. Note that in the latter case, the reported errors are errors on the mean bias measured from a set of best-fit SHAM lightcones and thus do not include clustering measurements errors from data. Despite the differences between the analyses already outlined at the beginning of this section, the predictions with error bars are in reasonable agreement. The set of results with incomplete error bars provides a qualitative cross-check. 


\section{Conclusions}
\label{sec:conclusion}

The sample of $\sim 270k$ ELGs collected by the DESI One-Percent survey in the redshift range between 0.8 and 1.6 (average redshift of 1.13) is used to study the ELG small-scale clustering in the HOD framework. Thanks to the high completeness of the sample, the clustering measurements can be pushed down to scales never probed before in redshift space, 0.04 Mpc/$h$ in $r_p$ for the projected correlation function $w_p$ and 0.17 Mpc$/h$ in separation $s$ for the two even multipoles of the 2PCF. A strong one-halo signal is observed at the smallest scales, below 0.2 Mpc$/h$ in $r_p$ and below 1 Mpc$/h$ in $s$. To correctly model the strong one-halo term signal requires putting close pairs of galaxies in small-mass halos. 

For central galaxies, we consider different prescriptions, a pure Gaussian distribution and three asymmetric ones, the strongest skewness being achieved with a log normal distribution. For satellites, we use a standard power law and do not require the presence of a central galaxy to put a satellite in the halo. Satellite positioning follow a NFW profile with a cut-off set at the halo virial radius, and we allow for velocity dispersion biased w.r.t that of the halo dark matter particles. Several extensions of these models are also explored.

In our baseline settings, whatever the different prescriptions for the central HOD, we achieve a good modelling of the measured clustering down to the smallest scales but obtain satellite HODs that decrease at large halo mass, contrary to expectations from semi-analytical ELG models. 
We recover satellite occupation distributions that agree with expectations if we introduce central-satellite conformity, that is if we require that satellite occupation is conditioned by the presence of central galaxies of the same type.

With or without conformity, whatever the  prescriptions for central HOD, satellite velocity dispersion and secondary biases, when the standard NFW profile is used for satellites, our modelling of the measured clustering, although good, exhibit residuals with a reproducible pattern between 0.1 and 1 Mpc$/h$, showing that extra pairs of galaxies are lacking in our predictions for this region.
A much better modelling is obtained with a modified NFW profile, allowing for ELG positioning outside of the halo virial radius, following a decreasing exponential law. With this prescription, we find that the measured ELG clustering clearly indicates that around $0.5\%$ of ELGs reside in the outskirts of halos. The significant improvement in the goodness of fit with the modified satellite profile leaves the other parameters of the HOD modelling unchanged.

Moreover, with or without conformity, and whatever the model for central galaxies, we find that the satellite velocity dispersion must be enhanced w.r.t. that of dark matter particles to correctly reproduce the measured clustering. We show that this modelling cannot be disentangled from a coherent satellite infall velocity inside halos. The velocity bias reaches $\sim1.6$ when our modified NFW profile for satellite positioning is used, and $\sim1.3$ otherwise.
Note that an increased velocity dispersion is coherent with the picture of ELGs residing in the outskirts of halos as recently-accreted sub-halos in these regions are expected to have higher velocities than the virial velocity of the halo.

The above findings are the main results of our work. With our best fitting HOD modelling, that is with central-satellite conformity, an extended NFW profile for satellite positioning and satellite velocity bias, the average halo mass of the ELG sample is $\log_{10} \left\langle M_h \right\rangle\sim 11.9$, the linear bias factor at a redshift of 1.1 is $\sim1.3$ and the fraction of galaxies which are not alone in their halos (the so-called one-halo component) is $\sim 7\%$. The fraction of satellites is $\sim 3\%$ but is highly dependent on the details of the HOD modelling, and would be $\sim 12\%$ without central-satellite conformity. 

We also investigate secondary biases and do not observe significant differences in our results when allowing for assembly bias as a function of halo concentration, local density or local density anisotropies. Although we report a slight improvement in the $\chi^2$ value for assembly bias as a function of halo concentration, this effect has a small impact on clustering statistics (almost indistinguishable).

Splitting the ELG sample in two redshift bins, from 0.8 to 1.1 and 1.1 to 1.6 moderately changes the HOD and derived parameters. We do see a slight change across redshift in terms of halo mass populated with ELGs (0.08 dex), which we do not consider as significant.

The above results are obtained using simulation boxes from the \absum suite generated at the baseline Planck 2018 cosmology but we investigate two other cosmologies, with higher $N_\mathrm{eff}$ and lower $\sigma_8$ values respectively. These moderate change in the simulation cosmology have no significant impact on the one-halo term fraction and most HOD parameters, except for $\log_{10} M_c$ and $f_{\sigma_v}$, and thus for the predicted average halo mass of the sample which varies at most by 0.08 dex, which again cannot be considered as significant. This effect may be related to the different $\sigma_8$ values in the three cosmologies tested. However, despite the change of cosmology, the data clustering can be modelled with similar goodness of fit. 

Finally, in the DESI framework, this study will be used to generate a large suite of accurate DESI-like mocks, varying the HOD models. These mocks will be useful to study the impact of observational systematic effects, test the corresponding mitigation algorithms and
to study the impact of the complexity of galaxy formation and evolution on cosmological inference.

\section*{Data availability}
All the material needed to reproduce the figures of this publication is available at the following address: \url{https://doi.org/10.5281/zenodo.7993413}

\acknowledgments

This material is based upon work supported by the U.S. Department of Energy (DOE), Office of Science, Office of High-Energy Physics, under Contract No. DE–AC02–05CH11231, and by the National Energy Research Scientific Computing Center, a DOE Office of Science User Facility under the same contract. Additional support for DESI was provided by the U.S. National Science Foundation (NSF), Division of Astronomical Sciences under Contract No. AST-0950945 to the NSF’s National Optical-Infrared Astronomy Research Laboratory; the Science and Technology Facilities Council of the United Kingdom; the Gordon and Betty Moore Foundation; the Heising-Simons Foundation; the French Alternative Energies and Atomic Energy Commission (CEA); the National Council of Science and Technology of Mexico (CONACYT); the Ministry of Science and Innovation of Spain (MICINN), and by the DESI Member Institutions: \url{https://www.desi.lbl.gov/collaborating-institutions}. Any opinions, findings, and conclusions or recommendations expressed in this material are those of the author(s) and do not necessarily reflect the views of the U. S. National Science Foundation, the U. S. Department of Energy, or any of the listed funding agencies.

The authors are honored to be permitted to conduct scientific research on Iolkam Du’ag (Kitt Peak), a mountain with particular significance to the Tohono O’odham Nation.

\bibliographystyle{JHEP}
\bibliography{references}{}

\begin{appendices}
\section{Proxies for $r_s$ and $r_{vir}$ in the NFW profile}
\label{app:proxies}

We further discuss our proxy choice for $r_s$ and $r_{vir}$ in the NFW profile used in our analysis. Figure~\ref{fig:Rs proxy} shows the predicted projected 2-point correlation function $w_p$ on scales $r_p < 0.4$ Mpc/$h$ for the same HOD model, changing the proxy for $r_{vir}$ and $r_s$. For $r_{vir}$, we test two different choices, either $r_{98}$, the radius of a sphere enclosing 98\% of the halo particles and $r_{so}$, the radius of a sphere containing the total halo mass $M_{vir}$, computed as the sum of the halo particle masses and 
expressed as an overdensity~$\Delta$: 
\begin{equation}
    \label{eq:Rvir}
    r_{so} \equiv \left(\frac{3}{4\pi} \frac{M_{vir}}{\Delta\rho_c(z) }\right)^{1/3}
\end{equation}
where $\rho_c$ is the critical density. The overdensity is provided for each \absum  snapshot, e.g. for the snapshot corresponding to the effective redshift $z=1.1$ of the ELG sample, $\Delta = 223$. For the $r_s$ proxy, we use the radius $r_x$ of a sphere encompassing different percentages of the halo particles, with $x = 50, 33, 25$ and $10 \%$. We compare the above predictions to that from a particle based mock (where the satellite assignment is based on particles inside the halo) for the same HOD model. The shaded grey region represents the $\pm1\sigma$ measurement error for the actual DESI ELG sample in the redshift range between  0.8 and 1.6. 
From this comparison, the proxy that best reproduces the particle based mock corresponds to $r_{vir}=r_{98}$ and $r_s=r_{25}$.  
 
\begin{figure}[htbp]
\centering
\begin{tabular}{cc}
\includegraphics[width=0.45\textwidth, height=6cm]{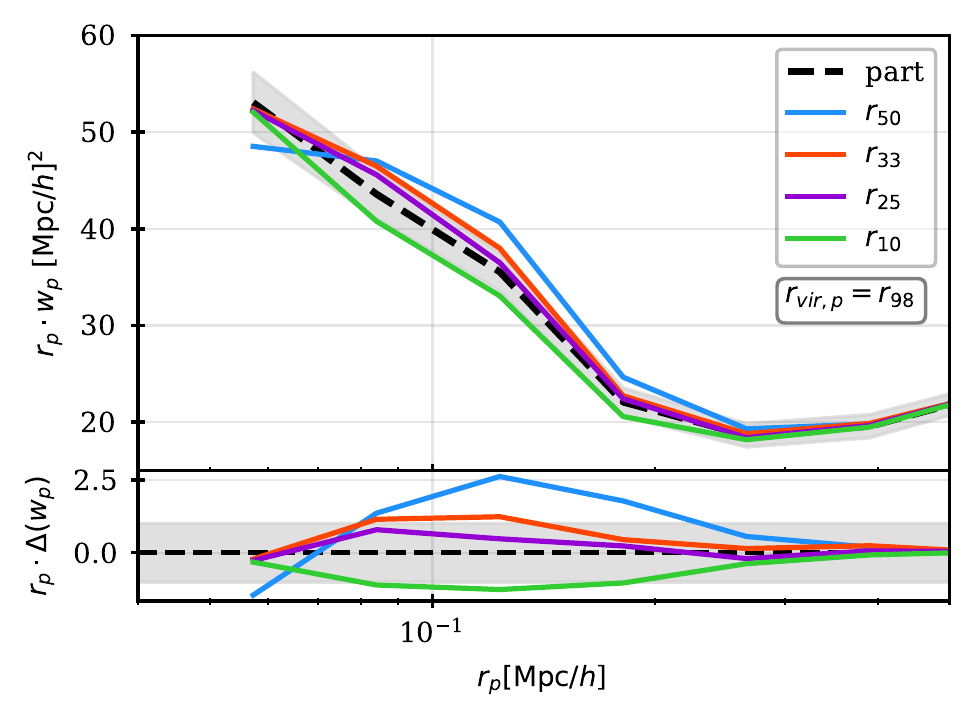}   &
\includegraphics[width=0.45\textwidth, height=6cm]{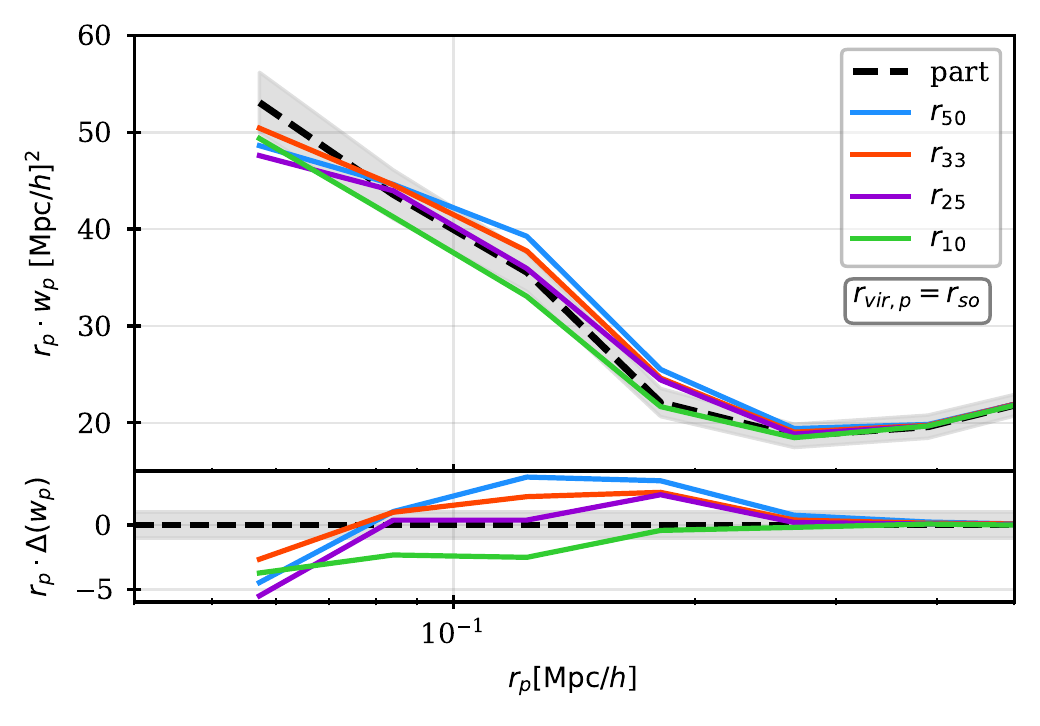}   \\
\end{tabular}
\caption{\label{fig:Rs proxy}
{\it Top:} Predicted $w_p$ clustering on scales $r_p < 0.4$ Mpc/$h$ for the same HOD model, using as a proxy for $r_{vir}$ either $r_{98}$ (left) or $r_{so}$ (right). Predictions for different proxies for $r_s$, corresponding to the radius of a sphere that contains $50, 33, 25$ and $10\%$ of the halo particles (in blue, red, purple and green, respectively) are compared to the clustering of one mock where the satellite assignment is based on DM particles (dashed black line). {\it Bottom:} $w_p$ difference between mocks with different $r_s$ proxies and the particle based mock, multiplied by $r_p$. The shaded grey area corresponds to the $\pm1\sigma$ error of DESI data as shown in Figure~\ref{fig:weights}.}
\label{fig:Rs proxy}
\end{figure}  

\section{Contour plots of the mHMQ fits}
\label{app:contours}
\begin{figure}[htbp]
\centering
\includegraphics[width=0.9\textwidth]{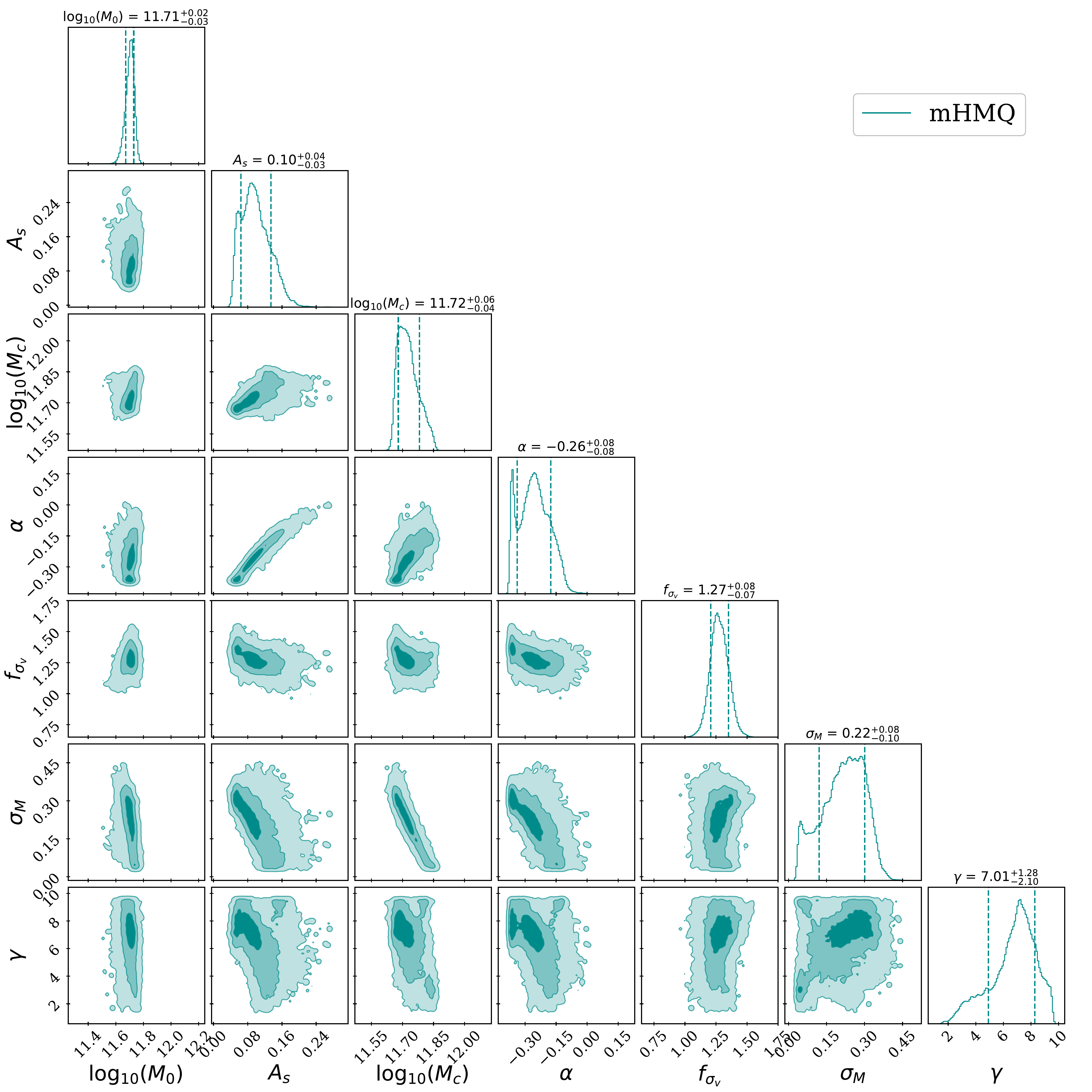}    
\caption{\label{fig:contour_mHMQ}  Contours  (at 1,2 and 3$\sigma$ level)
and marginalised 1D posteriors at final iteration obtained in the GP pipeline for the mHMQ fit  to the One-Percent DESI survey ELG data for the whole redshift bin $0.8<z<1.6$, without conformity bias between central and satellite galaxies.}
\end{figure}

\begin{figure}[htbp]
\centering
\includegraphics[width=0.9\textwidth]{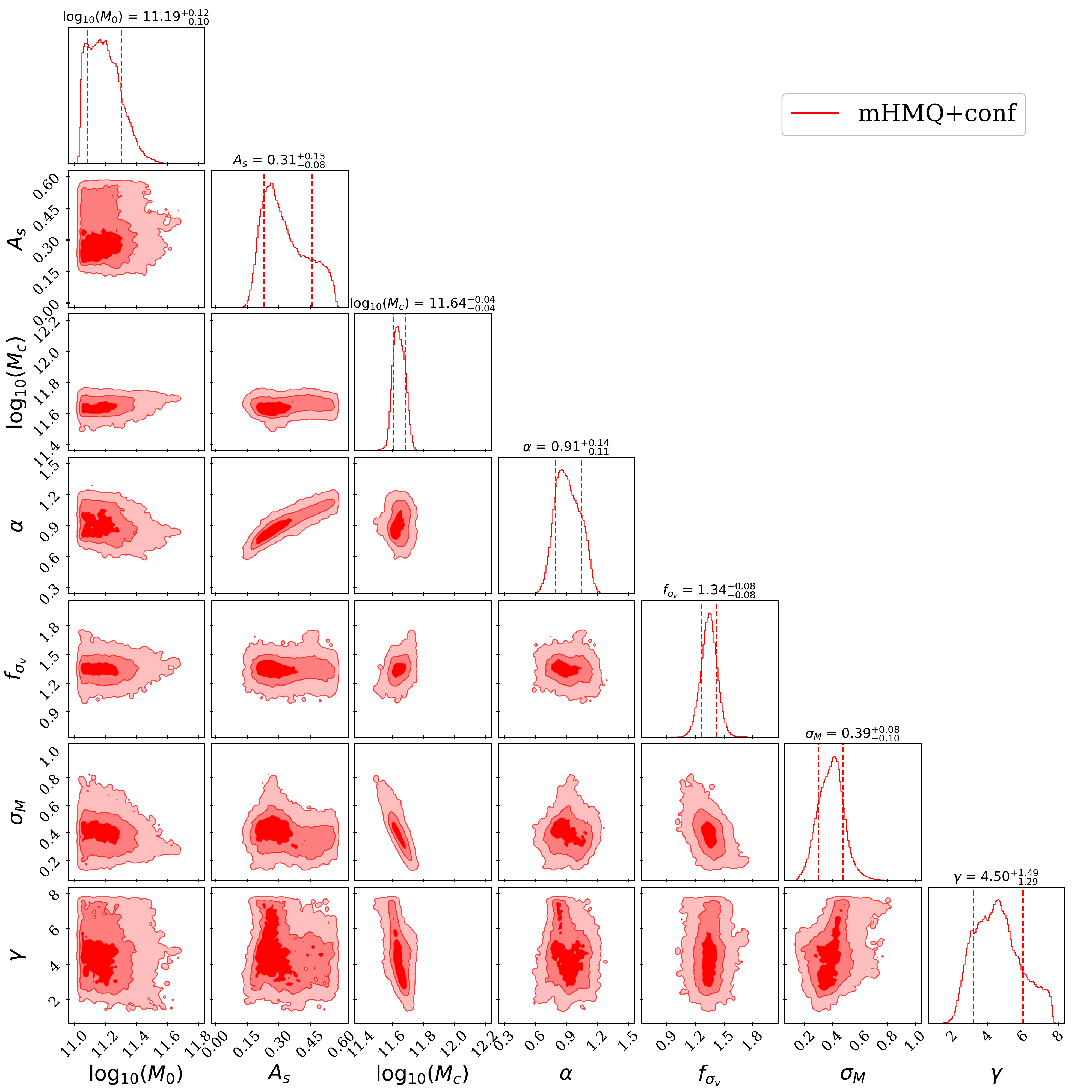}    
\caption{\label{fig:contour_mHMQ+conf} Same as Figure \ref{fig:contour_mHMQ} for the mHMQ model with strict conformity.}
\end{figure}

Figures~\ref{fig:contour_mHMQ} and~\ref{fig:contour_mHMQ+conf} show the contours obtained at final iteration in the Gaussian Process (GP) pipeline for mHMQ with and without strict conformity bias fits to the DESI One-Percent Survey ELG sample. Most contours are well enclosed in our prior ranges. The notable exception is $\gamma$ and $\log_{10} M_0$ for the conformity case. For $\log_{10} M_0$, the prior range is limited by the minimum halo mass available in the simulations, 
10.86 and the fact that $\log_{10} M_0$ is not constrained if its value is below the minimum mass of halos that can be populated with central galaxies. $\gamma$ is degenerated with $\sigma_M$ and has a weak impact on the shape of the HOD compared to $\sigma_M$.

The parameters we constrain the most are  $\alpha$ and its degeneracy with $A_s$,  $\log_{10} M_c$ and its degeneracy with $\sigma_M$, $f_{\sigma,v}$ and $\log_{10} M_0$ (only for the case without conformity for the latter two parameters).

\begin{figure}[htbp]
\centering
\includegraphics[width=\textwidth]{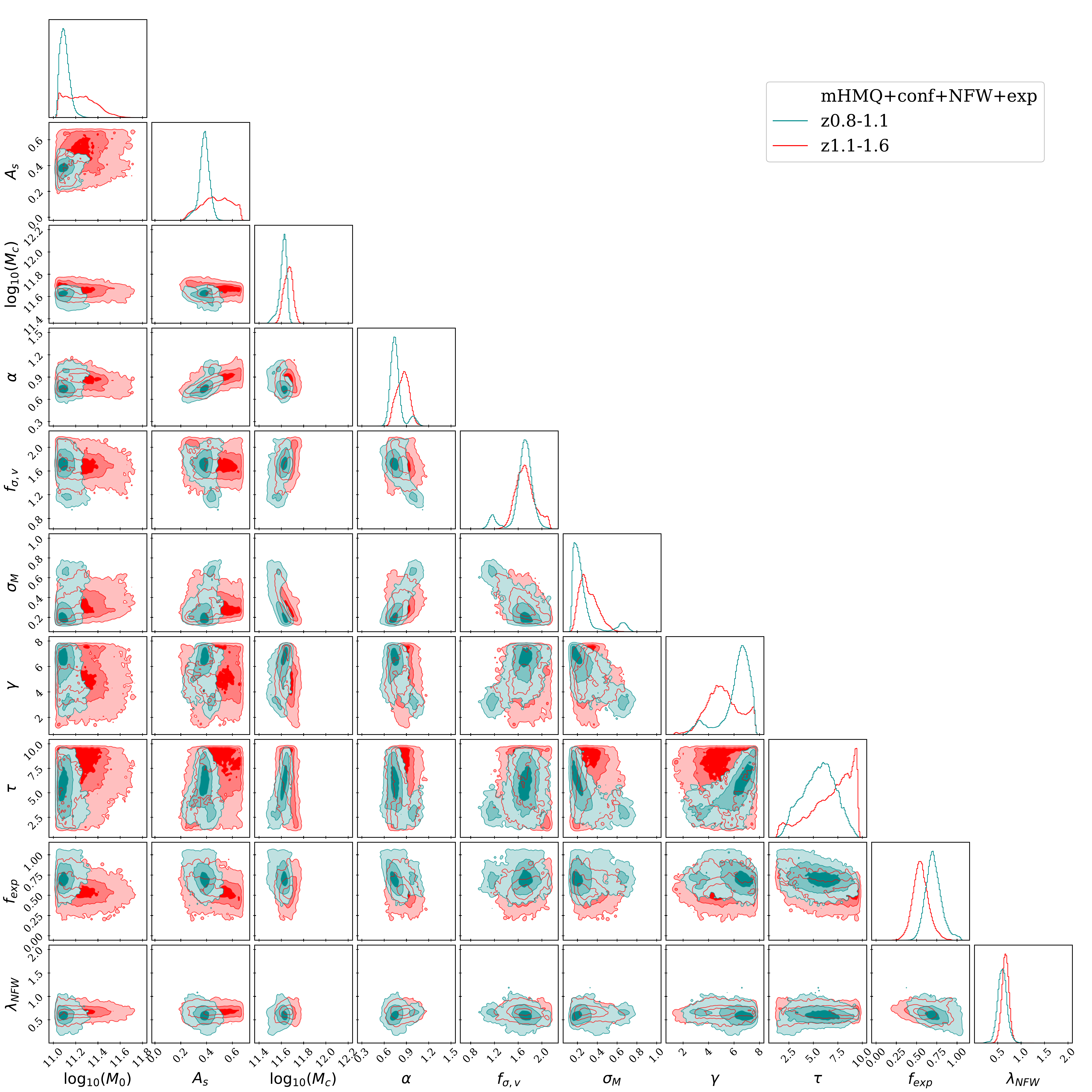}    
\caption{\label{fig:contour_zbins_mHMQ+conf} Contours (at 1,2 and 3$\sigma$ level) and marginalised 1D posteriors at final iteration obtained in the GP pipeline for the mHMQ fits to the One-Percent DESI survey ELG data with redshifts between $0.8<z<1.1$ in green and $1.1<z<1.6$ in red. The mHMQ model in this plot has strict conformity bias between central and satellite galaxies and a modified NFW profile for satellite positioning. This figure shows the small evolution of the HOD parameters with redshift.}
\end{figure}

Figure~\ref{fig:contour_zbins_mHMQ+conf} shows the contours obtained at final iteration in the Gaussian Process (GP) pipeline for mHMQ fits to the DESI One-Percent Survey ELG sample in the two redshift bins considered in this paper, $0.8<z<1.1$ and $1.1<z<1.6$. Strict conformity is applied as well as our modified NFW profile for satellite positioning. The HOD parameters are well constrained in the lower redshift bin, while the constraints are less stringent in the higher bin, where we  
constrain only $\log_{10} M_c$ and its degeneracy with $\sigma_M$, $\alpha$ and its degeneracy with $A_s$, $f_{\sigma_v}$ and two of the satellite profile parameters, $f_{exp}$ and $\lambda_{NFW}$.

\end{appendices}

\end{document}